\documentclass[fleqn,twoside]{article}
\usepackage{amsmath,amssymb,espcrc2,color,cite}

\usepackage{graphicx,soul}
\usepackage[figuresright]{rotating}

\newcommand{\url}[1]{{\tt #1}}

\newcommand{\gmt}{\ensuremath{(g-2)_\mu}}
\newcommand{\br}{{\rm BR}}
\newcommand{\bsg}{BR($b \to s \gamma$)}

\newcommand{\bmm}{\ensuremath{\br(B_s \to \mu^+\mu^-)}}
\newcommand{\ssi}{\ensuremath{\sigma^{\rm SI}_p}}
\newcommand{\ssd}{\ensuremath{\sigma^{\rm SD}_p}}
\newcommand{\Och}{\ensuremath{\Omega_\chi h^2}}

\newcommand{\MW}{M_W}

\newcommand{\Mh}{M_h}

\newcommand{\MA}{M_A}

\newcommand{\mt}{m_t}
\newcommand{\mgl}{m_{\tilde g}}

\newcommand{\neu}[1]{\tilde \chi^0_{#1}}
\newcommand{\mneu}[1]{m_{\tilde \chi^0_{#1}}}

\newcommand{\tb}{\tan\beta}

\newcommand{\gev}{\,\, \mathrm{GeV}}
\newcommand{\mev}{\,\, \mathrm{MeV}}

\definecolor{Orange}{named}{Orange}
\definecolor{Purple}{named}{Purple}

\newcommand{\ETslash}{/ \hspace{-.7em} E_T}
\newcommand{\pTslash}{/ \hspace{-.7em} p_T}

\graphicspath{{figs/}}

\title{\bf Supersymmetry and Dark Matter in Light of LHC 2010 and Xenon100 Data
 \\ \vspace{0.5em}}

\author{
{\bf O.~Buchmueller}\address[Imperial]
   {High\,Energy\,Physics\,Group,\,Blackett\,Laboratory,\,Imperial\,College,\,Prince\,Consort\,Road,\,London\,SW7\,2AZ,\,UK},
{\bf R.~Cavanaugh}\address[FNAL]
   {Fermi National Accelerator Laboratory, P.O. Box 500, 
    Batavia, Illinois 60510, USA}\hbox{$^{\rm ,}$}\address[UIC]
   {Physics Department, University of Illinois at Chicago, Chicago, 
    Illinois 60607-7059, USA},
{\bf D.~Colling}\addressmark[Imperial],
{\bf A.~De Roeck}\address[CERN]
   {CERN, CH--1211 Gen\`eve 23, Switzerland}\hbox{$^{\rm ,}$}\address[Antwerpen]
   {Antwerp University, B--2610 Wilrijk, Belgium},
 {\bf M.J.~Dolan}\address[IPPP]
{Institute for Particle Physics
     Phenomenology,\,University\,of\,Durham,\,South 
     Road,\,Durham\,DH1\,3LE,\,UK},
{\bf J.R.~Ellis}\addressmark[CERN]\hbox{$^{\rm ,}$}\address[KCL]{Theoretical Physics
  and Cosmology Group, Department of Physics, King's College London, London
  WC2R 2LS, UK}, 
{\bf H.~Fl\"acher}\address[Rochester]
   {Department of Physics and Astronomy, University of Rochester, 
    Rochester, New York 14627, USA},
{\bf S.~Heinemeyer}\address[Santander]
   {Instituto de F\'{\i}sica de Cantabria (CSIC-UC), 
    E--39005 Santander, Spain},
{\bf G.~Isidori}\address[Frascati]
   {INFN, Laboratori Nazionali di Frascati, Via E. Fermi 40, 
    I--00044 Frascati, Italy},
{\bf D.~Mart\'inez~Santos}\addressmark[CERN],
{\bf K.A.~Olive}\address[Minnesota] 
   {William\,I.\,Fine\,Theoretical\,Physics\,Institute,\,University\,of\,Minnesota,\,Minneapolis,\,Minnesota\,55455,\,USA},
{\bf S.~Rogerson}\addressmark[Imperial],
{\bf F.J.~Ronga}\address[ETHZ]
   {Institute for Particle Physics, ETH Z\"urich, CH--8093 Z\"urich, 
   Switzerland},
{\bf G.~Weiglein}\address[DESY]
   {DESY, Notkestrasse 85, D--22607 Hamburg, Germany}
}

\begin{document}

\begin{abstract}
We make frequentist analyses of the CMSSM, NUHM1, VCMSSM and mSUGRA
parameter spaces taking into account all the public results of searches for
supersymmetry using data from the 2010 LHC run and the Xenon100 direct search
for dark matter scattering. The LHC data set includes ATLAS and CMS searches for
jets + $\ETslash$ events (with or without leptons) and for the heavier
MSSM Higgs bosons, and the upper limit on \bmm\ including data from LHCb
as well as CDF and D\O. 
The absences of signals in the LHC data favour somewhat heavier mass
spectra than in our previous analyses of the CMSSM, NUHM1 and VCMSSM,
and somewhat smaller dark matter scattering cross sections, all close to or within
the pre-LHC 68\% CL ranges, but do not impact significantly the favoured
regions of the mSUGRA parameter space. We also discuss the impact of the
Xenon100 constraint on spin-independent dark matter scattering, stressing
the importance 
of taking into account the uncertainty in the $\pi$-nucleon $\sigma$
term $\Sigma_{\pi N}$, that affects the spin-independent scattering
matrix element, and we make predictions for spin-dependent dark matter scattering.
Finally, we discuss briefly the potential impact of the
updated predictions for sparticle masses in the CMSSM, NUHM1,
VCMSSM and mSUGRA on future $e^+e^-$ colliders.

\bigskip
\begin{center}
{\tt CERN-PH-TH/2011-129, DCPT/11/62, DESY 11-119, IPPP/11/31, \\
FTPI-MINN-11/12, KCL-PH-TH/2011-12, LCTS/2011-01, UMN-TH-3002/11}
\end{center}
\vspace{2.0cm}
\end{abstract}

\maketitle

\section{Introduction}
\label{sec:intro}

One of the most appealing possible extensions of the Standard Model (SM)
is supersymmetry (SUSY)~\cite{Haber:1984rc}.
It would stabilize the electroweak mass hierarchy and facilitate
grand unification, it predicts a relatively light Higgs boson that would
be consistent with the indications from precision electroweak data, it
offers a possible explanation of the apparent discrepancy 
between the experimental measurement of the anomalous magnetic
moment of the muon, \gmt, and the theoretical value calculated
within the SM, and the lightest supersymmetric particle (LSP) is a
plausible candidate for astrophysical  dark matter.

We have published results from frequentist analyses
of the minimal supersymmetric extension of the Standard Model (MSSM), 
using likelihood functions to take into account the
experimental, phenomenological and astrophysical constraints on SUSY. 
The unconstrained MSSM contains too many 
parameters for a full exploration of its parameter space to be possible
using present data, even including the current LHC data set of 
$\sim 35$/pb~\cite{mc5}.  
Therefore, we have focused on making estimates
within simplified versions of the MSSM,
specifically the constrained MSSM (the CMSSM)~\cite{cmssm1,mc1} in which soft
SUSY-breaking mass parameters are assumed to be universal at
the GUT scale, in the simplest generalization of this model in which the
universality is relaxed to allow non-universal Higgs masses
(the NUHM1)~\cite{nuhm1,mc2,mc3}, in a very constrained model in which the
supplementary relation $A_0 = B_0 + m_0$~%
\footnote{We recall that our convention~\cite{mc4,mc5} for the sign of $A_0$ is opposite to
that of {\tt SoftSUSY}.}
is imposed on trilinear and bilinear soft
SUSY-breaking masses in the CMSSM (the VCMSSM) \cite{vcmssm,mc4}, and in
minimal supergravity 
(mSUGRA) in which, in addition, the gravitino mass $m_{3/2}$ is set
equal to the common soft SUSY-breaking scalar  mass $m_0$
before renormalization~\cite{vcmssm,mc4}. In each case, we assume that the
LSP is the lightest neutralino $\neu{1}$.
More details on the model definitions can be found in~\cite{mc4}.

In a series of papers~\cite{mc1,mc2,mc3,mc4}~%
we have presented predictions for Higgs and
sparticle masses as well as for \bmm\ and the spin-independent dark
matter scattering cross section, $\ssi$,
and also for $\mt$ and 
$\MW$~\cite{mc35}. Most recently~\cite{mc5} we have included in global analyses the
results of an initial CMS search in multijet + $\ETslash$
channels (CMS $\alpha_T$)~\cite{cms0l-aT} and an ATLAS search in lepton
+ multijet + $\ETslash$ channels (ATLAS 1L)~\cite{ATLAS1l}.%
\footnote{Other analyses can be found in~\cite{otheranalyses},
where similar effects were found in the CMSSM and in gauge-mediated models.}
Incorporating these new results led
to upward shifts in the lower bounds on the gluino mass, $\mgl$, by 
$\sim 100 \gev$ in the models considered. Other masses connected
to $\mgl$, such as that of the lightest neutralino, $\mneu{1}$
(which we assume to provide the astrophysical cold dark matter
(CDM)~\cite{EHNOS}) 
also moved upward by corresponding amounts. This in turn led to somewhat
lower expectations for the spin-independent dark matter scattering
cross section $\ssi$ in the models considered~\cite{mc5}.

Subsequent to our analysis~\cite{mc5} of the implications of these initial
LHC searches for SUSY, LHC experiments have provided
several new constraints on SUSY using an integrated luminosity
of $\sim 35$/pb of data at 7~TeV. ATLAS has published the results of
a search in multijet + $\ETslash$ channels (ATLAS 0L)~\cite{ATLAS0l}
that has greater sensitivity in some regions to the types of gluino and squark 
pair-production events expected in the supersymmetric models discussed
here than did the earlier ATLAS 1L search~\cite{ATLAS1l}, and has also released results
obtained by combining the one- and zero-lepton searches~\cite{ATLAScombined}.
CMS has announced results from two other searches
in multijet + $\ETslash$ channels that improve the CMS $\alpha_T$
sensitivity also to gluino and squark 
production in the models discussed
here. Both ATLAS and CMS have also published the results
of searches for jets + $\ETslash$ events with $b$ tags~\cite{bjets},
and for multilepton + jets + $\ETslash$ events~\cite{leptons}.
In addition, CMS and ATLAS have published new upper limits on
the production of the heavier neutral MSSM Higgs bosons 
$H,A$~\cite{ATLASHA,cmsHA}, and LHCb has recently provided a new upper
limit on \bmm~\cite{lhcb-bmm}, of comparable sensitivity to previous 
results from CDF~\cite{cdf-bmm} and D\O~\cite{d0-bmm}. 

In parallel, the Xenon100 Collaboration has recently released results from a
search for direct spin-independent dark matter scattering with 100.9
live days of data using a fiducial 
target with a mass of 48~kg~\cite{Xenon100new}. As we see later, this provides 
constraints on the parameter spaces of supersymmetric models that complement
those provided by collider experiments~%
\footnote{See~\cite{Farina:2011bh} for discussions of the Xenon100 results in the context of various models including the CMSSM. Ref.~\cite{Profumo} compares LHC limits and the sensitivities of astrophysical searches for 
supersymmetric dark matter in specific CMSSM $(m_0, m_{1/2})$ planes for
fixed values of $\tan \beta$. Ref.~\cite{LMNW} discusses the interplay between Xenon100 and LHC searches
in the context of a no-scale flipped SU(5) model.}.

In this paper we combine these new
constraints in updated global frequentist analyses of the parameter
spaces of the CMSSM, NUHM1, VCMSSM and mSUGRA that take into account
the results of all the searches using 2010 LHC data as well as the new Xenon100
constraint on the spin-independent scattering cross section, $\ssi$. At each point in the parameter
spaces of these models, we construct a global likelihood
function using previous data on electroweak precision observables, \gmt\ and \bsg, and
applying the strongest of the new constraints
from searches for multijet + $\ETslash$ events, in combination with the constraints from
$H/A$ searches, \bmm\ and $\ssi$, via the implementations
described in the next section.

The ATLAS and CMS searches for multijet + $\ETslash$ events
provide  constraints in complementary regions of the $(m_0, m_{1/2})$
planes of these models, while the searches for heavier neutral MSSM
Higgs bosons provide a relevant constraint in the $(\MA, \tb)$
plane of the NUHM1. The LHCb search for \bmm, in combination with
the CDF and D\O\ searches, affects significantly
the likelihood function for this observable, with particular relevance for the
NUHM1. The 
best-fit points in our new fits including all these 2010 LHC
constraints and the limit from the Xenon100 experiment
are all close to or within the regions favoured by pre-LHC fits at the
68\% CL. The spectra are somewhat heavier in the cases of 
the CMSSM, NUHM1 and VCMSSM, whereas the best-fit mSUGRA 
spectrum is little changed. The Xenon100 upper limit on $\ssi$ has
little impact on the favoured regions of the VCMSSM and mSUGRA,
and the impact on the CMSSM and NUHM1 parameter spaces is
limited by the present experimental uncertainty in the hadronic
scattering matrix element, that is currently inherited primarily from the
uncertainty in the low-energy $\pi$-N $\sigma$ term, $\Sigma_{\pi N}$.
Based on the combination of 2010 LHC and Xenon100 constraints, we
present updated 
likelihood functions for sparticle masses and other observables
including $\mgl$, \bmm\ and $\ssi$. We also present predictions for
the spin-dependent scattering cross section, $\ssd$, that lie considerably
below the present experimental upper limits. Finally, as an offshoot of our analysis,
we discuss briefly the potential impact of our results on future $e^+e^-$
  colliders.

\section{Methodology}
\label{sec:meths}

Our analyses are performed using the 
{\tt MasterCode} framework~\cite{mc1,mc2,mc3,mc35,mc4,mc5,mc-web}. 
The analyses have been made in
a  frequentist approach, in which we construct a global likelihood
function with contributions from precision electroweak observables,
$B$-physics observables, \gmt\ and the astrophysical cold dark matter density
\Och\ as well as the limits from the direct LEP searches
for the Higgs boson and sparticles and, most recently, from sparticle searches at the LHC.
The model parameter spaces are sampled using 
Markov Chain Monte Carlo (MCMC) techniques described in our previous papers. 
Our previous MCMC samplings of the CMSSM and NUHM1 parameter spaces
each comprised some 25,000,000 points, whereas those of the VCMSSM and mSUGRA
include some 30,000,000 and 17,000,000 points, respectively.
For the purposes of this paper we have added a sample of some 5,000,000
CMSSM points with $m_0 < 600 \gev$ and $250 \gev < m_{1/2} < 800 \gev$,
designed to improve our understanding of the global likelihood function at
values of $m_{1/2}$ that are somewhat larger than the previous best-fit
values in our pre-LHC analysis of the CMSSM. 
This extra sampling had very little impact on our estimates of the
best-fit points and 68 and 95\% CL regions
extracted from the $\chi^2$ evaluation, confirming the adequacy of
our sampling in the parameter regions of interest.

The pre-LHC constraints are also treated similarly to our previous analyses,
see Ref.~\cite{mc5} for the most up-to-date description.
The numerical evaluation within the  
{\tt MasterCode}~\cite{mc1,mc2,mc3,mc35,mc4,mc5,mc-web},
combines
{\tt SoftSUSY}~\cite{Allanach:2001kg}~%
\footnote{In this paper we have upgraded from the version {\tt 2.0.11} used in earlier analyses to the
  new version {\tt 3.0.13}: we indicate below where this change
  affects our analysis.}%
, 
{\tt FeynHiggs}~\cite{Degrassi:2002fi,Heinemeyer:1998np,Heinemeyer:1998yj,Frank:2006yh},  
{\tt SuFla}~\cite{Isidori:2006pk,Isidori:2007jw},
{\tt SuperIso}~\cite{Mahmoudi:2008tp,Eriksson:2008cx}, 
a code for electroweak observables based 
on~\cite{Heinemeyer:2006px,Heinemeyer:2007bw} and
{\tt MicrOMEGAs}~\cite{Belanger:2006is} (with
{\tt DarkSUSY}~\cite{Gondolo:2005we} as an option not used in this paper), making
extensive use of the SUSY Les Houches
Accord~\cite{Skands:2003cj,Allanach:2008qq}.
The predictions we make for  \bmm\ using  {\tt MasterCode}
are checked for specific fit parameters using the
independent {\tt SSARD} code~\cite{SSARD}. In the analysis of $\ssi$
in this paper, we link a part of {\tt SSARD} to
{\tt MasterCode} to  take account of hadronic uncertainties in dark matter
scattering matrix elements, making cross-checks with {\tt MicrOMEGAs}.

The {\tt MasterCode} is designed in such a way that the constraints from
new observables can be taken into account and incorporated quickly and 
easily into the global likelihood function as `afterburners',
i.e., by adding the calculated contribution to the likelihood function
from the new 
observable and subsequently re-evaluating the global $\chi^2$ function.
The new ingredients in this analysis coming from 2010 LHC and other searches
are incorporated as just such `afterburners', via the implementations
described below.

\section{Implementations of 2010 LHC and other Constraints}
\label{sec:implems}

Studies by the LHC Collaborations have shown that multijet + $\ETslash$
constraints, with or without a single lepton, are relatively insensitive
to $\tb$ and $A_0$.  
Accordingly, we treat the ATLAS and CMS constraints on such signatures as independent
of $\tb$ and $A_0$, and regard their constraints in the $(m_0, m_{1/2})$ plane as
`universal'~\cite{ATLAS0l,ATLAS1l,cms0l-aT,cmstb}. At each point in this plane, we compare the strengths of
these ATLAS and CMS constraints, and
retain the stronger, not attempting to combine the constraints 
from different experiments.

The constraints due to CMS and ATLAS searches
for events containing two or more leptons~\cite{leptons} are in general less
sensitive than the constraints due to events with jet + $\ETslash$ and at most one lepton, in the models
considered here, and hence are not relevant for our evaluation of the global likelihood function.
Moreover, these searches including leptons are also more sensitive to  the value of $\tb$,
as are searches using $b$ tags~\cite{bjets}. 
Since the reaches of the latter searches
do not exceed those of the pure
multijet + $\ETslash$ searches, even at large $\tan \beta \sim 50$,
they also do not contribute to the global likelihood
function~\footnote{We note
in passing that LEP and Tevatron searches for sparticle pair-production also
do not contribute significantly to the global likelihood function, whereas the LEP
search for the lightest MSSM Higgs boson does contribute significantly.}.

\subsection*{\it ATLAS jets + $\ETslash$ + 0, 1 lepton analyses}

We treat the ATLAS analyses of events with multiple jets, zero or one lepton and $\ETslash$ (ATLAS 0L, 
ATLAS1L)~\cite{ATLAS0l,ATLAS1l,ATLAScombined} as follows. 
ATLAS reports the combined results of these searches as a 95\% CL
exclusion contour in the $(m_0, m_{1/2})$ plane 
for $\tb = 3$ and $A_0 = 0$~%
\footnote{As mentioned above, this contour is not very sensitive
to these choices: see the discussions
in~\cite{ATLAS0l,ATLAS1l,ATLAScombined}.}%
. As seen in~\cite{ATLAScombined}, the ATLAS 0L
analysis provides the dominant constraint on $m_{1/2}$ for $m_0 < 300 \gev$. Moreover, Fig.~17d 
of~\cite{ATLASextra} shows that the ATLAS 0L search with the greatest impact on
the parameter spaces of the CMSSM, NUHM1 and VCMSSM
is ATLAS search D ($\ge 3$ jets with
leading $p_T > 120 \gev$, other jets with $p_T > 40 \gev$, 
$\ETslash > 100 \gev$, $\Delta \phi({\rm jet}, \pTslash) > 0.4$,
$m_{\rm eff} > 1000 \gev$, $\ETslash/m_{\rm eff} > 0.25$).

Two events were observed in ATLAS 0L search D, to be compared with the
number of $2.5 \pm 1.0 {~}^{+1.0}_{-0.4} \pm 0.2$ events expected due to 
SM backgrounds~%
\footnote{These errors are due to the uncorrelated systematic uncertainty (including also the jet energy resolution and lepton efficiencies), the jet energy scale, and the luminosity, respectively.} 
We interpret this as a `signal' of $- 0.5 \pm 2.2$
events, corresponding to a 95\% CL upper limit of 3.8 events. This
corresponds to the quoted 95\% CL upper limit of 0.11~pb and the 35/pb
of integrated luminosity analyzed by ATLAS, and reproduces approximately
the 95\% CL contour for search D shown in Fig.~17d
of~\cite{ATLASextra}. This figure also reports the numbers of events
expected in ATLAS search D for points with various different values of
$(m_0, m_{1/2})$. We calculate the corresponding numbers of effective
deviations $\sigma_{\rm eff}$ from the observed `signal', and construct
a map of the 
deviations for intermediate values of $(m_0, m_{1/2})$ by interpolating
between these values. At larger values of $(m_0, m_{1/2})$, where expected
event numbers are not provided, we scale the event numbers $\propto
M^{-4}$, where $M \equiv \sqrt{m_0^2 + m_{1/2}^2}$, following~\cite{mc5}
and consistent with previous ATLAS  
studies. We then estimate the corresponding numbers of effective
deviations $\sigma_{\rm eff}$ from the observed `signal' using the same
prescription as above, and use this to calculate the corresponding value
of $\chi^2$.  

For $m_0 > 300 \gev$, the best available ATLAS constraint on $ m_{1/2}$ comes 
from a combination of the ATLAS 0L and ATLAS 1L analyses. To estimate
 the corresponding contribution to the likelihood function at larger $(m_0, m_{1/2})$,
we again use $M^{-4}$ scaling to estimate the expected numbers of events.

We evaluate the overall ATLAS contribution to $\chi^2$ for each of the
points in our  
samples of the CMSSM, NUHM1, VCMSSM and mSUGRA parameter spaces by
combining these treatments of the ATLAS searches at small and large $m_0$.

\subsection*{\it CMS multijet + $\ETslash$ analyses}

Following the initial $\alpha_T$ analysis~\cite{cms0l-aT} that we analyzed previously~\cite{mc5},
results from an additional CMS multijet + $\ETslash$
analysis have been released (CMS MHT)~\cite{MHT} 
which has greater sensitivity in the $(m_0, m_{1/2})$
plane. The CMS MHT analysis also imposes stronger constraints in the
$(m_0, m_{1/2})$ plane than does the ATLAS combined analysis~\cite{ATLAScombined}
when $m_0 > 600 \gev$, 
so we now analyze its results in more detail. The limit obtained in this 
search is very close to the median expected limit, corresponding
to a difference between the numbers of events observed and expected from
background that is negligible compared to the $\sigma_{\rm eff}$ for the
number of background events. We therefore approximate the impact of this
search outside its nominal 95\% CL contour again by assuming that the number
of effective $\sigma$ is simply proportional to the number of signal
events expected at any given supersymmetric point, which we assume to be
$\propto M^{-4}$, following~\cite{mc5}, and we then calculate the
corresponding $\chi^2$ penalty.

\subsection*{\it Combining information of ATLAS and CMS analyses}

In our implementation of the combination of these constraints, for each
supersymmetric point we compare the contributions to $\chi^2$ from the
ATLAS and CMS MHT searches calculated as described above, 
and retain just the larger of the two $\chi^2$ penalties, dropping the
contribution from the lesser constraint. 
This procedure is conservative, but any non-trivial combination of
the constraints would require an understanding of the common systematic
uncertainties that is currently unavailable, and would be justified only
if the ATLAS and CMS collaborations provided additional information
making possible more detailed modelling of their likelihood functions.

We note in passing that both CMS and ATLAS have published limits on
simplified models based on the above searches. These limits are not
directly applicable to the classes of supersymmetric models considered
here since, for example, they consider cases in which $m_{\tilde q} \gg
\mgl \gg \mneu{1}$ and gluinos decay exclusively to ${\bar q} q
\neu{1}$, whereas in the models considered here 
other gluino decay modes are also important.

\subsection*{\it LHC searches for $H/A \to \tau^+ \tau^-$}

The ATLAS and CMS Collaborations have also released the results of
searches for heavier MSSM Higgs bosons $H/A$, produced mainly via 
$b \bar b \to H/A$ and decaying to $\tau^+ \tau^-$ pairs~\cite{ATLASHA,cmsHA}. 
The stronger of these constraints is provided by the CMS
Collaboration, which we implement as follows. The CMS Collaboration has provided
model-independent limits on the $H/A$ production cross section times $\tau^+ \tau^-$ branching
ratio ($\sigma \times$BR) at the 68\%, 95\% and 99.7\%~CLs as functions of
$\MA$~\cite{nikitenko}, corresponding to a one-dimensional $\chi^2$ contribution of
1, 3.84, and 9, respectively. For each fixed value of $\MA$, we assume that the
$\chi^2$ penalty for other values of $\sigma \times$BR may be approximated
by the functional form $\Delta \chi^2 \propto (\sigma \times$BR$)^{p(\MA)}$, 
normalized to unity on the 68\% CL line and fitting the power $p(\MA)$
independently for each value of $\MA$ (typical values are $\sim 1.3$).
The existing CMS bounds on $b \bar b \to H/A \to \tau^+ \tau^-$ are expected to impact
significantly only the NUHM1 scenario,
where relatively low values of $\MA$ and high values of $\tan \beta$ lie within the region
allowed by other constraints at the 95\%~CL. Therefore, we have evaluated
$\sigma \times$BR for a representative grid of points in the NUHM1 by 
using the the SM
result for $\sigma(b\bar b \to H_{\rm SM})$~\cite{bbhnnlo} 
modified by the effective NUHM1 couplings obtained from {\tt FeynHiggs},
which we also use to calculate the branching ratio for the decay to $\tau^+\tau^-$. 
A factor of two is included to take into account the production of the
CP-even~$H$ and the CP-odd~$A$ boson, which have approximately the same
production cross section and decay widths in the relevant parameter space,
$\MA \ge 150 \gev$ and large $\tb$. 
We have then checked that $\sigma \times$BR for fixed $\MA$
has a dependence $\sim \tan^2\beta$ in the parameter regions of interest.
Using the value of $\sigma \times$BR calculated in this way for each
point in the NUHM1 parameter space, we then apply the $\chi^2$ penalty
estimated as described above as an afterburner in our global fit.

\subsection*{\it LHCb, CDF and D\O\ searches for $B_s \to \mu^+ \mu^-$}

The paper by LHCb~\cite{lhcb-bmm} provides 95\% and 90\% upper limits on
\bmm\ of 56 and $43 \times 10^{-9}$,
to be compared with the Standard Model prediction of
$(3.2 \pm 0.2) \times 10^{-9}$.
These limits are similar to the ones provided by CDF~\cite{cdf-bmm} and D\O~\cite{d0-bmm}, and a combination
of the results from the three experiments provides a stronger constraint on \bmm. 
In order to make such a combination, we first performed approximate
studies, based on the signal and background expectations in each experiment, and comparing
with the observed pattern of events, generating toy experiments that reproduce their quoted 90\% CL upper limits.
The toy LHCb experiment was constructed using the infomation shown in Table~3 of~\cite{lhcb-bmm}.
The toy CDF experiment was based on the information given in Table II of~\cite{cdf-bmm},
combined with the invariant mass resolution, normalization factors and averaged 
Neural Network efficiencies quoted in the text. 
In order to match exactly the observed 90\% limit quoted by CDF, a small difference in the Neural
Network efficiencies between the CMU-CMU and CMU-CMX channels~\cite{cdf-bmm} was introduced.
Finally, the toy D\O\ experiment was based on Fig.~4 of~\cite{d0-bmm}, together with
the invariant mass resolution and normalization factor quoted in the text.
These toy experiments also reproduce the quoted 95\% CL limits, giving some support to this approximate treatment.
The Tevatron results were afterwards recomputed using the latest world average $f_d/f_s = 3.71\pm0.47$~\cite{HFAG}, for consistency with the LHCb analysis.
The results of the three experiments were combined using the $CL_{s}$ method, treating
the error on  $f_d/f_s$ and the branching ratio of $B^+\to J/\psi(\mu^+\mu^-)K^+$
as systematic errors common to the three experiments.
The combined likelihood function yields formal upper limits of $20 (24) \times 10^{-9}$ at the 90(95)\% CL:
our global fit uses the full likelihood function calculated using the above
experimental information to beyond the 99\% CL.

\subsection*{\it Xenon100 search for dark matter scattering}

Finally, we implement the constraint imposed by the direct upper limit
on dark matter scattering given by the Xenon100
experiment~\cite{Xenon100new}. Its results are presented as a 95\% CL upper
limit on the spin-independent cross section as a function of $\mneu{1}$, 
under assumptions for the local halo density and the dark matter
velocity distribution that are described in~\cite{Xenon100new} and have
uncertainties that are small compared to that in the spin-independent
scattering matrix element discussed below~\cite{hadunc,Ellis:2008hf}. 
The Xenon100 Collaboration report the observation of 3 events in
100.9 live days within a fiducial detector with a mass of 48~kg, in a
range of recoil energies where $1.8 \pm 0.6$ events were expected~%
\footnote{The probability for such a Poisson background process to yield 3 or 
more events is 28\%, so this observation does not constitute a
significant signal. }. Using this information,
we have constructed a model for the Xenon100 contribution to the
global $\chi^2$ likelihood function as a function of the number of events using the $CL_s$ method,
which is quite similar to a Gaussian function with mean 1.2 and standard deviation 3.2 events. 
Our model for the Xenon100 likelihood function yields a 90\% CL upper limit of 6.1 events so,
for any given value of $\mneu{1}$, we assume that the 90\% CL upper limit on
$\ssi$ quoted in~\cite{Xenon100new} corresponds to 6.1 events,
and use simple scaling to estimate the event numbers corresponding to other values of $\ssi$.
We then use the Gaussian model for the Xenon100 $\chi^2$ function to estimate the
contribution of this experiment to the global likelihood function for other $\ssi$ values.
We note that, because of the insignificant `excess' of 1.2 events in the Xenon100 data,
there is a contribution $\Delta \chi^2 \sim 0.3$ to the global likelihood function
at small values of \ssi~%
\footnote{The predicted values of \ssi\ at the post-2010-LHC best-fit points are all smaller
than preferred by this `excess', so they all receive $\Delta \chi^2 \sim 0.3$ from the
Xenon100 data, as seen in the Table~\ref{tab:compare}. For this reason,
the lower 68\% CL limits on \ssi are 
essentially unchanged when the Xenon100 data are incorporated in the fits.}.

In order to translate this estimate into contributions to the
global likelihood functions for various supersymmetric models,
we must take account of the uncertainty in the calculation of $\ssi$ for fixed
supersymmetric model parameters. The dominant uncertainty is that in
the determination of the strange quark
scalar density in the nucleon, $\langle N | {\bar s} s | N \rangle$,
which is induced principally by the experimental uncertainty in the $\pi$-nucleon
$\sigma$ term, $\Sigma_{\pi N} \equiv 1/2(m_u + m_d)\langle N|{\bar u}u + {\bar d} d|N \rangle$:
\begin{equation}
y \equiv \frac{2 \langle N | {\bar s} s | N \rangle}{\langle N|{\bar u}u + {\bar d} d|N \rangle}
= 1 - \frac{\sigma_0}{\Sigma_{\pi N}} ,
\label{y}
\end{equation}
where $\sigma_0 \equiv 1/2 (m_u + m_d) \langle N | {\bar u} u + {\bar d} d - 2 {\bar s} s | N\rangle = 36 \pm 7~\mev$~\cite{sigma0}
is estimated from baryon octet mass splittings.
Estimates of $\Sigma_{\pi N}$ ranging from $\sigma_0$ (corresponding to $y=0$) up to a value as large as 
$64 \pm 8 \mev$ have been given in the literature~\cite{Pavan:2001wz} (and even larger values
cannot be excluded~\cite{Pavan2011}), whereas a recent
analysis based on lattice calculations~\cite{Young:2009zb} would suggest a lower value:
$\Sigma_{\pi N} \sim 40 \mev$~\cite{Giedt:2009mr}.
Here we span the plausible
range by using as our default $\Sigma_{\pi N} = 50 \pm 14 \mev$, while
also showing some results for $\Sigma_{\pi N} = 64 \pm 8 \mev$~%
\footnote{The estimated uncertainty
in $\sigma_0 = 36 \pm 7 \mev$ is also included in our analysis, as
are the smaller uncertainties associated with the quark masses.}.

The uncertainty in
$\Sigma_{\pi N}$ is quite significant for our analysis, since it
corresponds to an uncertainty in the spin-independent cross section for
fixed supersymmetric model parameters of a factor of 5 or more. 
We plea again for an effort to reduce this uncertainty by a new campaign
of experimental measurements and/or lattice QCD calculations.

\section{Impacts of the LHC and Xenon100 Constraints}

\subsection*{\it $(m_0, m_{1/2})$ planes}

\begin{figure*}[htb!]
\resizebox{8cm}{!}{\includegraphics{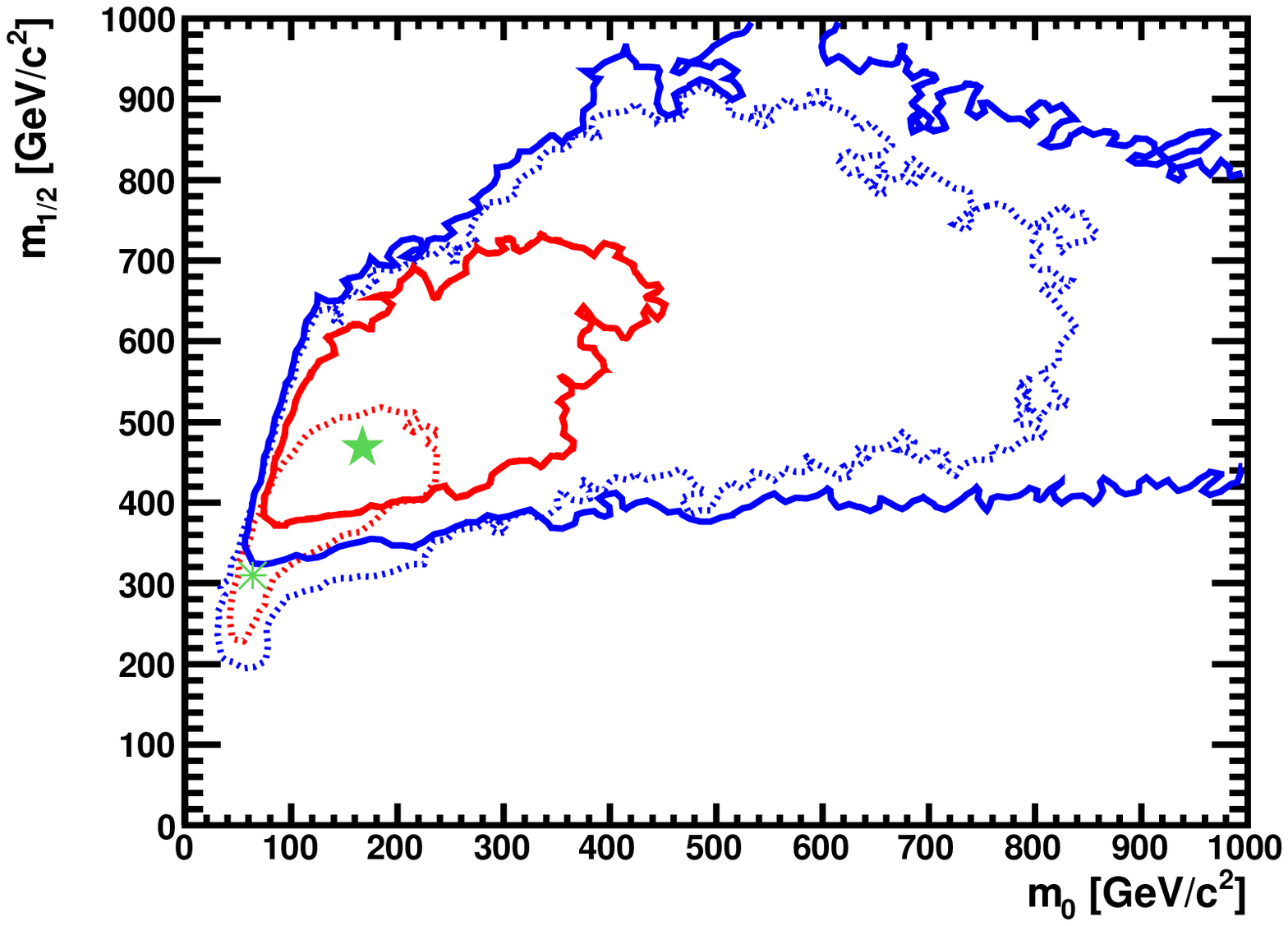}}
\resizebox{8cm}{!}{\includegraphics{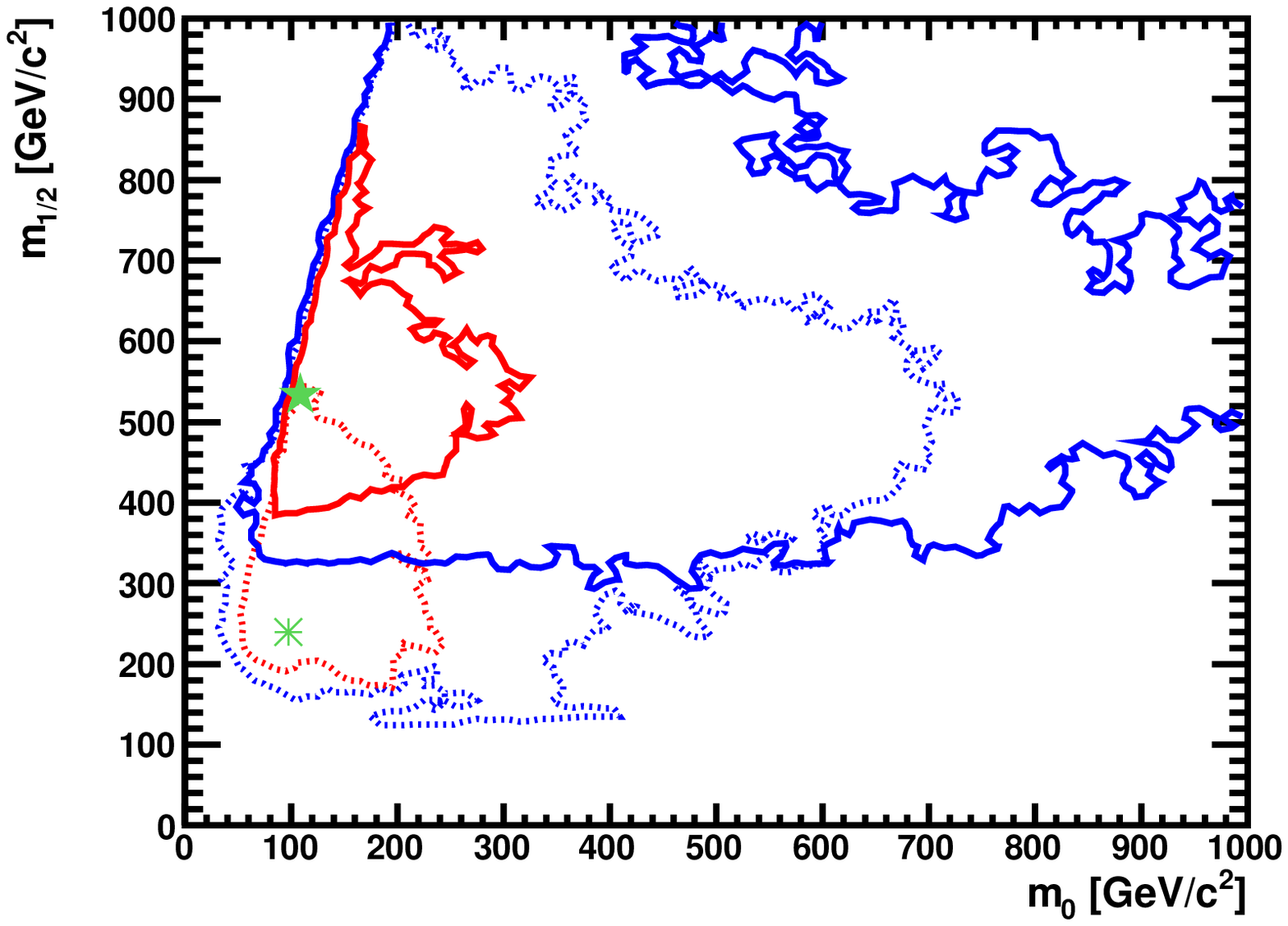}}
\resizebox{8cm}{!}{\includegraphics{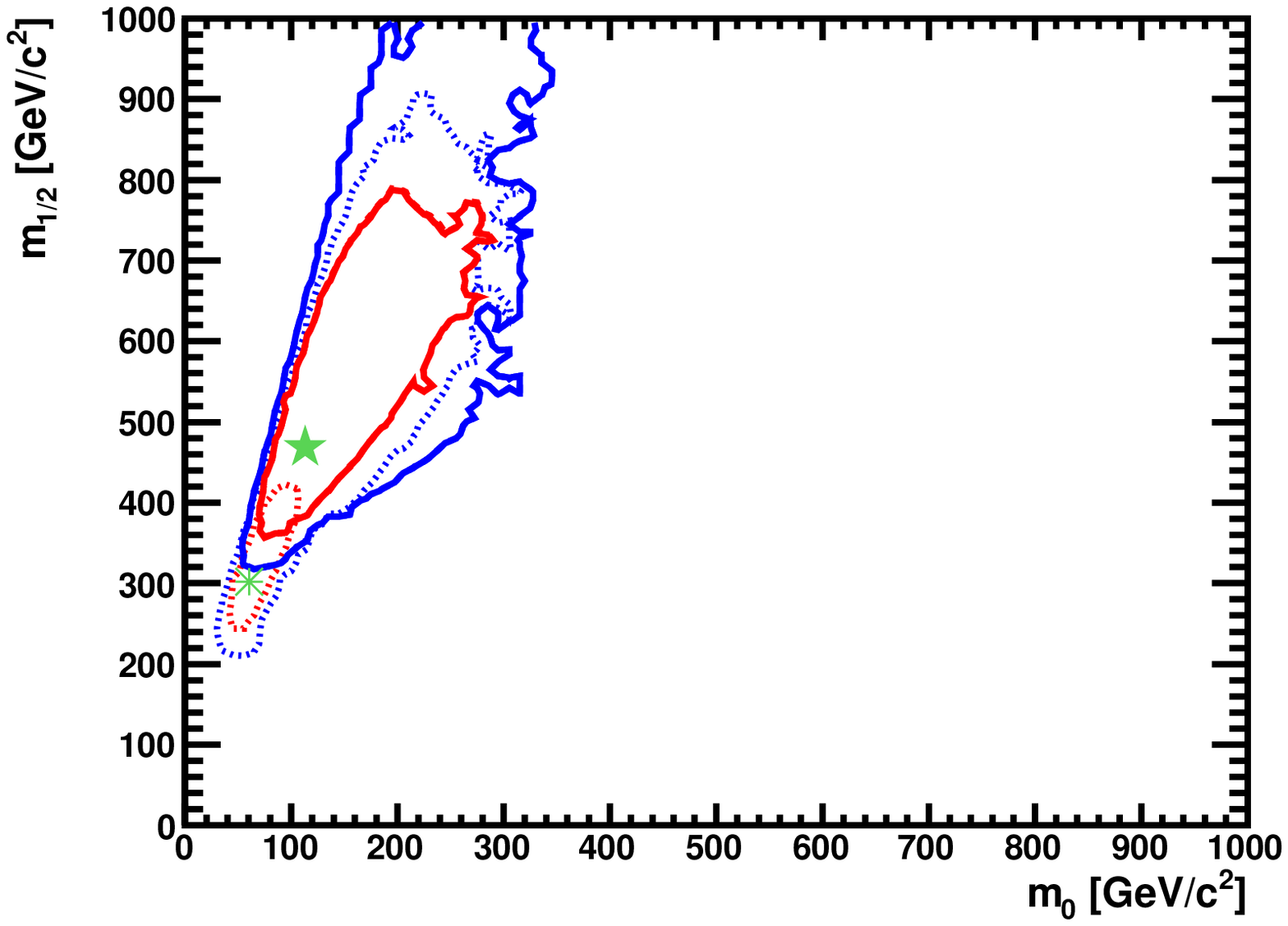}}
\resizebox{8cm}{!}{\includegraphics{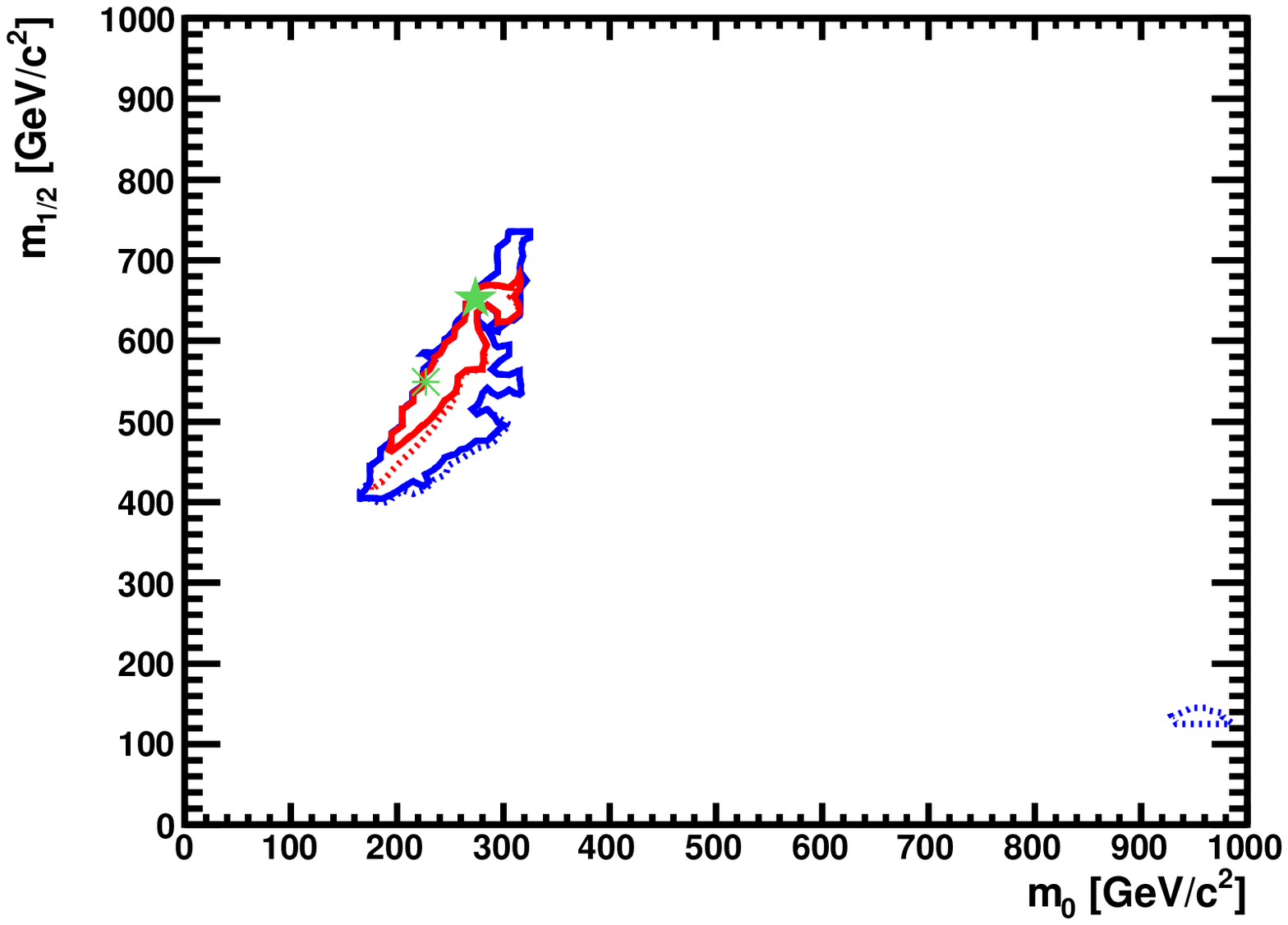}}
\vspace{-1cm}
\caption{\it The $(m_0, m_{1/2})$ planes in the CMSSM (upper left), the
  NUHM1 (upper right), the VCMSSM (lower left) and mSUGRA (lower
  right). In each plane, the best-fit point after incorporation of the
  2010 LHC and Xenon100 constraints is indicated by a filled green star, and the
  pre-LHC fit by an open star. The 68 and 95\% CL regions are indicated
  by red and blue contours, respectively, the solid lines including the
  2010 LHC and Xenon100 data, 
  and the dotted lines showing the pre-LHC fits. 
}
\label{fig:m0m12}
\end{figure*}

We display in Fig.~\ref{fig:m0m12} the $(m_0, m_{1/2})$ planes for the
CMSSM (upper left), NUHM1 (upper right), VCMSSM (lower left) and mSUGRA
(lower right), driven by the ATLAS 0L and CMS MHT constraints but also
taking into account the other 2010 LHC constraints discussed 
above, as well as the Xenon100 constraint. 
In these and subsequent plots, we show in all panels best-fit points
(in green), 68 and 95\% CL regions (red and blue lines,
respectively). Our pre-LHC results, taken from \cite{mc5}, are displayed
as `snowflakes' and dotted lines, and our post-2010-LHC/Xenon100 results are
displayed as full stars and solid lines~%
\footnote{Our pre-LHC
results differ slightly from those given in~\cite{mc5} as we use updated software including {\tt SoftSUSY 3.0.13}.}.

Pre-LHC, the most important lower limits on $m_{1/2}$ in these models were indirect, 
being provided by the lower limit on $\Mh$ from LEP, which had considerably greater impact 
in these models than did the direct sparticle searches at LEP and the Tevatron. 
In each of the
CMSSM, the NUHM1 and the VCMSSM the direct 2010 LHC constraints push the best-fit
values of $m_{1/2}$ to significantly higher values, as well as their 68 and 95\% CL ranges~\footnote{On 
the other hand, the best-fit mSUGRA point
is raised somewhat less, due to the different form of the global $\chi^2$ function.}, whereas the effect
of Xenon100 is not visible in this projection of the model parameter spaces.  
Thus the direct 2010 LHC limits are constraining
these models 
substantially more strongly than the LEP Higgs constraint.

\begin{figure*}[htb!]
\resizebox{8cm}{!}{\includegraphics{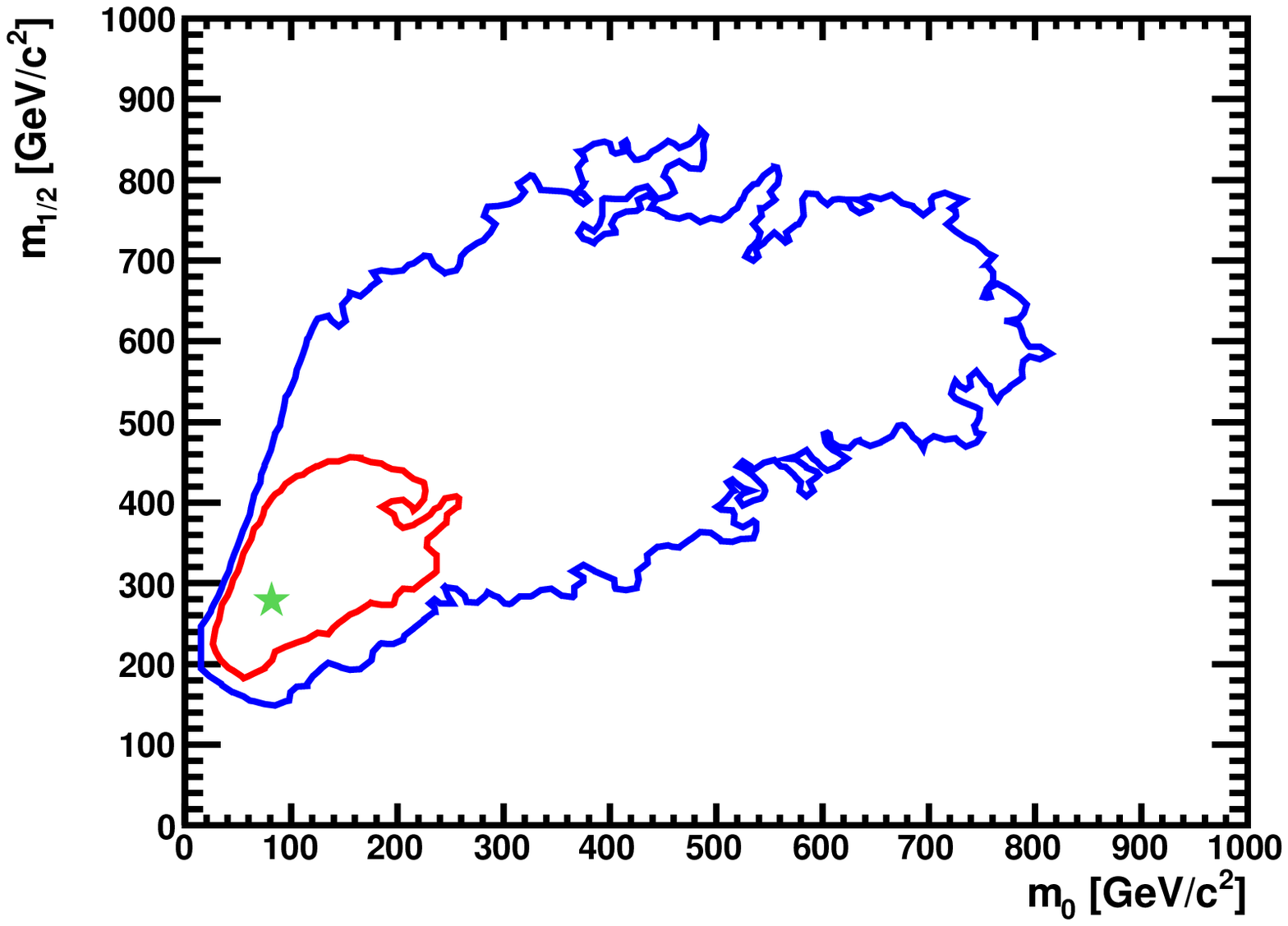}}
\resizebox{8cm}{!}{\includegraphics{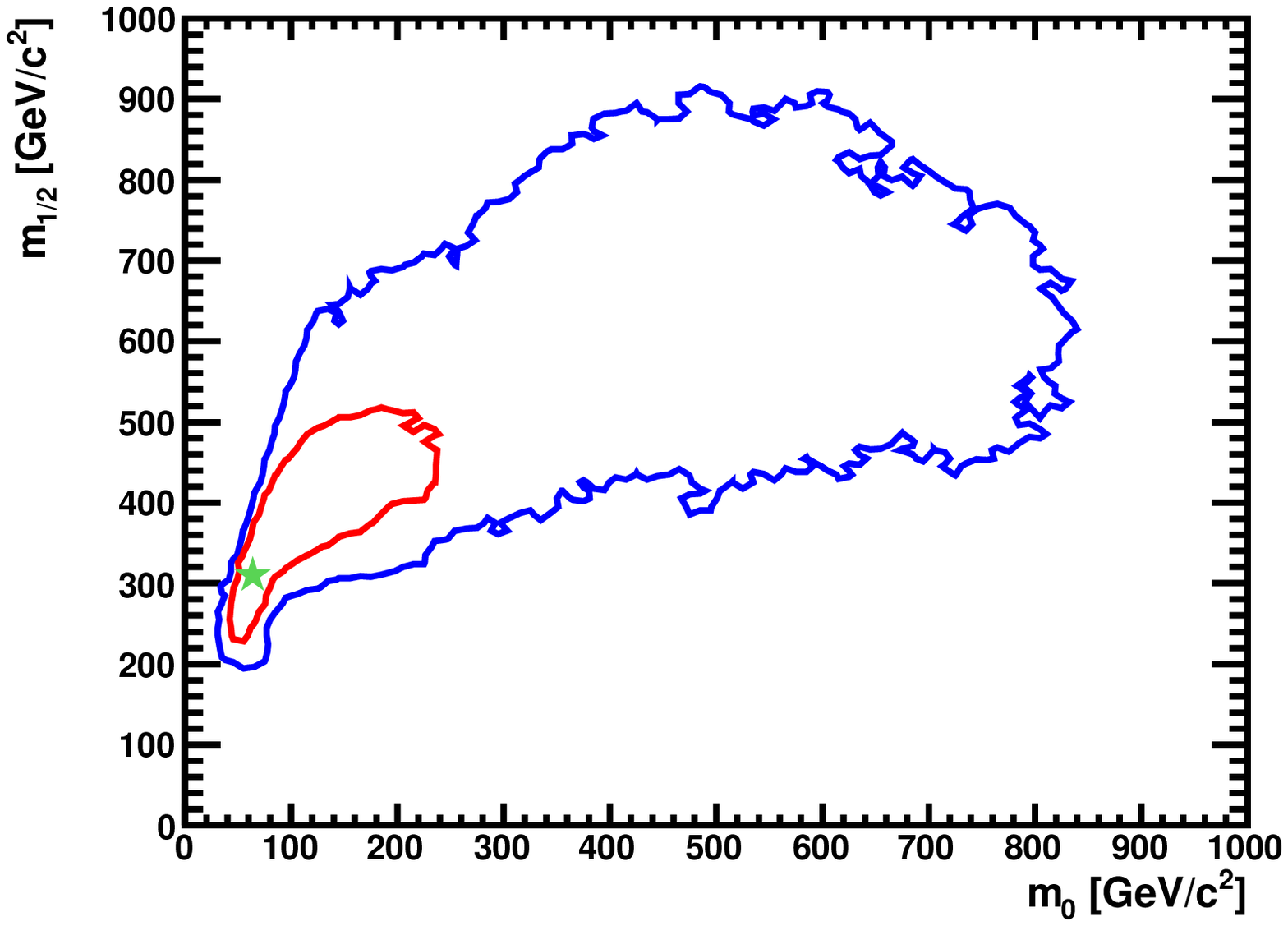}}\\
\resizebox{8cm}{!}{\includegraphics{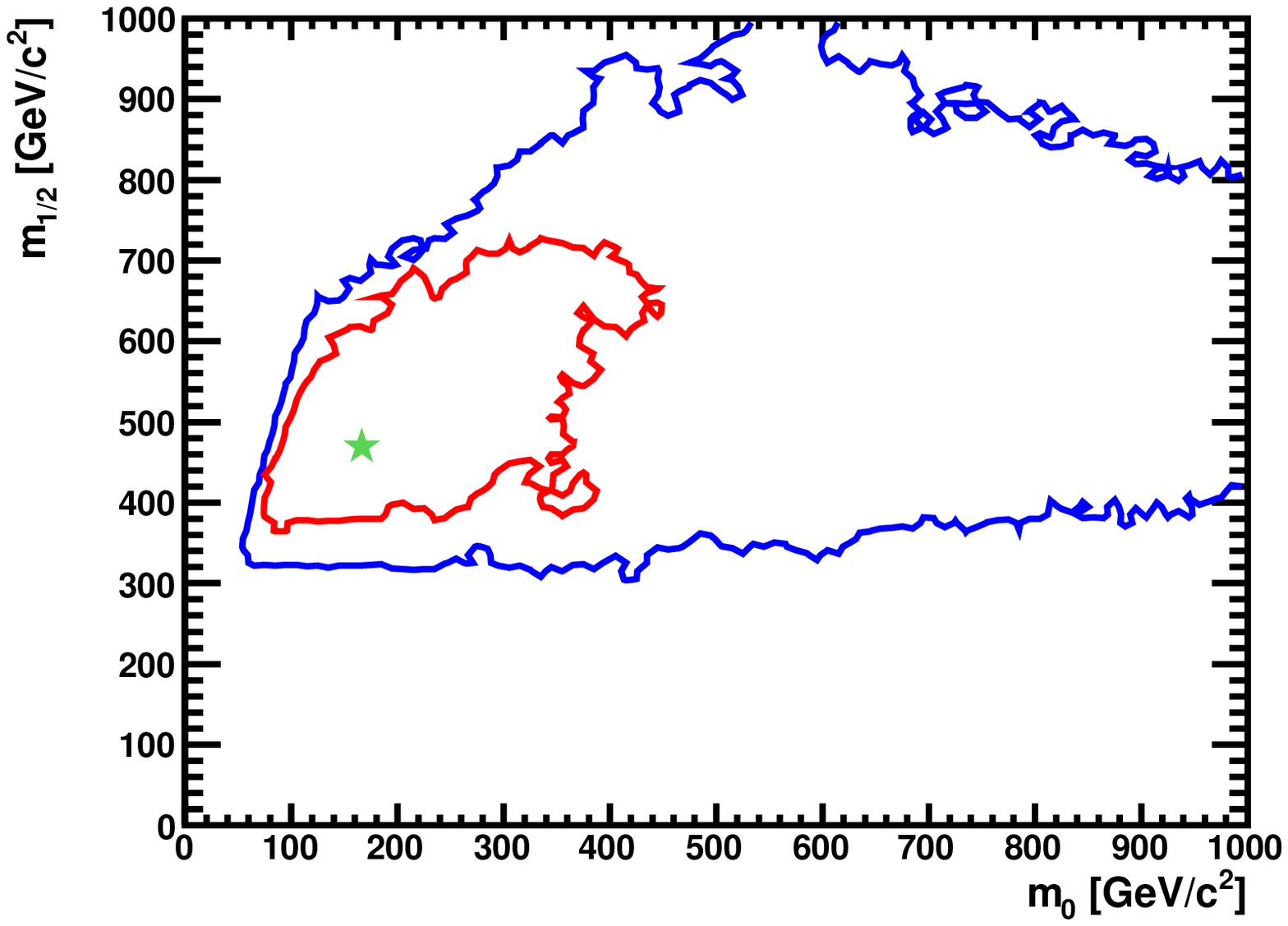}}
\resizebox{8cm}{!}{\includegraphics{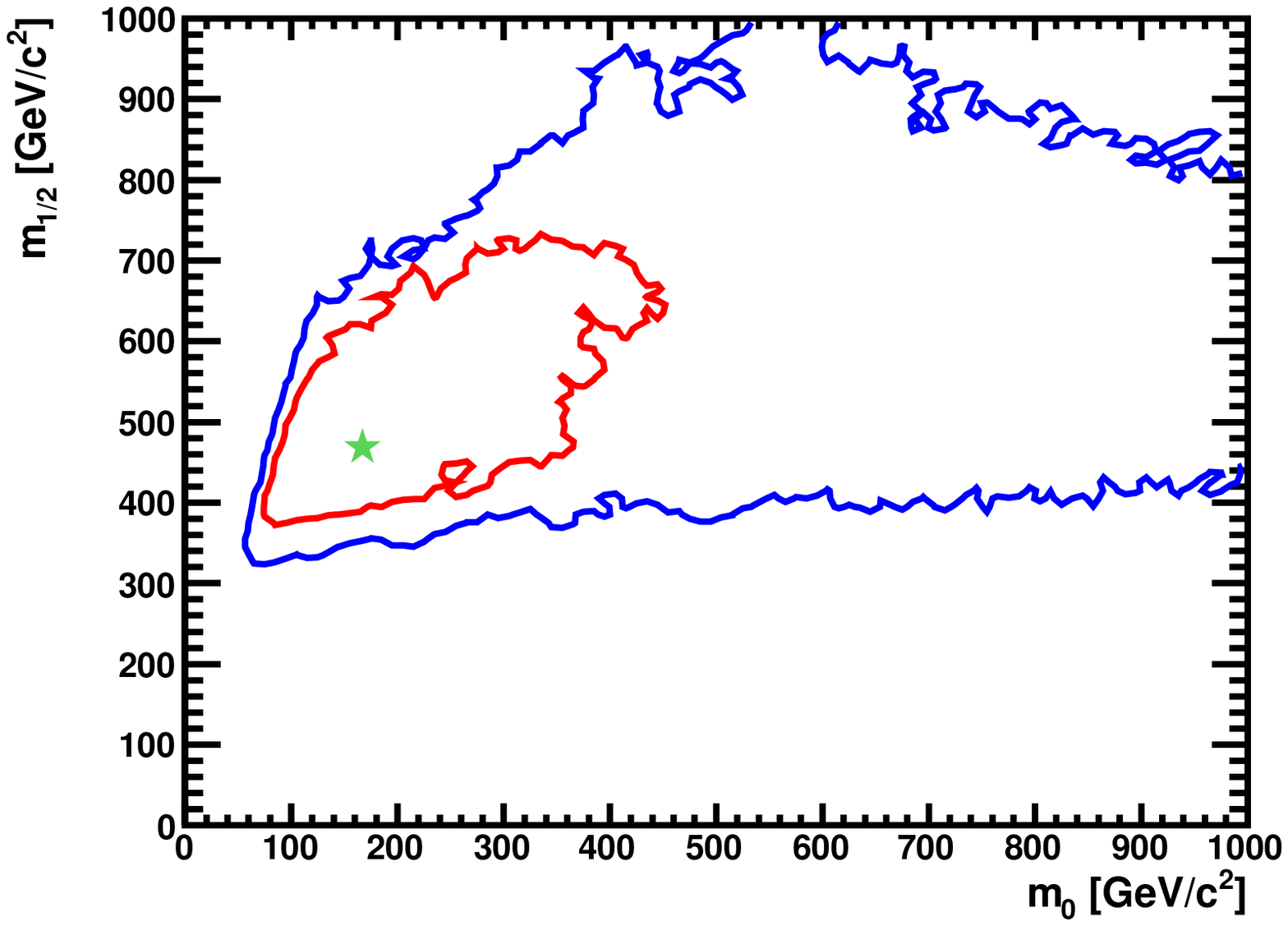}}
\vspace{-1cm}
\caption{\it The $(m_0, m_{1/2})$ planes in the CMSSM with neither
the LEP Higgs constraints nor the LHC constraints applied (upper left), with LEP but
without the LHC (upper right), without LEP but with the LHC (lower left) and with both
LEP and the LHC (lower right). In each plane, the best-fit point is indicated by a filled 
green star, and the 68 and 95\% CL regions are indicated
by red and blue contours, respectively.
}
\label{fig:Higgseffect}
\end{figure*}

This can be seen explicitly in the panels of Fig.~\ref{fig:Higgseffect}, which compare the
effects of the LEP Higgs and 2010 LHC constraints on the CMSSM. The upper left panel shows the
best-fit point, 68\% and 95\% CL contours without applying either the LEP or the 2010 LHC
constraints, and the upper right panel shows the effect of applying the LEP Higgs constraint but {\it not} the
2010 LHC constraints. We note that LEP moves the best-fit
from $m_{1/2} \sim 270 \gev$ to $\sim 320 \gev$ while the 95\% CL contour at large $m_0$ and $m_{1/2}$
expands slightly, reflecting the small rise in the minimum of $\chi^2$. The lower left panel shows the
best-fit point, 68\% and 95\% CL contours applying the 2010 LHC
constraints {\it without} the LEP Higgs constraint. The best-fit point now moves to $m_{1/2} \sim 470 \gev$,
and the 95\% CL contour moves correspondingly much further out. Finally, the lower right panel
shows the effect of applying the LEP Higgs constraint
as well. We see that the best fit remains essentially unchanged
at $m_{1/2} \sim 470 \gev$, and the 95\% CL contour is little affected at large $m_0$ and $m_{1/2}$.
In summary, applying the LEP Higgs constraint increases $m_{1/2}$ by $\sim 50 \gev$
in the absence of the LHC constraints, and only marginally if they {\it are} applied, whereas the LHC constraints
increase $m_{1/2}$ by $\sim 200 \gev$ in the absence of the LEP Higgs constraint, and by $\sim 150 \gev$
if it {\it is} applied. Correspondingly, the effects of LHC on the 95\% CL contour are much greater than
those of the LEP Higgs constraint.

As seen in Fig.~\ref{fig:m0m12}, the effects of the LHC on the best-fit values
of $m_0$ are smaller, though there are significant increases in the
CMSSM and VCMSSM that are correlated with the increases in $m_{1/2}$. We
note that in all models the new best-fit point lies within or close to
the border of the pre-LHC 68\% CL contour, 
indicating that there is no significant tension between
the LHC constraints and prior indications on the scale of supersymmetry breaking.
 Nevertheless, in all cases other than mSUGRA, the pre-LHC best fit points
are now excluded at the 95\% CL. Furthermore. 
 the 2010 LHC constraints exclude roughly half of the pre-LHC 68\%
CL regions in the CMSSM and VCMSSM, and most of the pre-LHC 68\% CL region in the NUHM1.
However, the LHC has yet to make any significant inroad into even the 95\% CL
region of mSUGRA~%
\footnote{The
raggedness of  the CL contours should be regarded as indicative of the uncertainties in our analysis.
Recall also that, as already mentioned, our pre-LHC
results differ slightly from those given in~\cite{mc5}, as updated
software was used, in particular {\tt SoftSUSY 3.0.13}.}.

\begin{table*}[!tbh]
\renewcommand{\arraystretch}{1.2}
\begin{center}
\begin{tabular}{|c||c|c|c|c|c|c||c|c|} \hline
Model & Minimum & Probability & $m_{1/2}$ & $m_0$ 
      & $A_0$ & $\tb$ & $\Mh$ (GeV) \\
& $\chi^2$/dof & & (GeV) & (GeV) & (GeV) & & (no LEP)\\ \hline \hline
CMSSM pre-LHC     & 22.5/19 & 26\% & $310_{-50}^{+120}$ & $60_{-10}^{+90}$ 
      & $-60^{+410}_{-840}$ & $10_{-4}^{+10}$ & 108.6\\
post-2010-LHC &  26.1/19  &   13\%    &  $470_{-70}^{+140}$  & $170_{-80}^{+330}$  
      &  $-780_{-820}^{+1410}$ &  $22_{-13}^{+27}$  &   115.7 \\
post-Xenon ($50\pm14$)  &  26.2/20  &   16\%    &  $470_{-70}^{+140}$  & $170_{-80}^{+330}$  
      &  $-780_{-820}^{+1410}$  &  $22_{-13}^{+27}$  &   115.7 \\
\hline
NUHM1 pre-LHC     & 20.5/17 & 25\% & $240_{-50}^{+150}$ & $100_{-40}^{+70}$ 
      & $920^{+360}_{-1260}$ & $7_{-2}^{+11}$ & 119.4 \\
post-2010-LHC &  24.1/18  &  15\%    &  $530_{-90}^{+220}$  &  $110_{-20}^{+80}$  
      &  $ -370_{-1000}^{+1070}$  & $27_{-10}^{+24}$  &  117.9 \\
post-Xenon ($50\pm14$)  &  24.2/19  &  19\%    &  $530_{-90}^{+220}$  &  $110_{-20}^{+80}$  
      &  $ -370_{-1000}^{+1070}$  & $27_{-10}^{+24}$  &  117.9 \\
\hline
VCMSSM pre-LHC    & 22.6/20 & 31\% & $300_{-40}^{+60}$ &  $60_{-10}^{+20}$ 
      &   $30^{+50}_{-30}$ & $8_{-1}^{+3}$ & 110.0 \\
post-2010-LHC &  27.9/20  &  11\%    &  $470_{-80}^{+150}$  &  $110_{-30}^{+110}$  
      &  $120^{+300}_{-190}$  & $13_{-8}^{+14}$  &  115.0 \\
post-Xenon ($50\pm14$)  &  28.1/21  &  14\%    &  $470_{-80}^{+150}$  &  $110_{-30}^{+110}$  
      &  $120^{+300}_{-190}$  & $13_{-8}^{+14}$  &  115.0 \\
\hline
mSUGRA pre-LHC    & 29.4/19 & 6.0\% & $550_{-90}^{+170}$ & $230_{-40}^{+80}$  
      & $430_{-90}^{+190}$ & $28_{-2}^{+5}$ & 107.8 \\
post-2010-LHC &  30.2/20  &   6.7\%     & $650_{-130}^{+70}$  & $270_{-50}^{+50}$ 
      & $530_{-130}^{+130}$ & $30_{-3}^{+4}$ &  122.2 \\
post-Xenon ($50\pm14$) &  30.3/21  &  8.6\%      & $650_{-130}^{+70}$  & $270_{-50}^{+50}$ 
      & $530_{-130}^{+130}$ & $30_{-3}^{+4}$ &  122.2 \\
\hline
\end{tabular}
\caption{\it Comparison of the best-fit points found in the pre-LHC
  analysis in the CMSSM, the NUHM1, the VCMSSM and the coannihilation
  region of mSUGRA~\cite{mc2,mc3,mc4,mc5}, 
  and our latest results incorporating the CMS, 
  ATLAS, LHCb, CDF, D\O\ and Xenon100 constraints.
  We also include the minimum
  value of $\chi^2$ and the fit probability in each scenario, 
   as well as the predictions for $\Mh$ {\it without imposing} the LEP
  constraint. 
}
\label{tab:compare}
\end{center}
\end{table*}

In Table~\ref{tab:compare} we compare the post-2010-LHC/Xenon100
best-fit points found in this paper 
with pre-LHC 
results~\cite{mc5} in the CMSSM, NUHM1, VCMSSM and mSUGRA
(in the latter case, only the best fit in the coannihilation region is reported). In addition to
the minimum value of $\chi^2$, the number of degrees of freedom,  and the fit 
probability in each scenario, we include the values of $m_{1/2}, m_0, A_0$ and
$\tb$ at  all the best-fit points, 
as well as the respective one-dimensional 68\% CL ranges,
and the predictions for $\Mh$ if the LEP Higgs
constraint is neglected. We note again that the
2010 LHC constraints are significantly stronger than those from previous sparticle searches
and the LEP Higgs limit, resulting in significant increases in the best-fit values of $m_{1/2}$
and smaller increases in $m_0$ in the CMSSM, NUHM1 and VCMSSM. 
We note also significant increases in the best-fit values of $\tan \beta$ in these models,
which are required by the \gmt\ constraint in order to compensate for the larger values
of $m_{1/2}$ and $m_0$. In the case of the VCMSSM, the scope for increasing $\tan \beta$
is restricted by the condition that $A_0 = B_0 + m_0$, which is largely responsible for the
relatively large increase in $\chi^2$ post-2010-LHC.%
\footnote{We recall that our convention~\cite{mc4,mc5} for the sign of $A_0$ is opposite to
that of {\tt SoftSUSY}.} The values of $A_0$ are poorly constrained in all the models,
and we have checked that there is not a strong dependence of the $\chi^2$ of the NUHM1
on the non-universality between the soft supersymmetry-breaking contributions to the Higgs
and sfermion masses, though small values of the former are somewhat preferred.
We see that the minimum values of $\chi^2$ have been increased by the inclusion of
the 2010 LHC data, in particular. These increases result in some decreases in the overall probabilities,
though insufficient to call the models into question. 
The Xenon100 constraint causes only small changes in the best-fit parameters of the models studied,
as well as small increases in the $\chi^2$ values
and a corresponding small {\em increase} in the probability.

Since the constraint that most disfavours large supersymmetry-breaking masses is \gmt, 
and since it is the interplay between this and the advancing LHC constraints that pushes the
best fits towards larger values of $\tan \beta$, we have investigated the effect of dropping this
constraint altogether. This possibility was explored previously using the pre-LHC data set
in~\cite{mc3}, where it was found that the large-$m_0$ focus-point region was slightly
disfavoured in the CMSSM and NUHM1, even when dropping the \gmt\ constraint, by a 
combination of other observables including $\MW$, in particular. Now, when
\gmt\ is dropped, using the 2010 LHC 
data set (whether the Xenon100 constraint is included, or not) we find a 
secondary minimum in the focus-point 
region that is disfavoured in the CMSSM by $\Delta \chi^2 \sim 1.0$ , whereas
this region was disfavoured by $\Delta \chi^2 \sim 1.6$ when \gmt\ was
dropped from the pre-LHC data set. In the case of the NUHM1, we do not find a clear
secondary minimum in the focus-point region when \gmt\ is dropped post-2010-LHC.

\subsection*{\it $(\tb, m_{1/2})$ planes}

\begin{figure*}[htb!]
\resizebox{8cm}{!}{\includegraphics{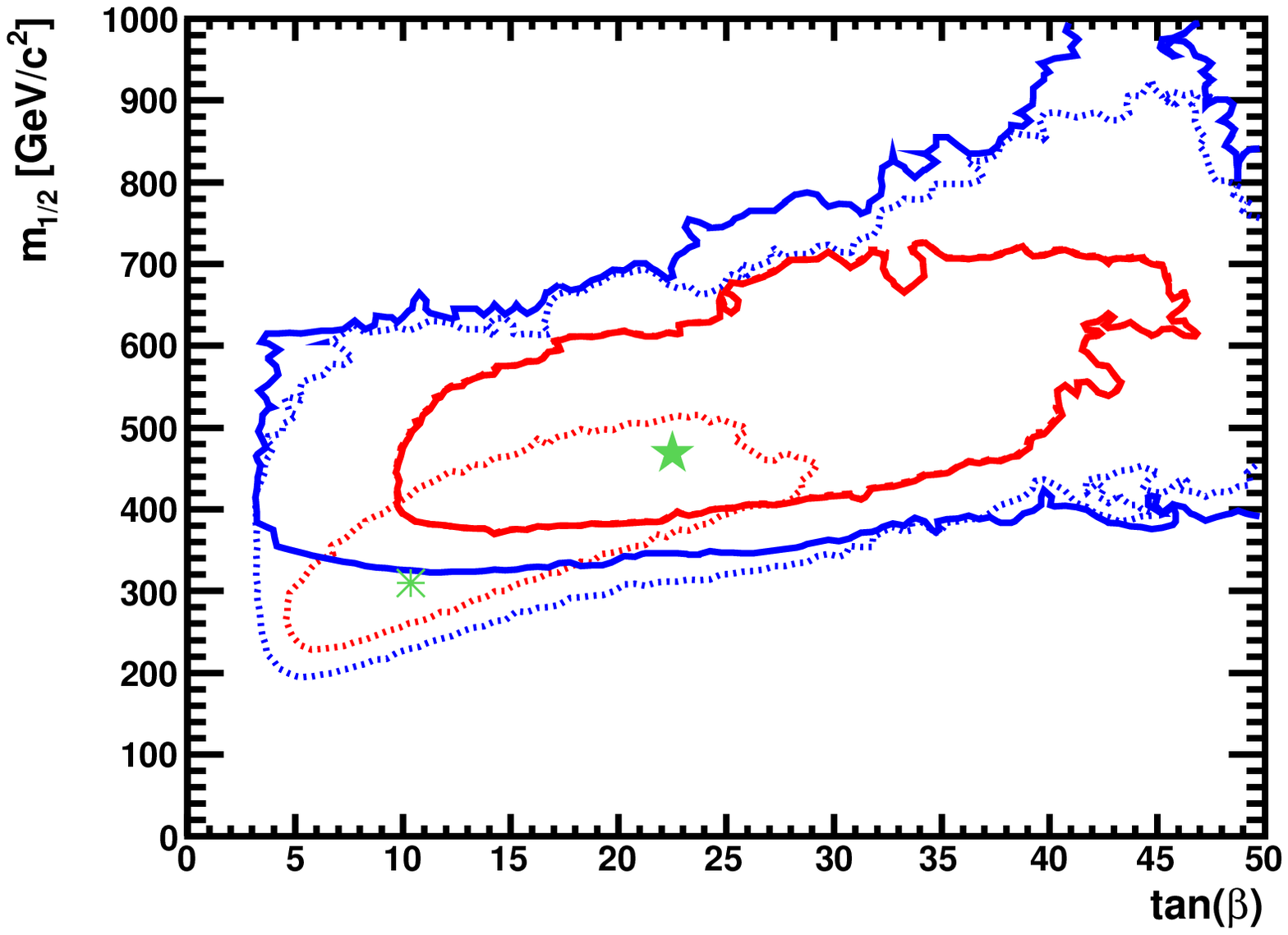}}
\resizebox{8cm}{!}{\includegraphics{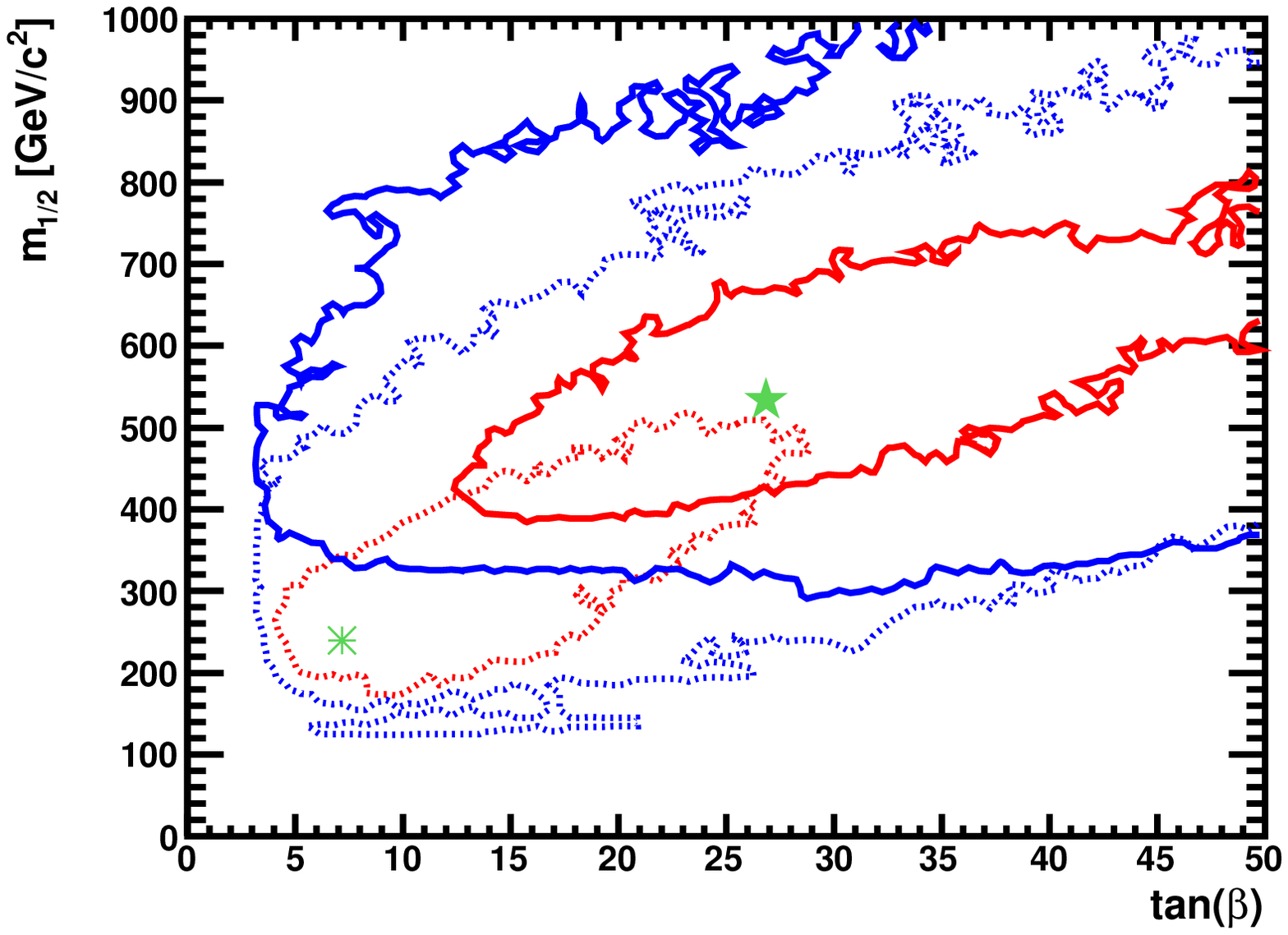}}
\resizebox{8cm}{!}{\includegraphics{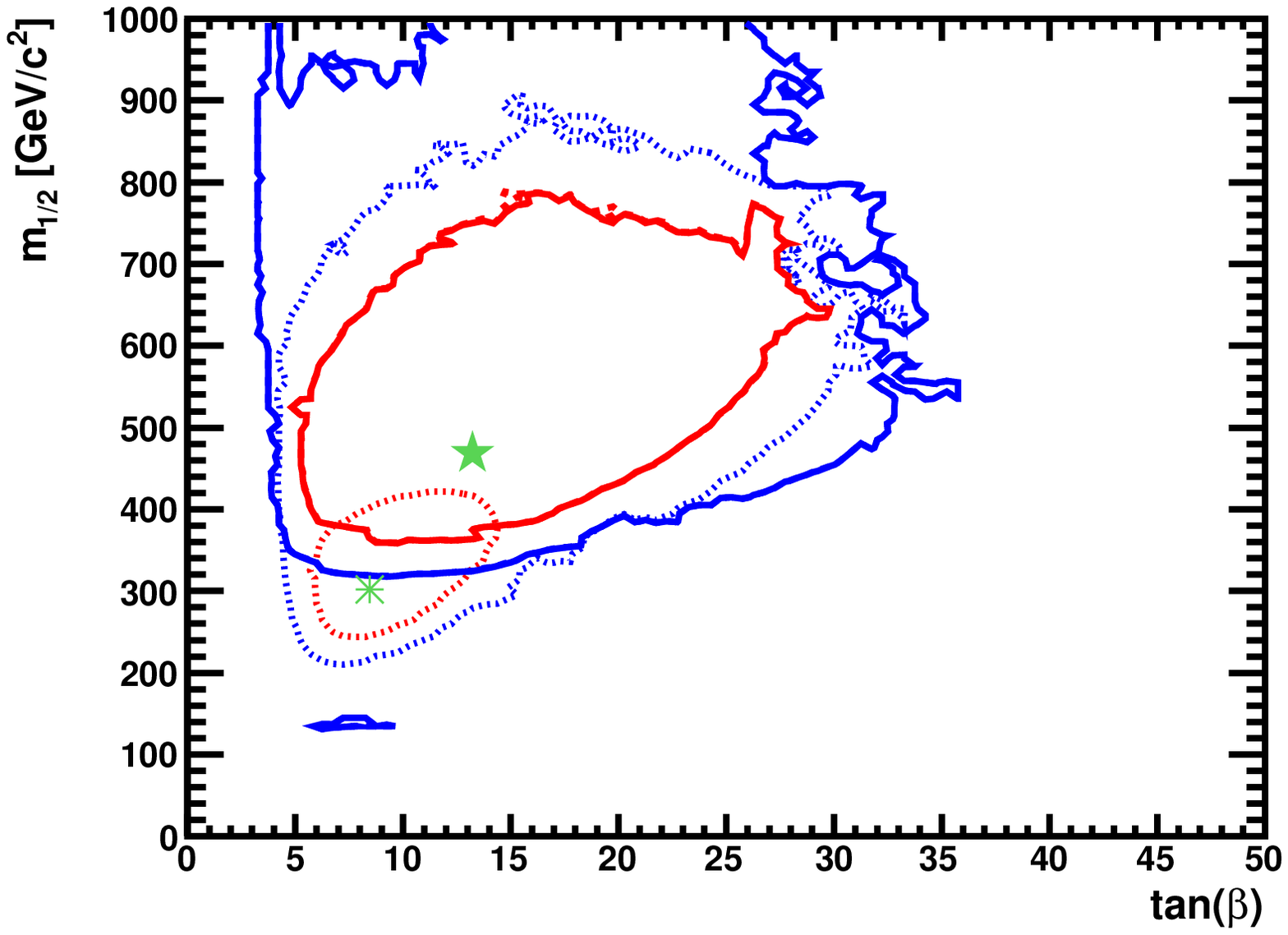}}
\resizebox{8cm}{!}{\includegraphics{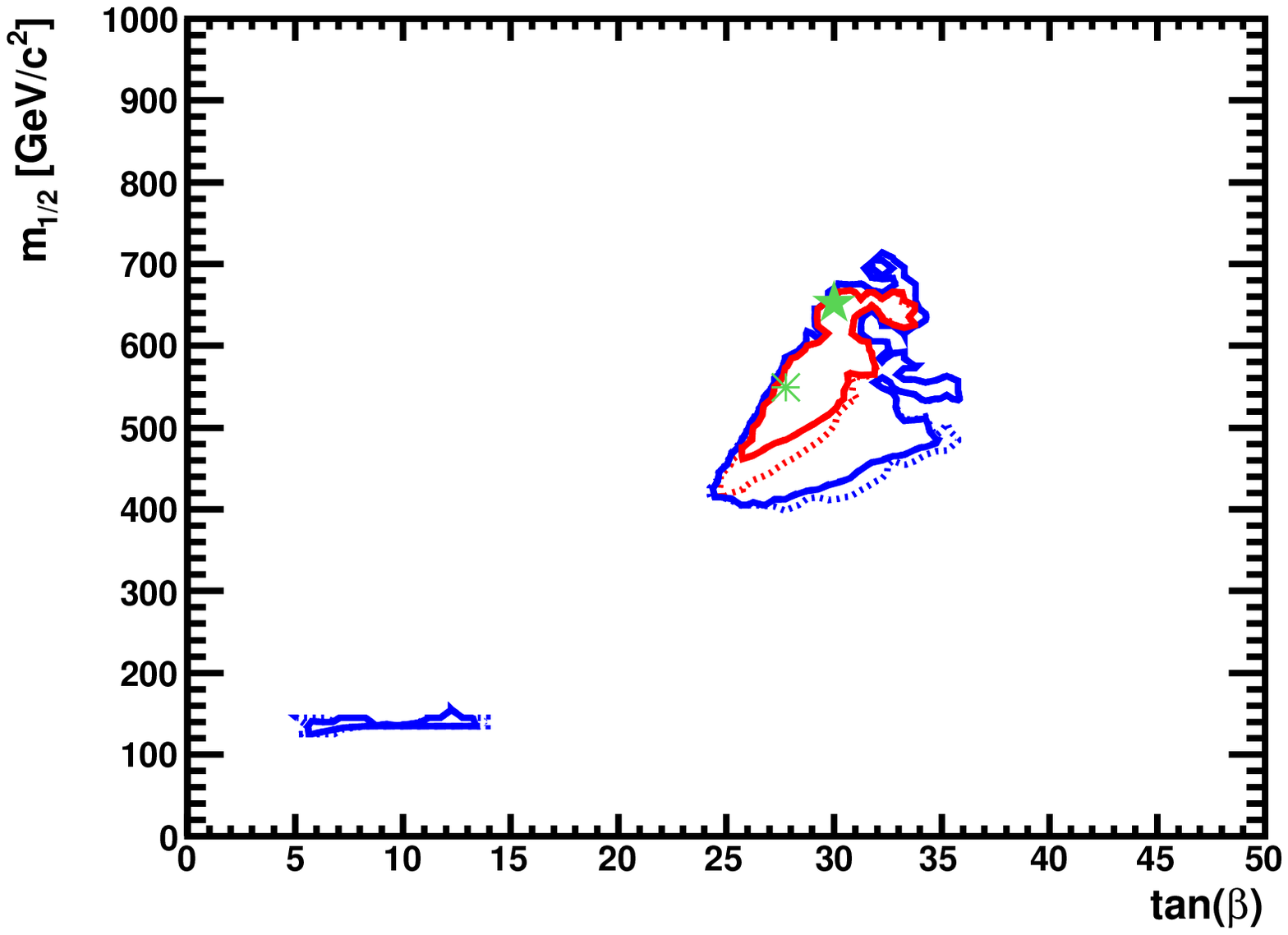}}
\vspace{-1cm}
\caption{\it The $(\tb, m_{1/2})$ planes in the CMSSM (upper left), the
  NUHM1 (upper right), the VCMSSM (lower left) and mSUGRA (lower
  right). In each plane, the best-fit point after incorporation of the
  2010 LHC and Xenon100 constraints is indicated by a filled green star, and the
  pre-LHC fit by an open star. The 68 and 95\% CL regions are indicated
  by red and blue contours, respectively, the solid lines including the
  2010 LHC and Xenon100 data, and the dotted lines including only the pre-LHC data.} 
\label{fig:tBm12}
\end{figure*}

In Fig.~\ref{fig:tBm12} we display the $(\tb, m_{1/2})$ planes for the
CMSSM (upper left), NUHM1 (upper right), VCMSSM (lower left) and mSUGRA
(lower right). We see again that the best-fit point and likelihood
contours in mSUGRA are only mildly affected by the LHC data,
whereas there are 
significant increases in the best-fit values of $\tb$ in the CMSSM,
NUHM1 and VCMSSM that are correlated with the increases in $m_{1/2}$. 
As already commented in~\cite{mc5}, these increases may be understood from
the interplay of the LHC and \gmt\ constraints. It is well known that for fixed $\tan \beta$
\gmt\ favours an elliptical band in the $(m_0, m_{1/2})$ plane that moves
to larger mass values as $\tb$ increases. Hence the pressure of the LHC
towards larger values of $m_{1/2}$ is reconciled with \gmt\ by
increasing $\tb$. It is apparent from the upper panels 
and the 68\% CL ranges given in the Table that the
constraints on the possible values of $\tb$ in the CMSSM and NUHM1 were
quite weak pre-LHC, and are still not very strong. 
In the lower left panel, we see that in the VCMSSM the best-fit value of $\tb$ has increased and its range has
broadened considerably post-2010-LHC~%
\footnote{Many early LHC analyses assumed $\tb = 3$ as a default. It is
apparent from Fig.~\ref{fig:tBm12} that such low values were
disfavoured even pre-LHC, and that a more plausible default choice
post-2010-LHC would be $\tb = 10$ or more.}. 

\subsection*{\it $(\MA, \tb)$ planes}

\begin{figure*}[htb!]
\resizebox{8cm}{!}{\includegraphics{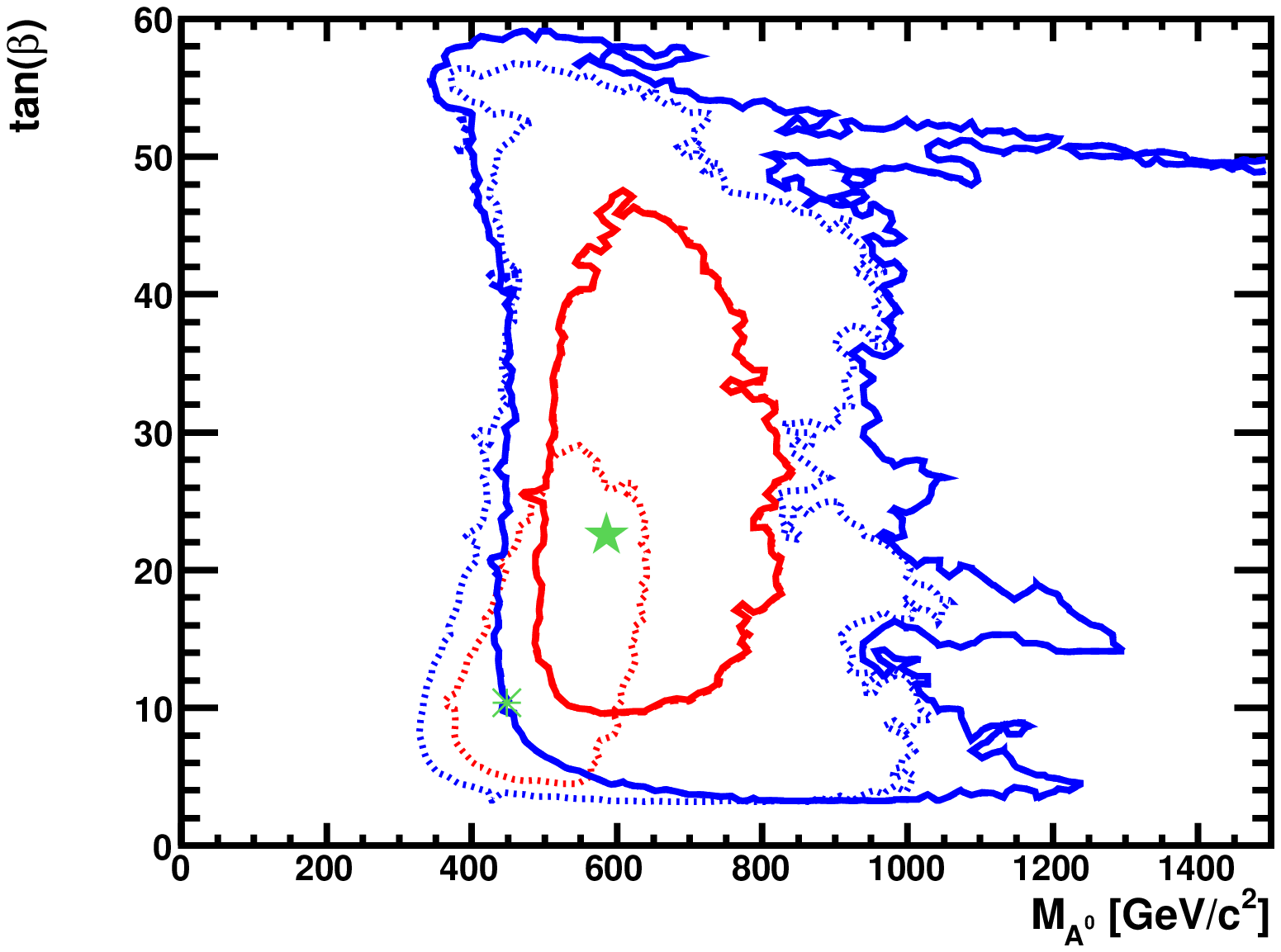}}
\resizebox{8cm}{!}{\includegraphics{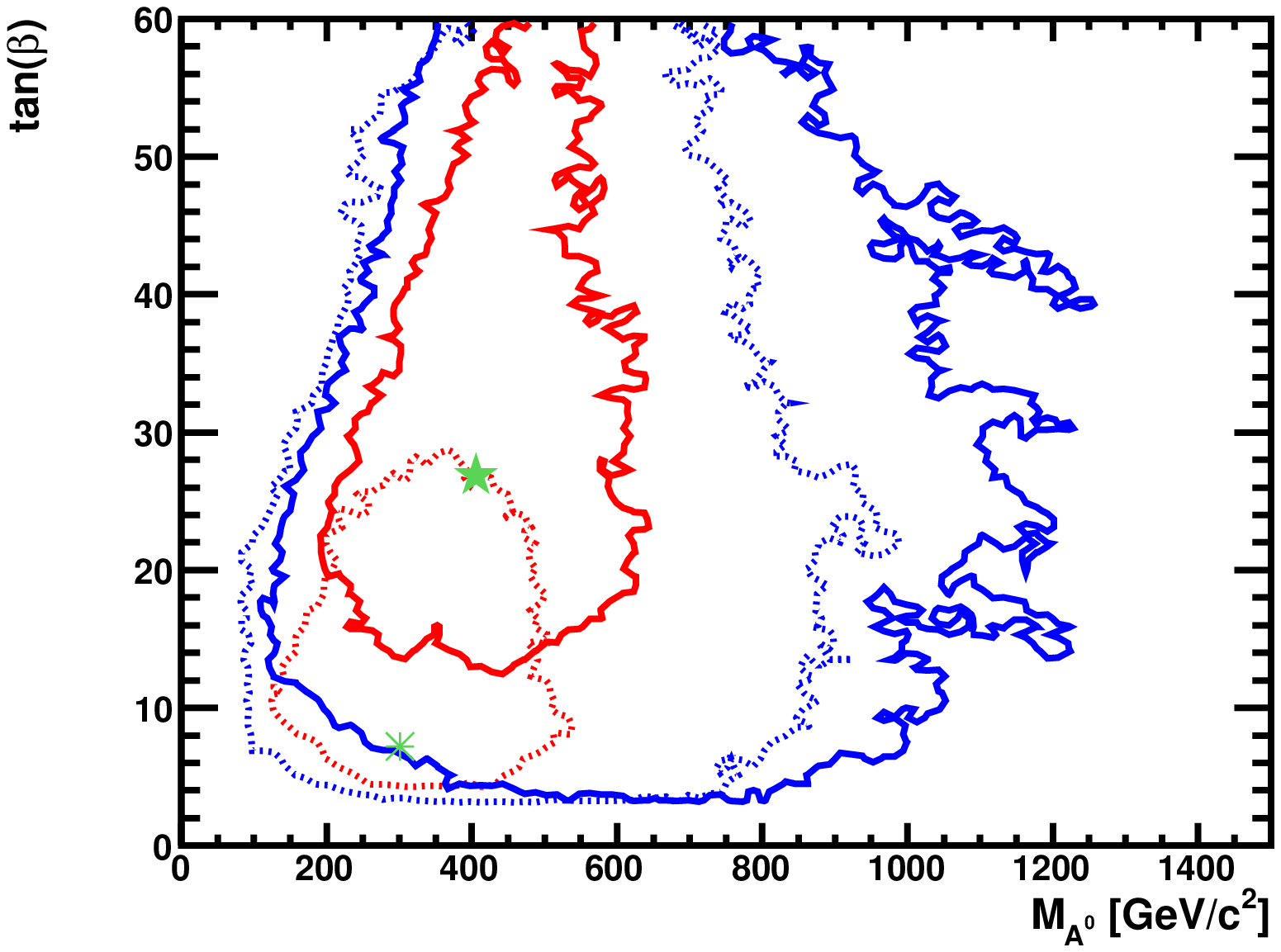}}
\resizebox{8cm}{!}{\includegraphics{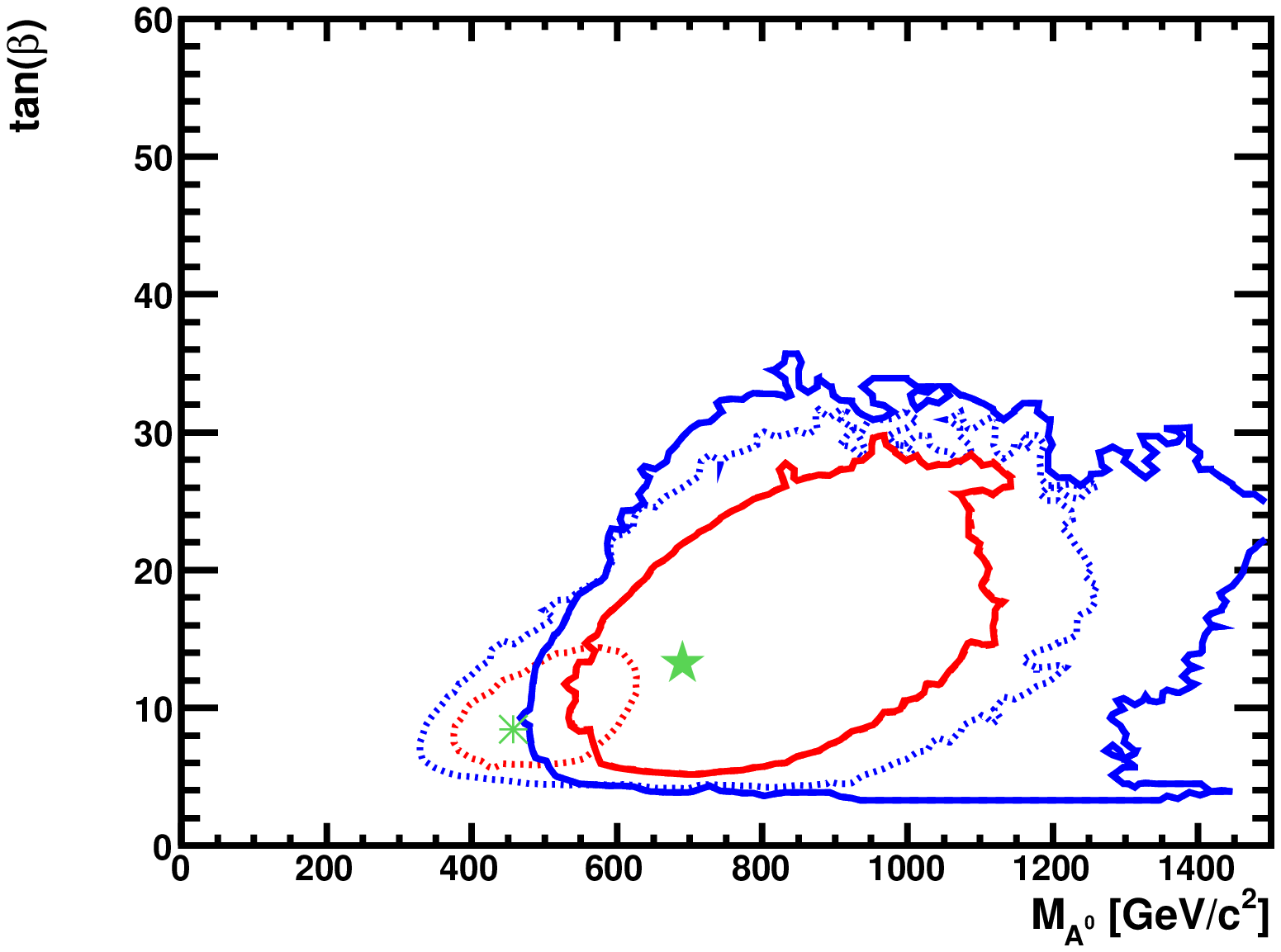}}
\resizebox{8cm}{!}{\includegraphics{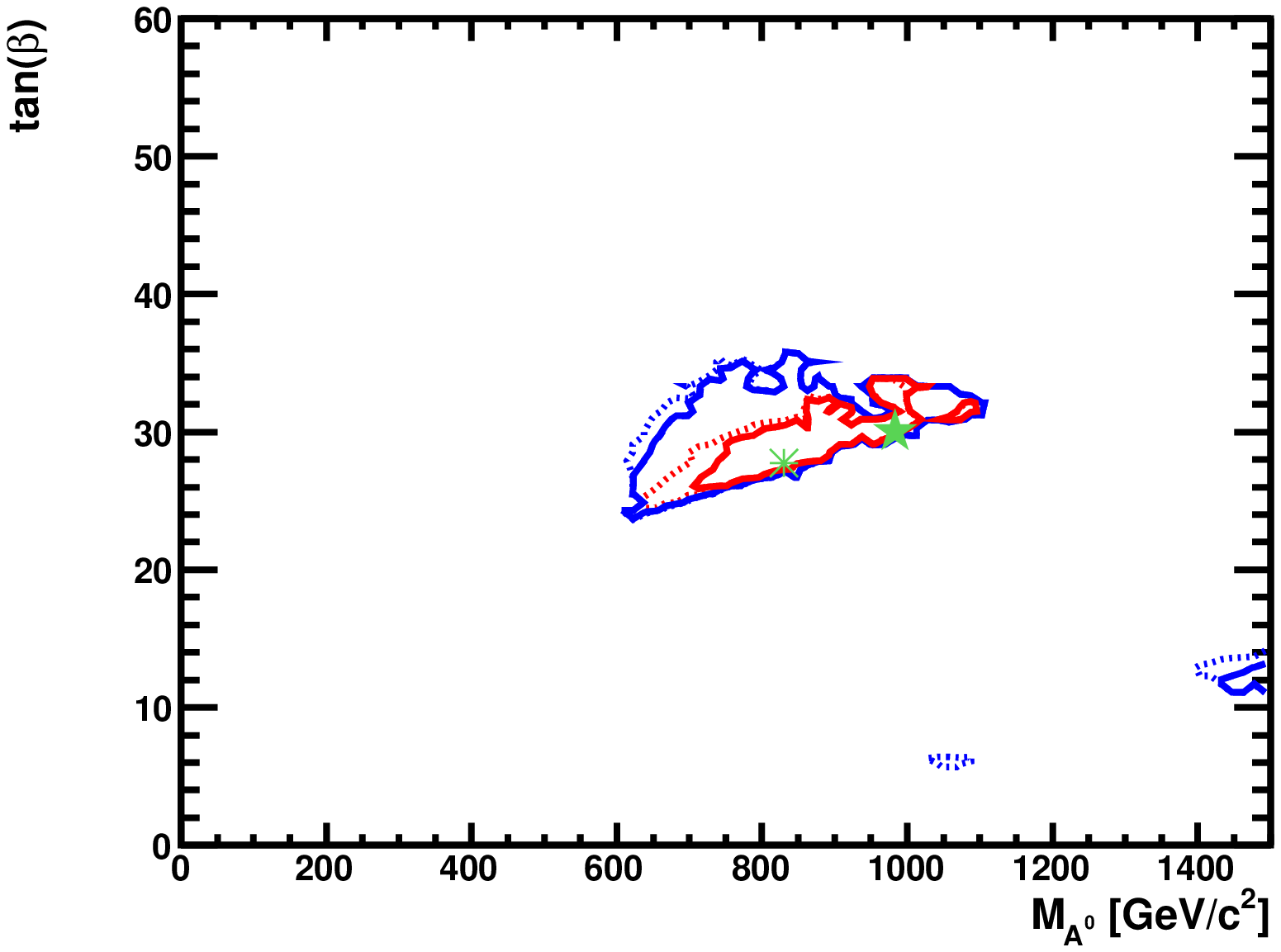}}
\vspace{-1cm}
\caption{\it The $(\MA, \tb)$ planes in the CMSSM (upper left), the
  NUHM1 (upper right), the VCMSSM (lower left) and mSUGRA (lower
  right). In each plane, the best-fit point after incorporation of the
  2010 LHC constraints is indicated by a filled green star, and the
  pre-LHC fit by an open star. The 68 and 95\% CL regions are indicated
  by red and blue contours, respectively, the solid lines including the
  2010 LHC data, and the dotted lines including only the pre-LHC data.} 
\label{fig:tBmA}
\end{figure*}

We display in Fig.~\ref{fig:tBmA} the corresponding best-fit points and
68 and 95\%~CL regions in the $(\MA, \tb)$ planes for the CMSSM, NUHM1,
VCMSSM and mSUGRA including the 2010 LHC and Xenon100 constraints. The LHC 
$b \bar b \to H/A \to \tau^+ \tau^-$ constraint has some impact in the NUHM1,
where a small part of the upper left region of the NUHM1 $(\MA, \tb)$
plane has been disfavoured by this new constraint, whereas the previous
Tevatron constraints on $H/A$ production had not impacted significantly
the parameter spaces of any of the models.

\begin{figure*}[htb!]
\resizebox{8cm}{!}{\includegraphics{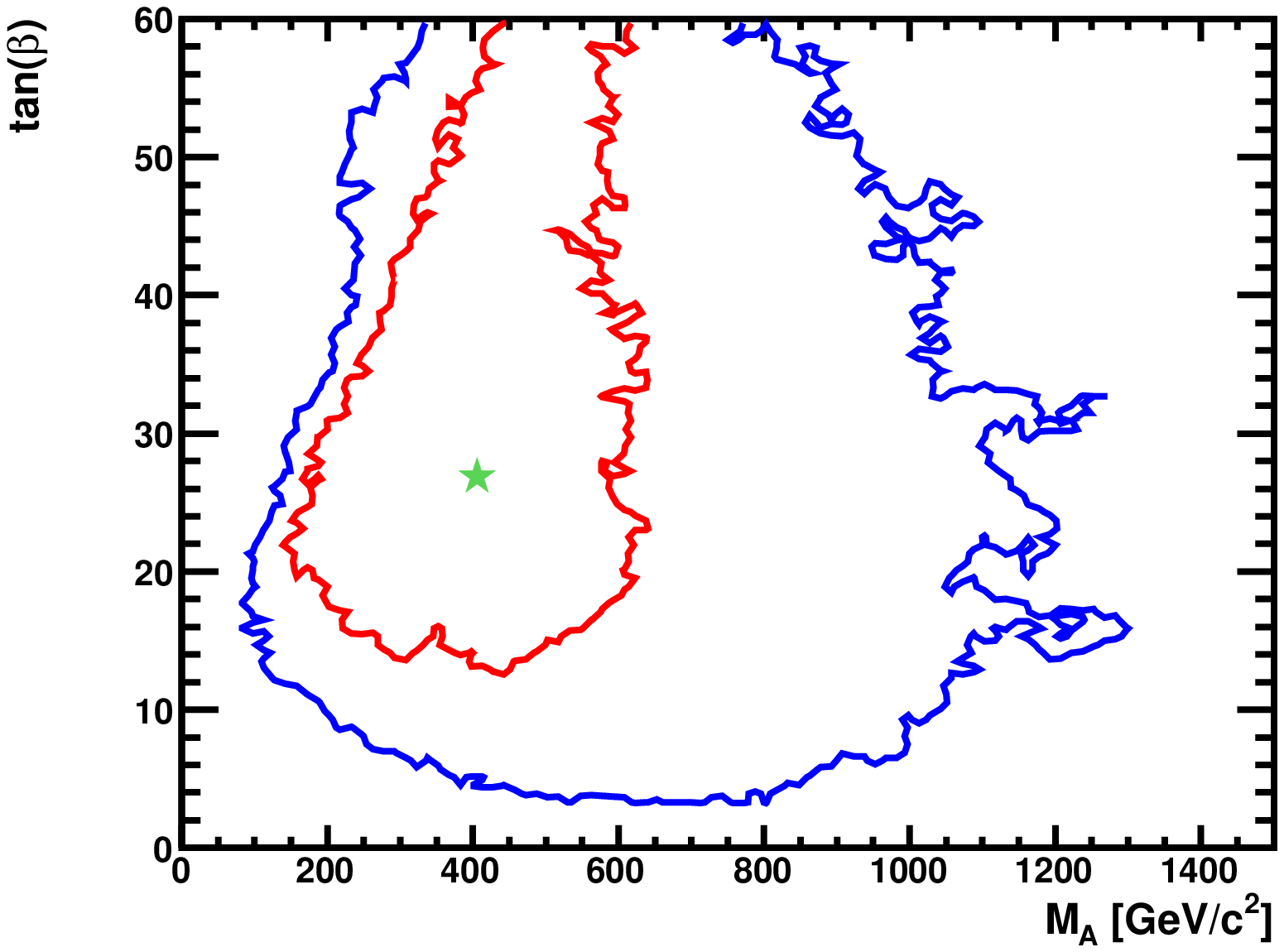}}
\resizebox{8cm}{!}{\includegraphics{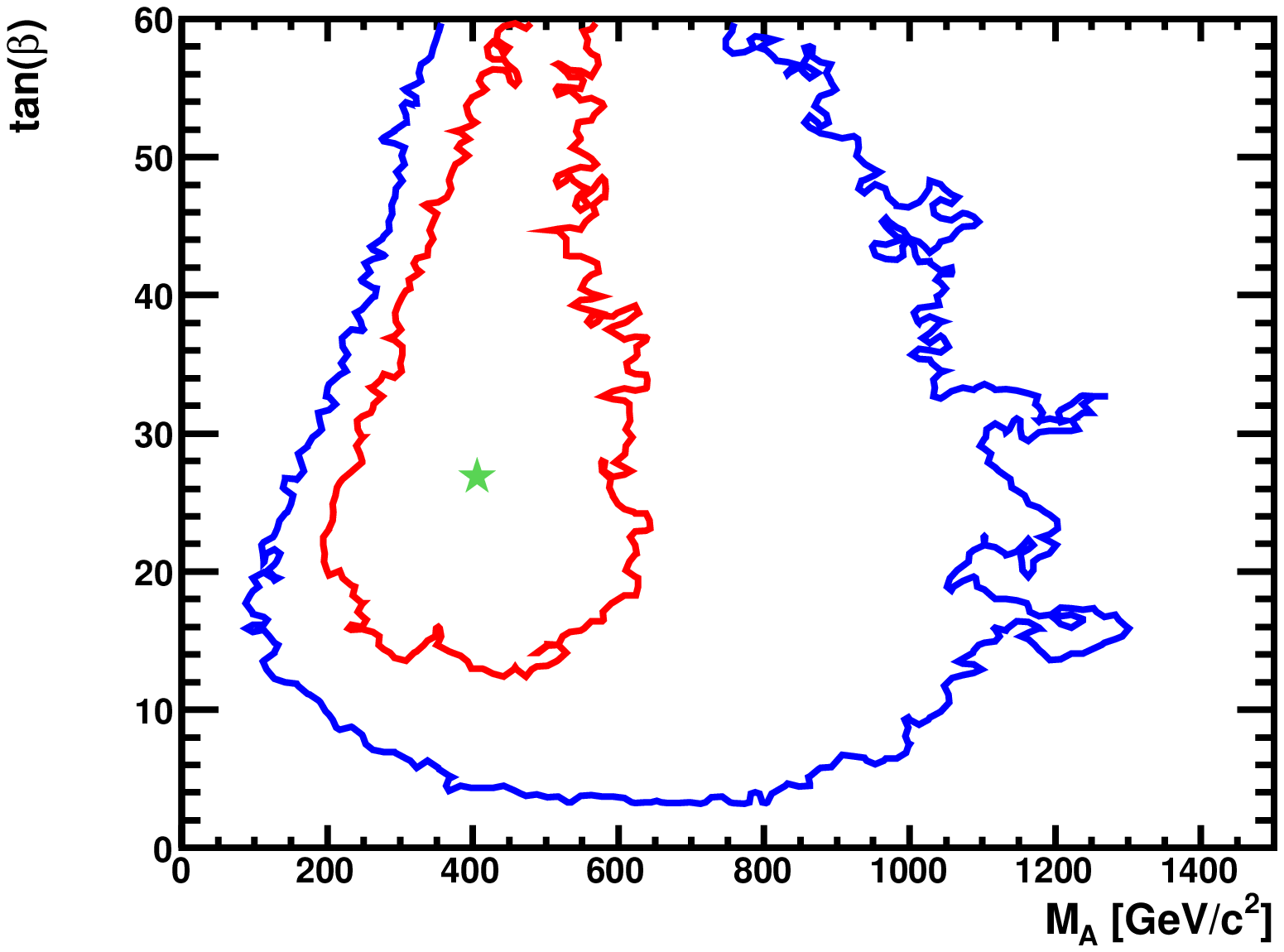}}
\vspace{-1cm}
\caption{\it The $(\MA, \tan \beta)$ planes in the NUHM1 with neither
the CMS $H/A \to \tau^+ \tau^-$ constraint nor the \bmm\ constraint applied
(left) and
with both the 
$H/A$ and the \bmm\ constraints (right). In each plane, the best-fit point is indicated by a filled 
green star, and the 68 and 95\% CL regions are indicated
by red and blue contours, respectively.
}
\label{fig:HttBsmmeffect}
\end{figure*}

Fig.~\ref{fig:HttBsmmeffect} illustrates the effects of the CMS $H/A$
constraint and the LHCb/CDF/D\O\ \bmm\ constraint on the $(\MA, \tb)$ plane
in the NUHM1. The other LHC constraints are
applied in all panels, but not the Xenon100 constraint. 
The left panel {\it drops} both the $H/A \to \tau^+ \tau^-$ and \bmm\
constraints, and the right panel includes {\it both} constraints, and
we note two principal effects. One is a contraction in the 68\% CL region at lower $\MA$,
resulting in the 68\% CL lower limit on $\MA$ increasing from $\sim 150 \gev$
to $\sim 200 \gev$, which is due to the $H/A$ constraint. The other effect is
some erosion of the 68\% CL region at large $\tan \beta > 50$, reducing the
upper limit on $\MA$ from $\sim 600 \gev$ to $\sim 550 \gev$, which is due to the
\bmm\ constraint. However, we observe that the location of the best-fit point at
$(\MA, \tan \beta) \sim (400 \gev, 26)$ is quite insensitive to these constraints,
indicating that they are not yet attacking the `heartland' of the NUHM1 parameter space.

\subsection*{\it $(A_0/m_0, \tan \beta)$ planes}

\begin{figure*}[htb!]
\resizebox{8cm}{!}{\includegraphics{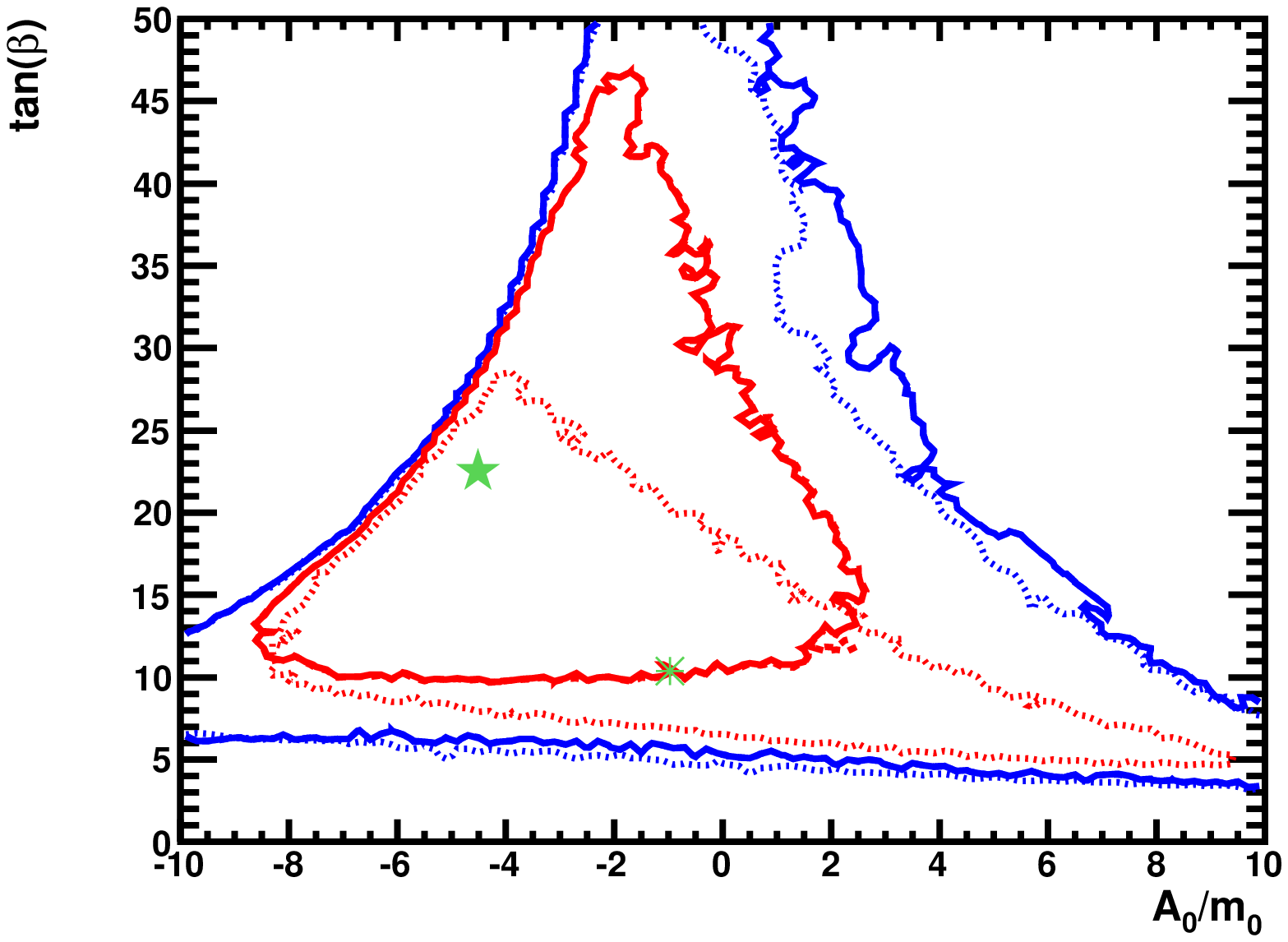}}
\resizebox{8cm}{!}{\includegraphics{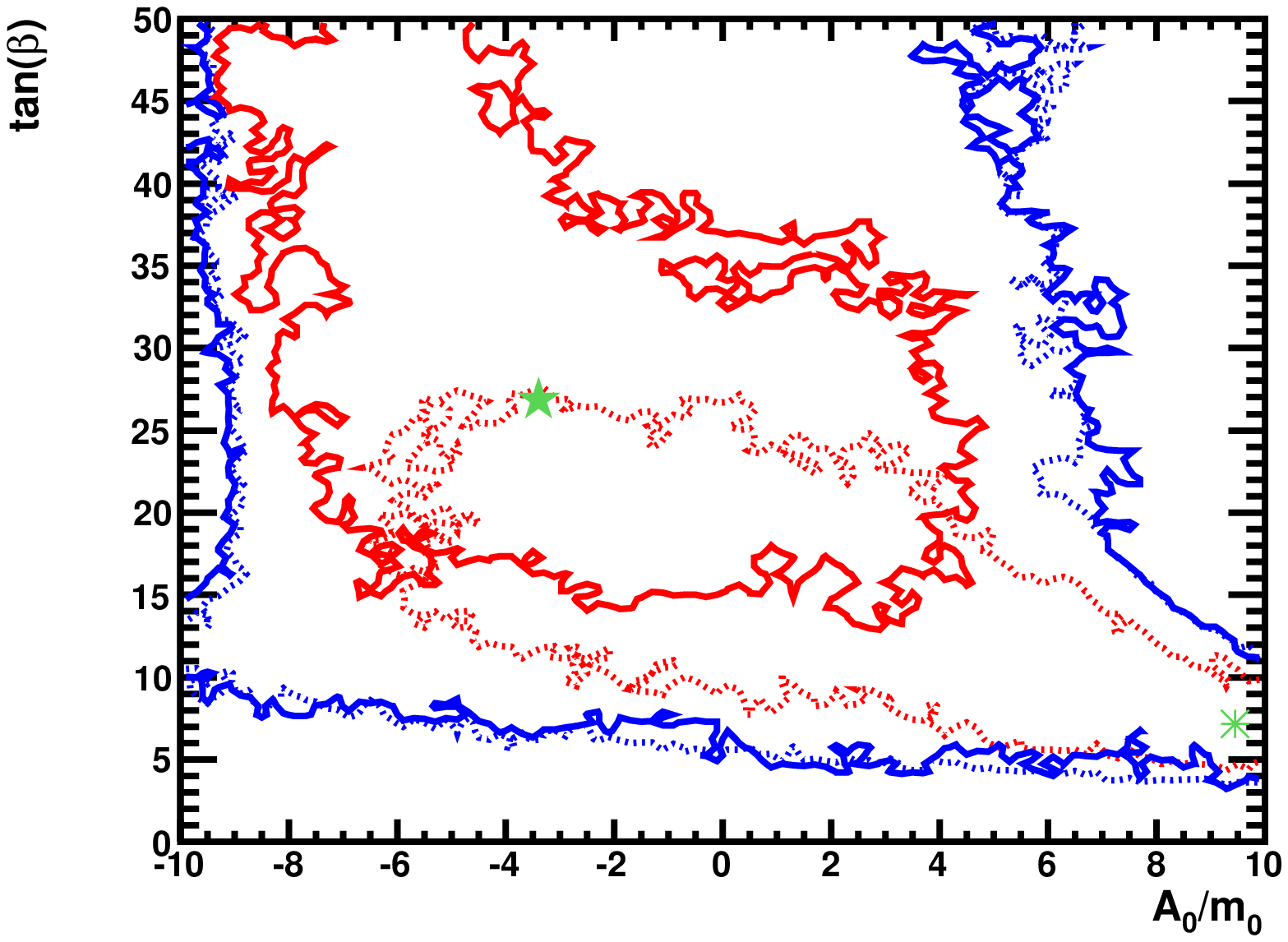}}
\resizebox{8cm}{!}{\includegraphics{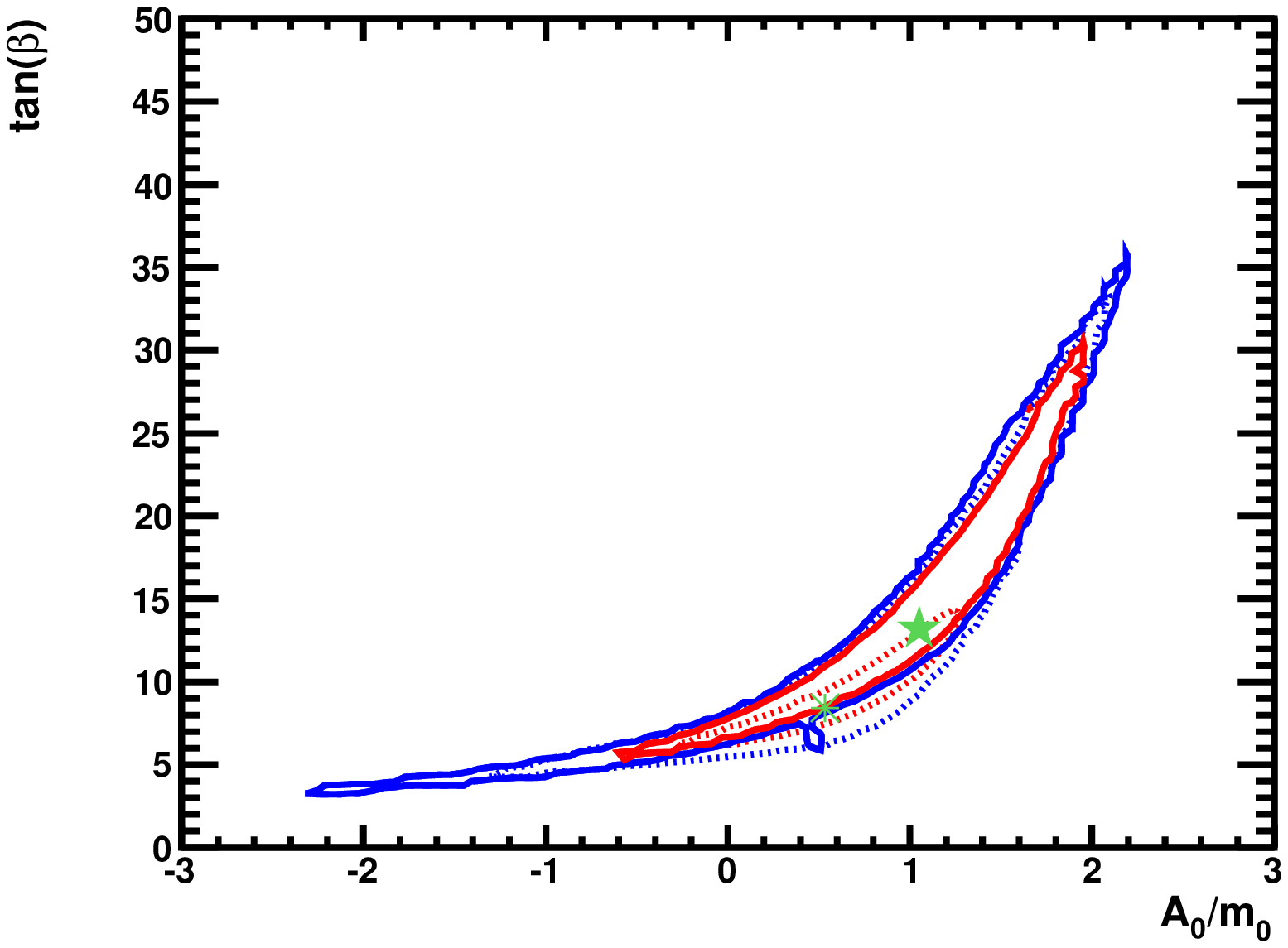}}
\resizebox{8cm}{!}{\includegraphics{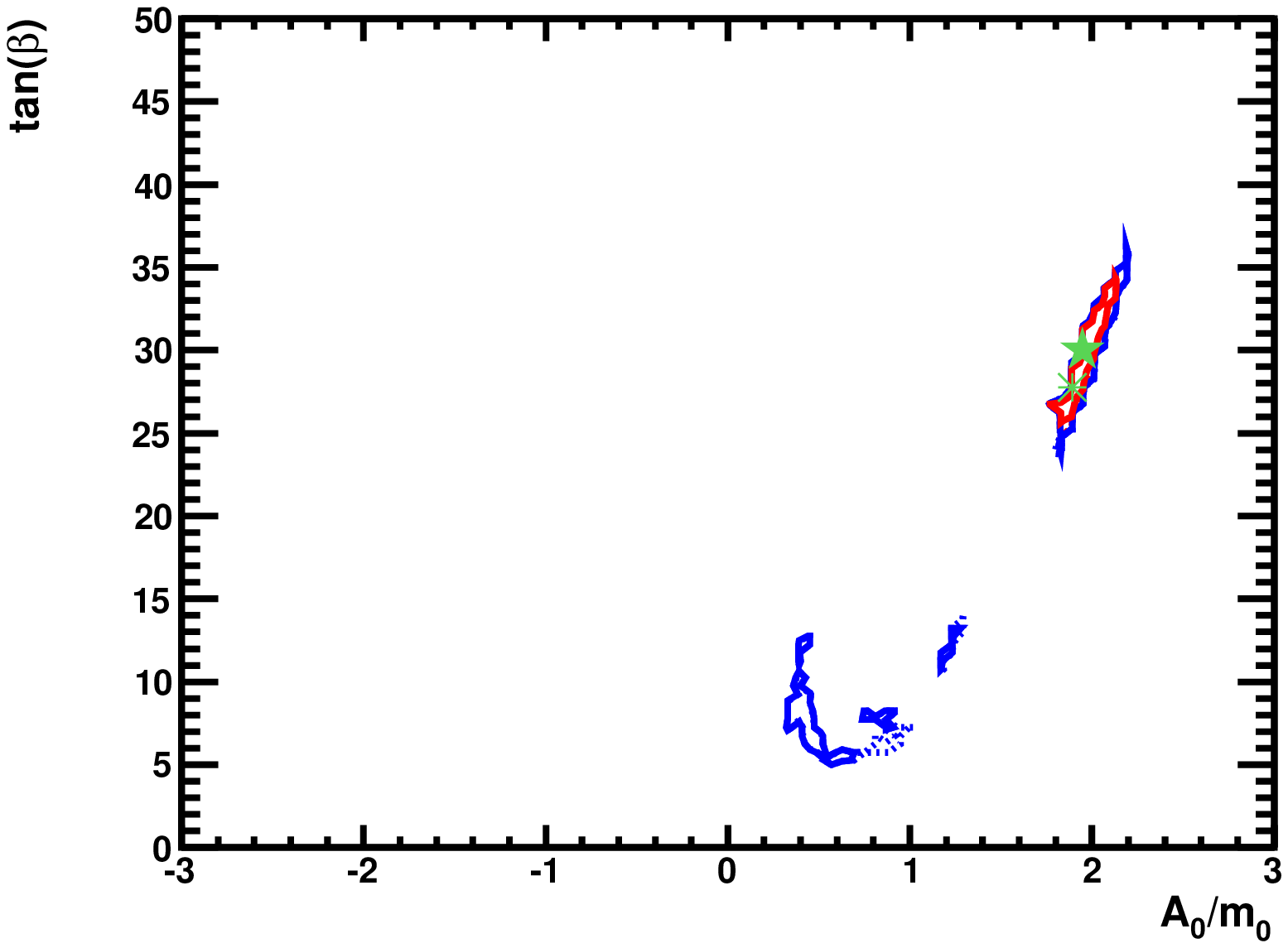}}
\vspace{-1cm}
\caption{\it The $(A_0/m_0, \tan \beta)$ planes
in the CMSSM (upper left panel), in the NUHM1 (upper right panel), 
in the VCMSSM (lower left panel) and in mSUGRA (lower right panel).
In each plane, the best-fit point after incorporation of the
2010 LHC and Xenon100 constraints is indicated by a filled green star, and the
pre-LHC fit by an open star. The 68 and 95\% CL regions are indicated
by red and blue contours, respectively, the solid lines including the
2010 LHC and Xenon100 data, and the dotted lines including only the pre-LHC data.
}
\label{fig:A0m0tB}
\end{figure*}

Fig.~\ref{fig:A0m0tB} displays the $(A_0/m_0, \tan \beta)$ planes
in the CMSSM (upper left panel), in the NUHM1 (upper right panel), 
in the VCMSSM (lower left panel) and in mSUGRA (lower right panel).
We see that the effect of the 2010 LHC constraints in the CMSSM is to
push the preferred region towards negative values of $A_0/m_0$,
largely as a result of the push towards larger values of $\tan \beta$
required to reconcile the LHC data with \gmt. The effects of the available
constraints in this plane are weaker in the NUHM1, particularly for
larger values of $\tan \beta$. In the cases of the VCMSSM and mSUGRA,
we see that the $(A_0/m_0, \tan \beta)$ planes are qualitatively similar
to those shown in~\cite{mc4}, the main difference being a shift of the
best-fit point to larger $A_0/m_0$ and $\tan \beta$.

\subsection*{\it Gluino mass}

\begin{figure*}[htb!]
\resizebox{8cm}{!}{\includegraphics{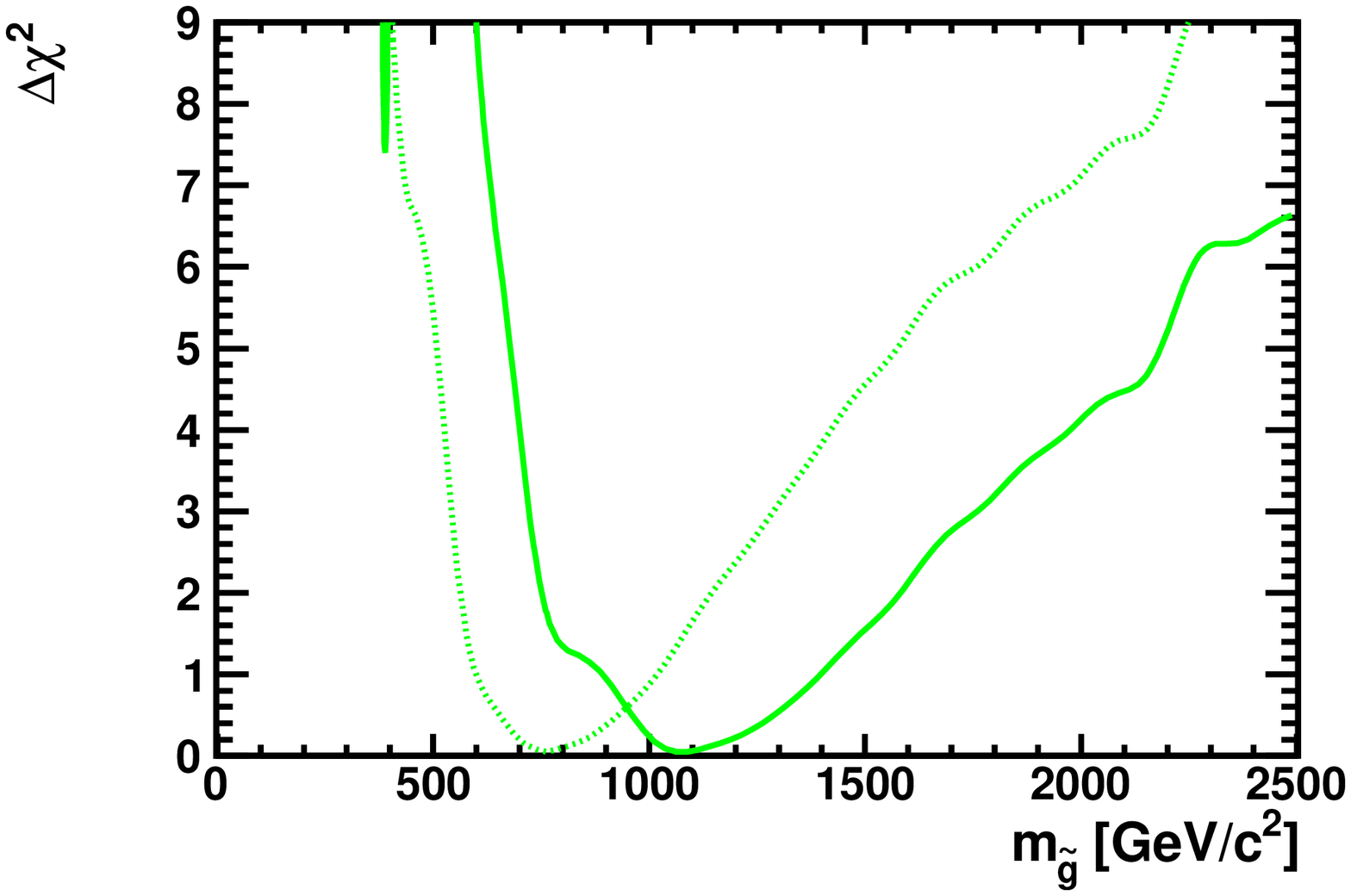}}
\resizebox{8cm}{!}{\includegraphics{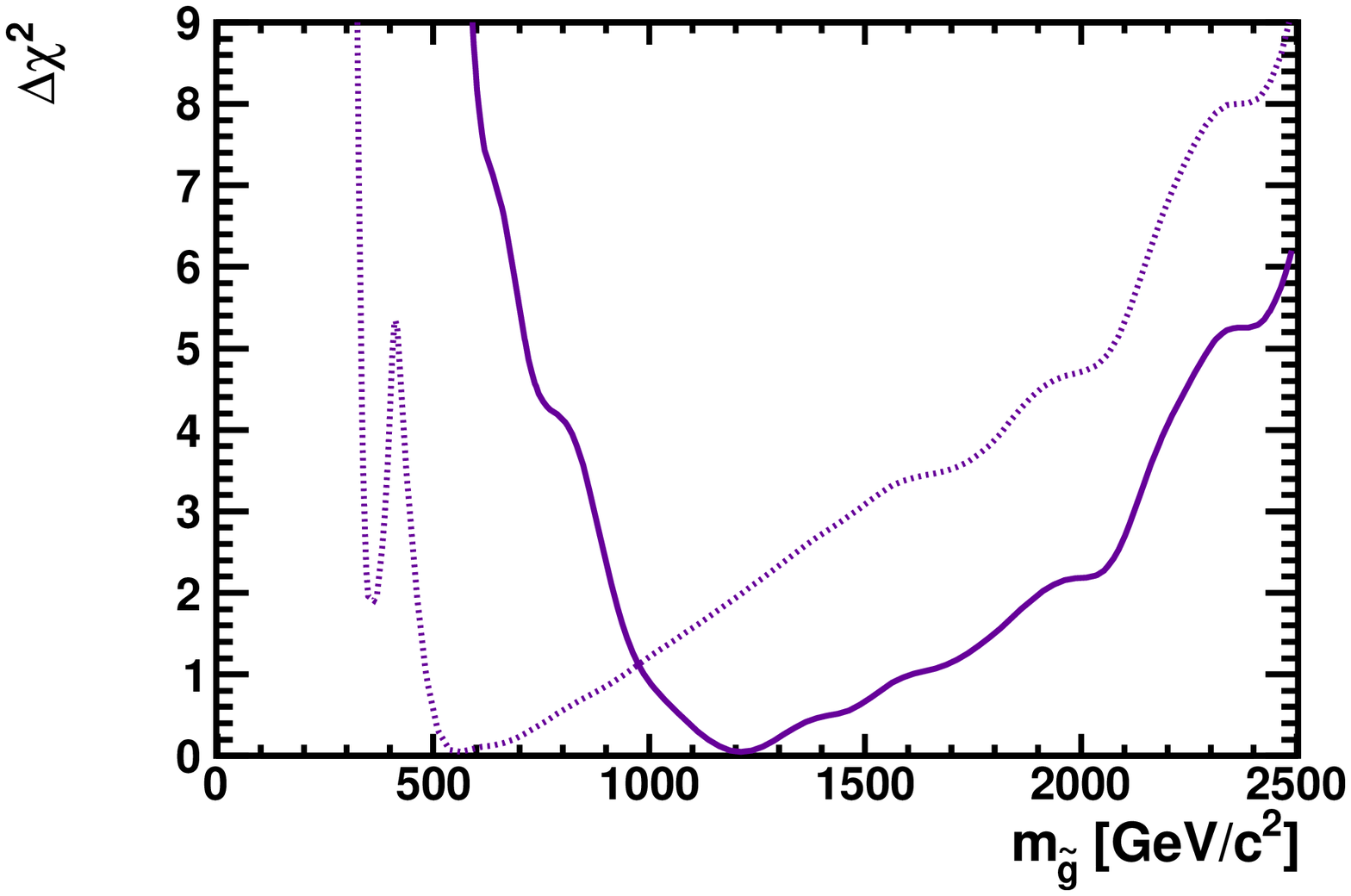}}
\resizebox{8cm}{!}{\includegraphics{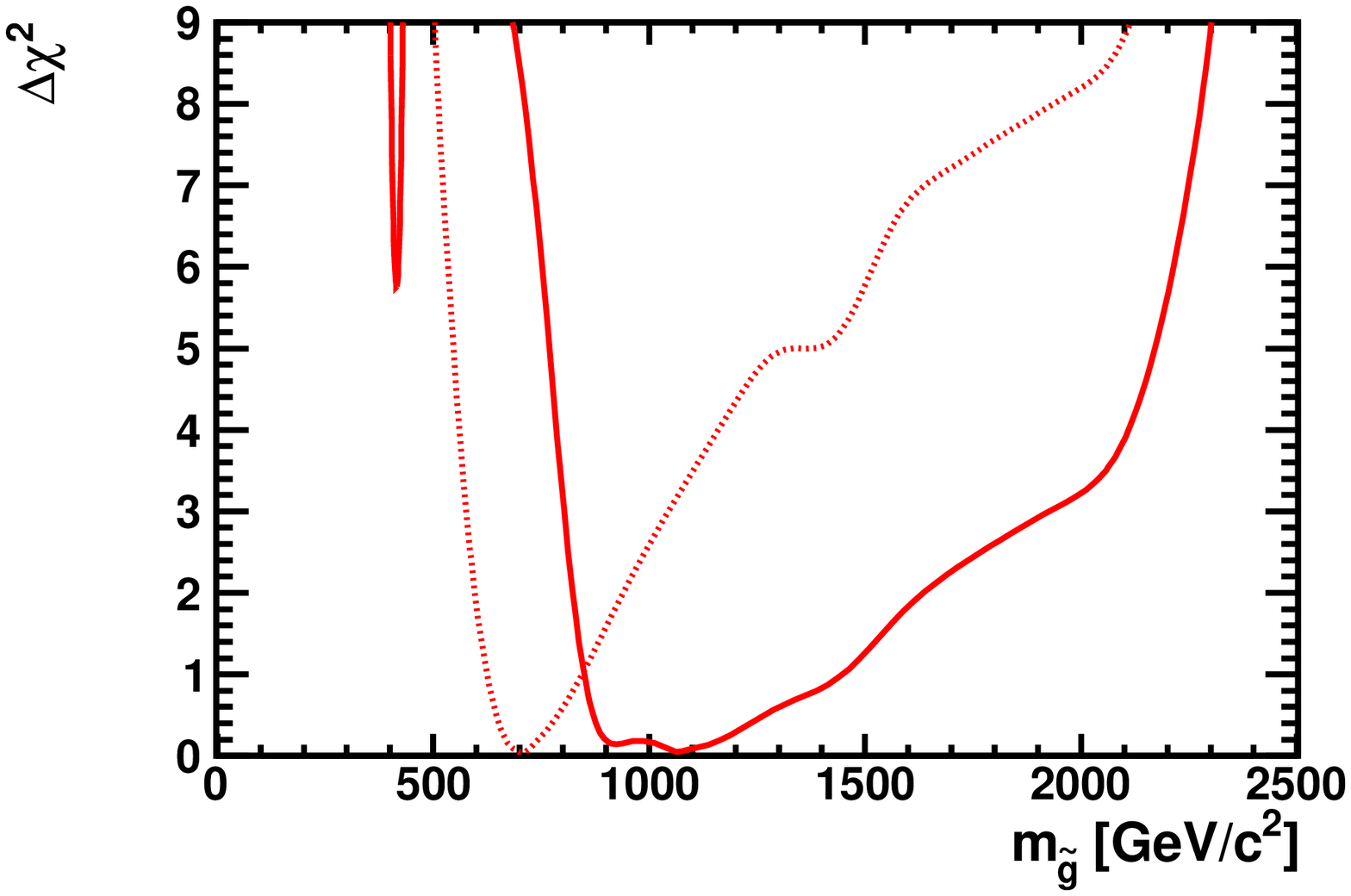}}
\resizebox{8cm}{!}{\includegraphics{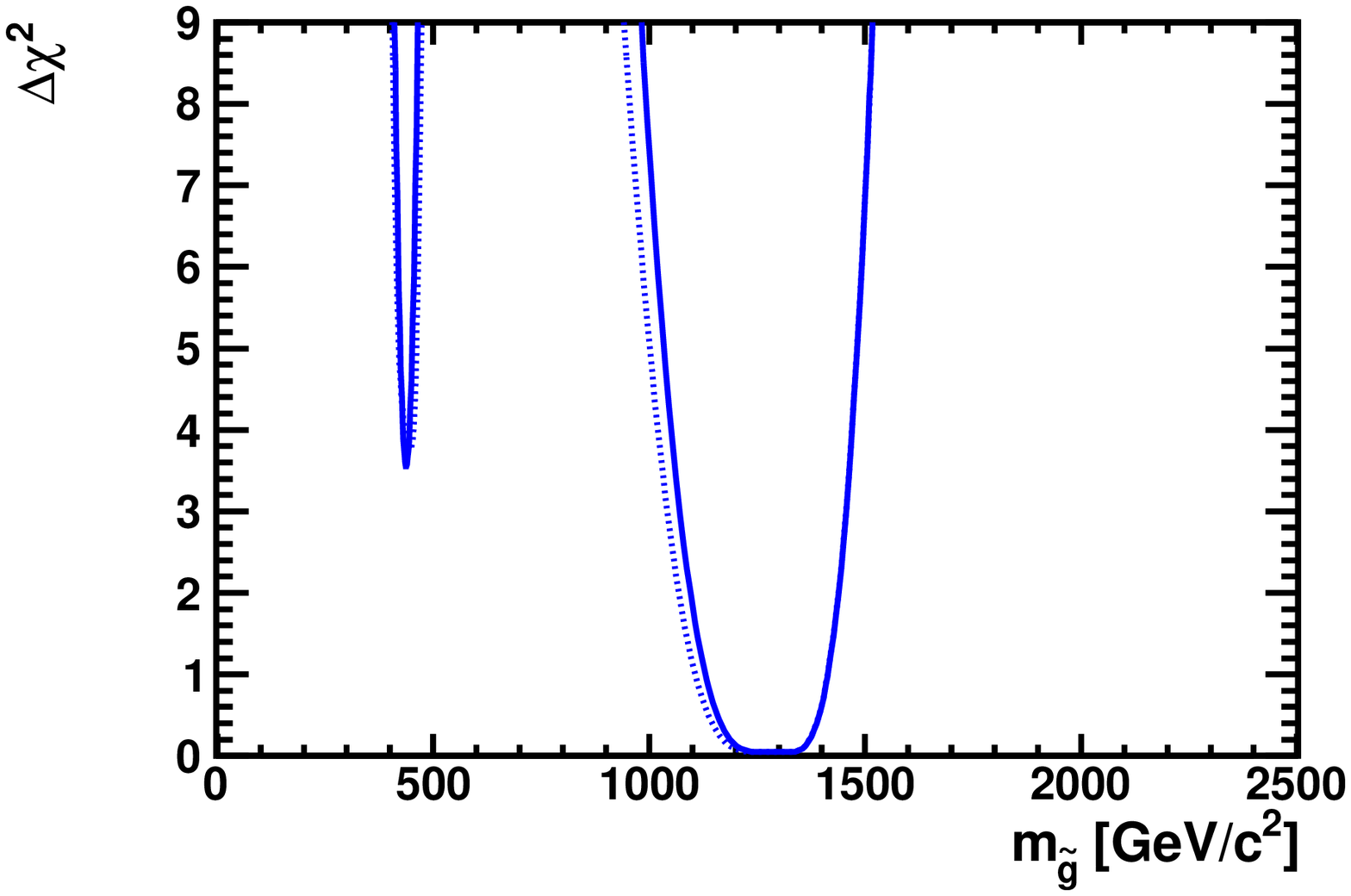}}
\vspace{-1cm}
\caption{\it The $\chi^2$ likelihood functions of $\mgl$ in the CMSSM (upper left), the NUHM1 (upper right),
the VCMSSM (lower left) and mSUGRA (lower right). The dashed
curves are derived from the pre-LHC data set, and the solid curves
include all the 2010 LHC constraints. In each case the 
value of $\Delta\chi^2$ relative to the respective best-fit point is displayed.
}
\label{fig:mgl}
\end{figure*}

Fig.~\ref{fig:mgl} illustrates the impacts of the 2010 LHC data on the
$\chi^2$ likelihood functions for $\mgl$ in the different models. 
The plots display the $\Delta\chi^2$ contributions of the different
fits relative to the respective best-fit points. The
pre-LHC likelihood functions are shown as dotted lines, and the
post-2010-LHC likelihood functions as solid lines. In each of the CMSSM,
NUHM1 and VCMSSM, the general effect of the 2010 LHC data is to increase
the preferred value of $\mgl$ by $\sim 300 \gev$ beyond our pre-LHC analyses~\cite{mc4}, reaching 
$\sim 1000 - 1300 \gev$, which is also some 100~GeV beyond
the results of our previous analyses using the initial CMS
$\alpha_T$ and ATLAS 1L searches~\cite{mc5}, whereas there is no significant effect on the likelihood function for
$\mgl$ in mSUGRA. 
Since the plots display the relative $\Delta\chi^2$ contributions, 
the differences in the overall $\chi^2$ 
between the pre- and
post-2010-LHC minima of $\sim 4 (6)$ in the CMSSM/NUHM1 (VCMSSM)
are responsible for the differences between the pre- and post-2010-LHC likelihood functions at
large $\mgl$, where the LHC constraints have no effect on the absolute
values of the $\chi^2$ functions. These new normalizations of $\chi^2$ 
are also responsible for the appearances of high-lying secondary minima at 
$\mgl \sim$ $400 \gev$ in the CMSSM and
VCMSSM, respectively, which was previously located out of sight at
$\Delta \chi^2 > 9$ for the CMSSM but has now dropped into view. These secondary
minima, like that for mSUGRA, are compatible with the astrophysical cold
dark matter density constraint thanks to rapid annihilation through a 
direct-channel light $h$ pole. The primary minima are located in the 
${\tilde \tau_1} - \neu{1}$ coannihilation regions, whereas the
focus-point regions are strongly disfavoured in our 
analysis, and not seen in any panel of Fig.~\ref{fig:mgl}.

\subsection*{\it \bmm}

\begin{figure*}[htb!]
\resizebox{8cm}{!}{\includegraphics{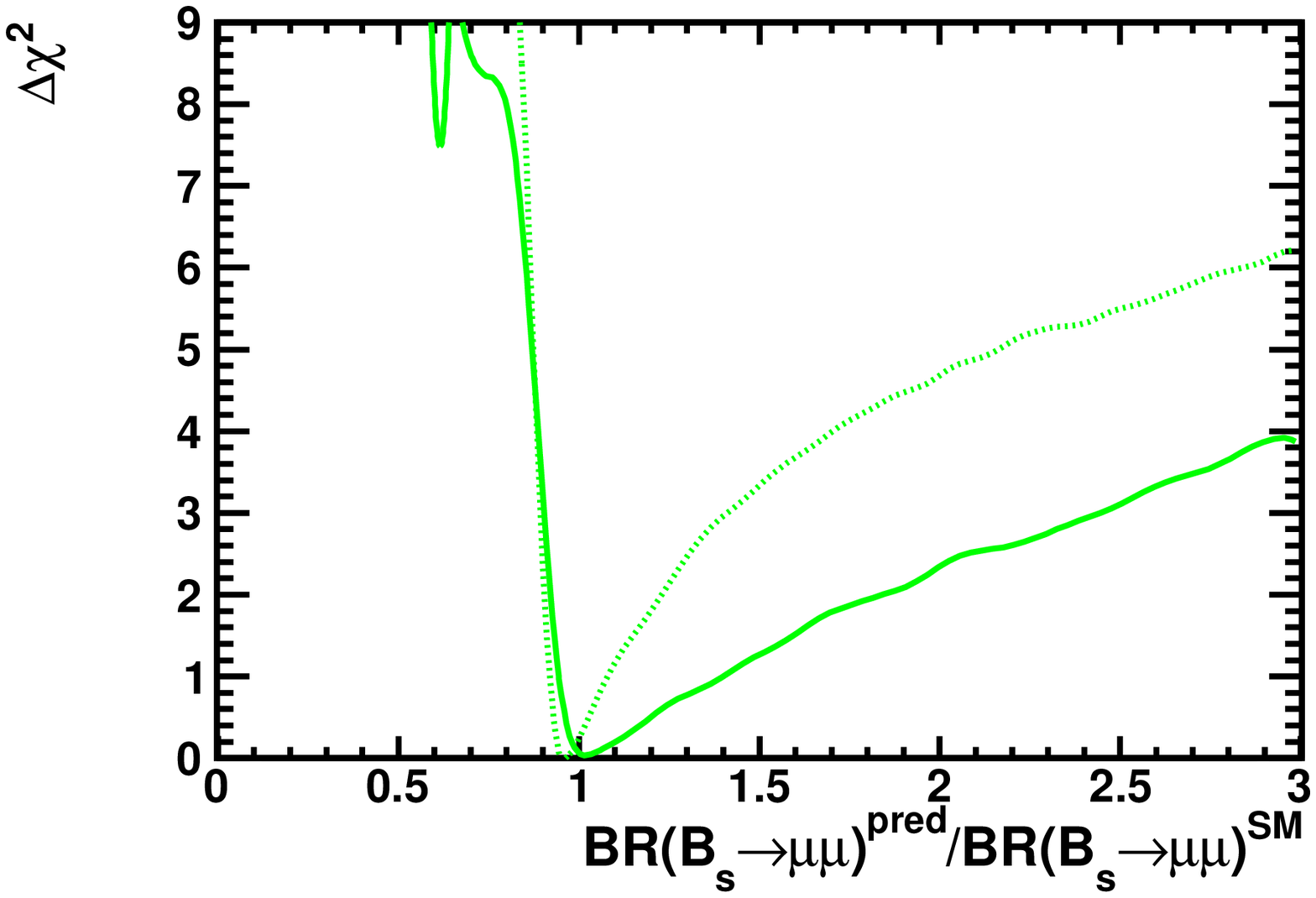}}
\resizebox{8cm}{!}{\includegraphics{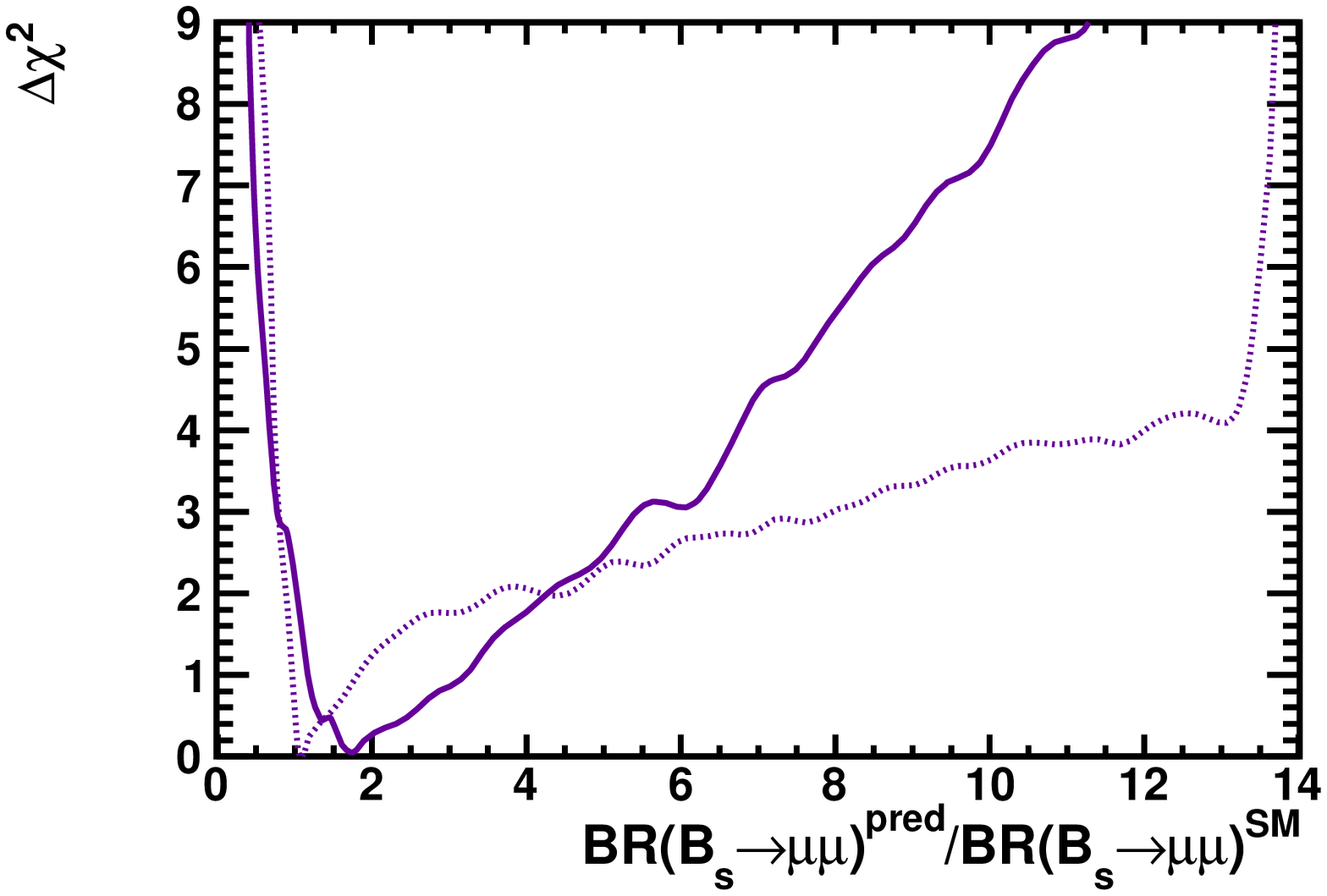}}
\resizebox{8cm}{!}{\includegraphics{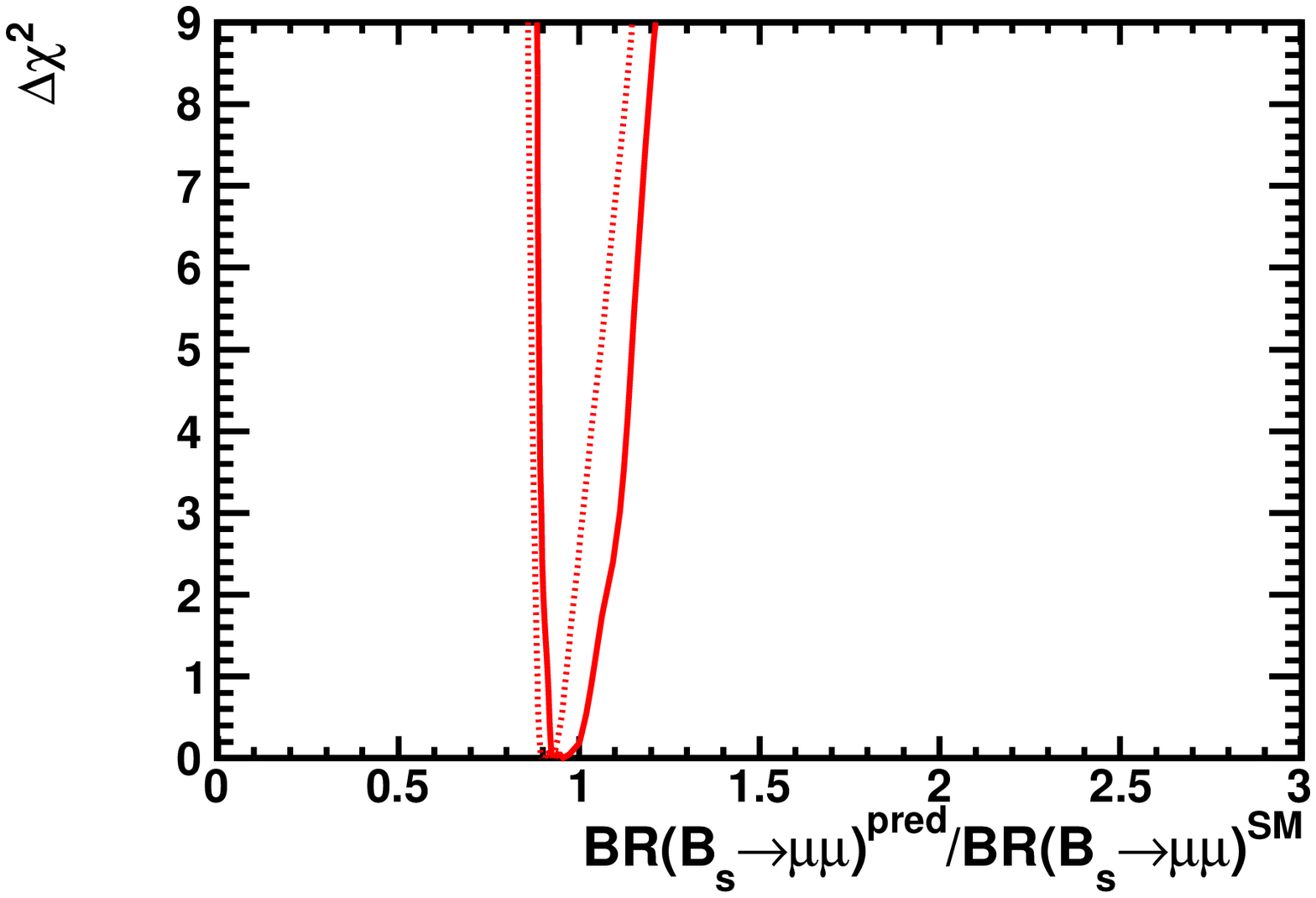}}
\resizebox{8cm}{!}{\includegraphics{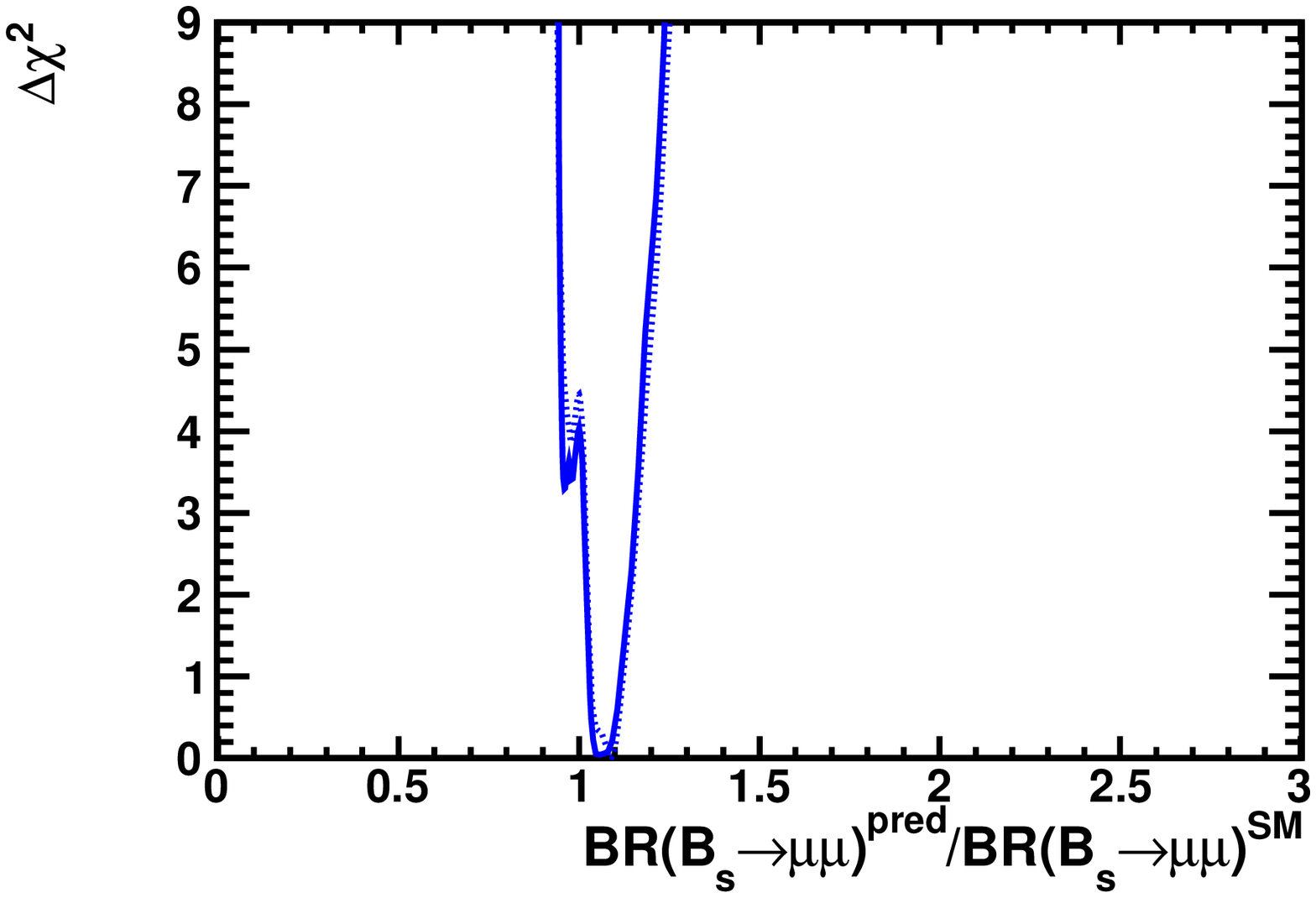}}
\vspace{-1cm}
\caption{\it The $\chi^2$ likelihood functions of \bmm\ relative to the
  SM prediction in the CMSSM (upper left), the NUHM1 (upper right), 
the VCMSSM (lower left) and mSUGRA (lower right). The dashed
curves are derived from the pre-LHC data set, and the solid curves
include all the 2010 LHC constraints. In each case the 
value of $\Delta\chi^2$ relative to the respective best-fit point is displayed.
}
\label{fig:bsmumu}
\end{figure*}

Fig.~\ref{fig:bsmumu} displays the post-2010-LHC likelihood functions for \bmm,
normalized to the SM prediction, where we see two principal effects.
In the CMSSM and, to some extent, also in the VCMSSM (upper and lower left, respectively), 
values of \bmm\ exceeding the SM prediction are less disfavoured than in the pre-LHC case.
This effect has a twofold origin. On the one hand, the LHC data disfavour a region 
of parameter space where a negative interference between SM and non-SM amplitudes 
gave rise to \bmm\ slightly below the SM prediction. 
On the other hand, the LHC data
increase the minimum of $\chi^2$ significantly, and in this figure we 
show the $\Delta\chi^2$ contribution
relative to the respective best-fit point.
Since the absolute values of $\chi^2$ at large \bmm\ are
essentially unchanged by the LHC data, the difference between the values of
$\chi^2$ at the minimum and at large \bmm\ are also reduced by $\sim 4 (6)$ in the CMSSM/NUHM1 (VCMSSM).
In the NUHM1 case we observe another important effect: here 
values of \bmm\ much greater than the SM value (by a factor more than about 6) 
are now more disfavoured than in our previous analysis. This is due to the implementation 
of the LHCb, CDF and D\O~constraints on \bmm\ that increases substantially 
the $\chi^2$ values at large \bmm.
Nevertheless, we stress that a value of \bmm\ that is substantially larger than the
SM value is still more likely in the NUHM1 than in the other models (note the
different horizontal scale used for the NUHM1).

\subsection*{\it Light Higgs mass predictions}

\begin{figure*}[htb!]
\resizebox{8cm}{!}{\includegraphics{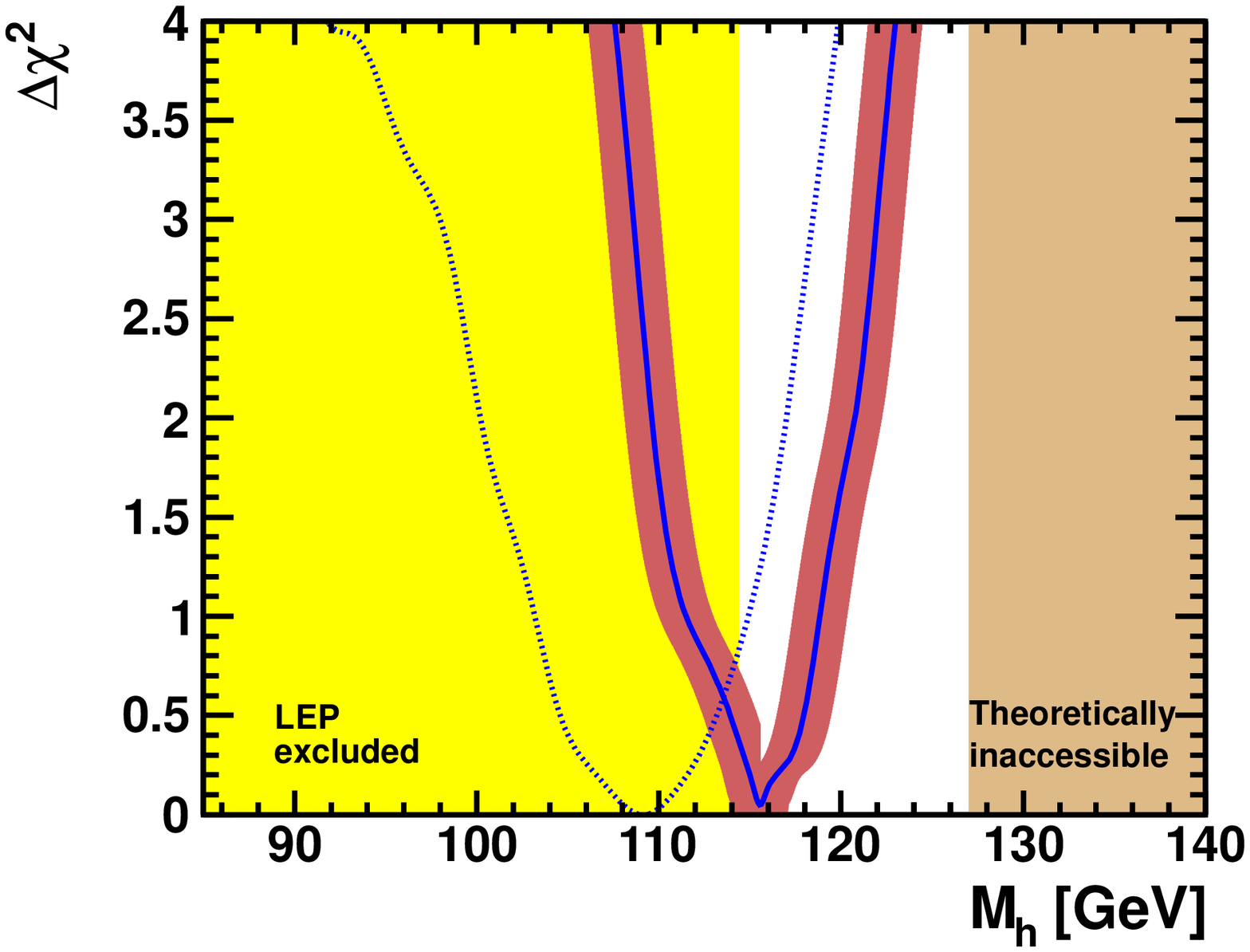}}
\resizebox{8cm}{!}{\includegraphics{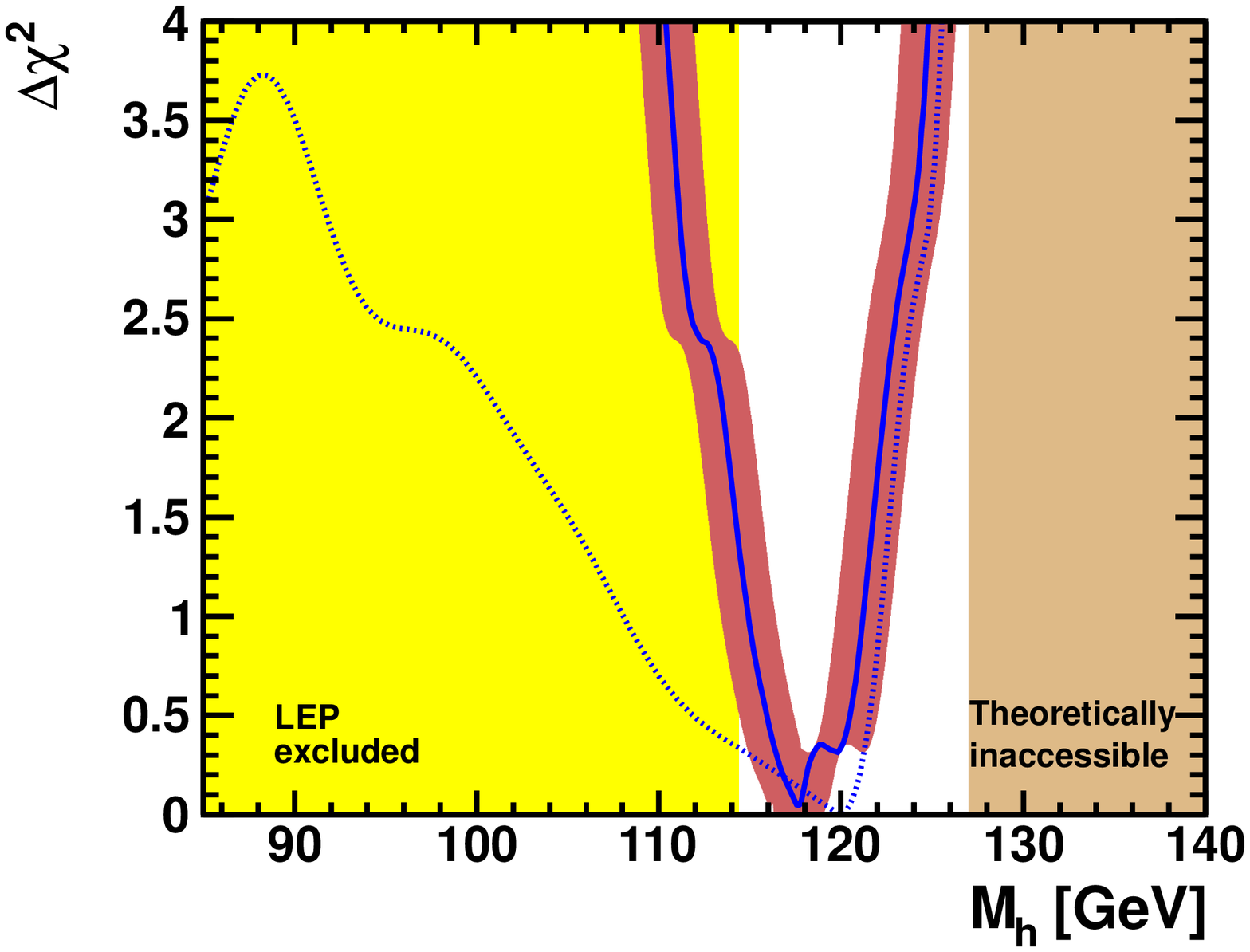}}
\resizebox{8cm}{!}{\includegraphics{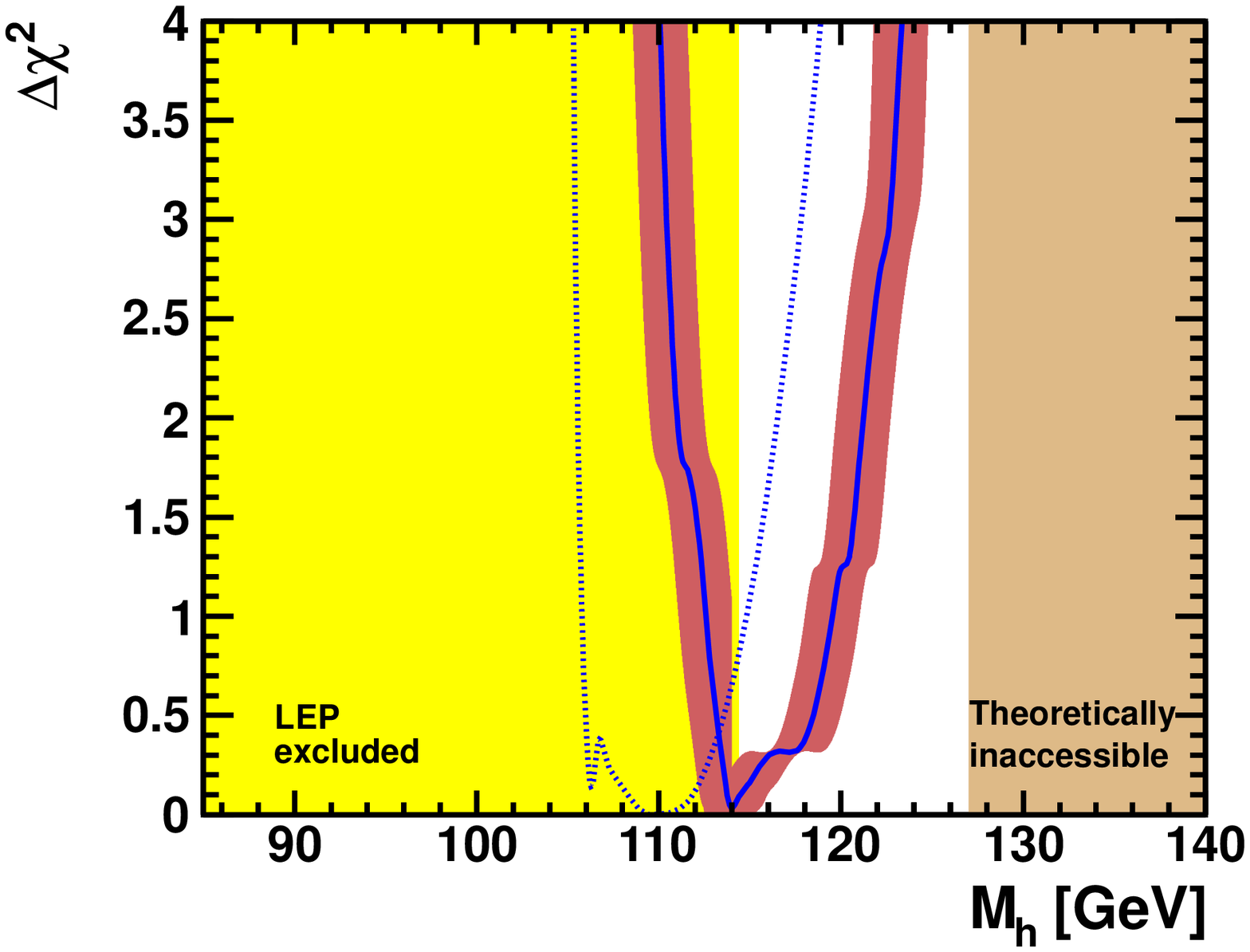}}
\resizebox{8cm}{!}{\includegraphics{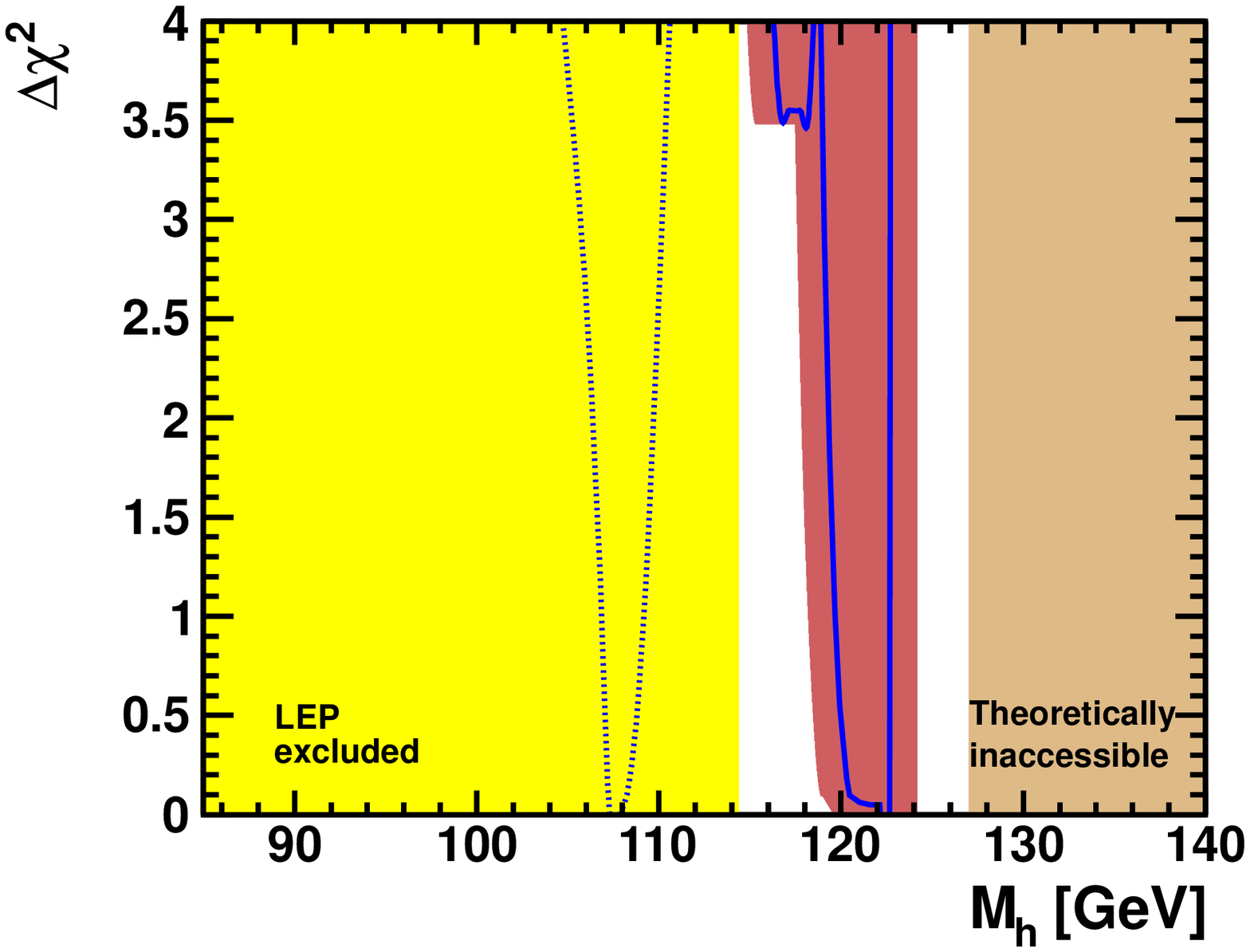}}\\
\caption{\it The one-parameter $\chi^2$ likelihood functions for the
  lightest MSSM Higgs mass $\Mh$ in the CMSSM (upper left),
NUHM1 (top right), VCMSSM (lower left) and mSUGRA (lower right). In each
  panel, we show the $\chi^2$ functions of the post-2010-LHC/Xenon100
  constraints as 
  solid lines, with a red band indicating the estimated theoretical
  uncertainty in the calculation of $\Mh$
of $\sim 1.5 \gev$, and the pre-LHC $\chi^2$ function is shown as a
  dashed line.}
\label{fig:mh}
\end{figure*}

In Fig.~\ref{fig:mh} the one-parameter $\chi^2$ functions for 
the lightest MSSM Higgs mass $\Mh$ in the CMSSM, NUHM1, VCMSSM 
and mSUGRA are shown. In this figure we {\it do not\/} include the
direct limits from 
LEP~\cite{Barate:2003sz,Schael:2006cr} or the Tevatron, so as to 
illustrate whether there is a conflict between these 
limits and the predictions of supersymmetric models.
For each model we display the new likelihood functions corresponding to
the post-2010-LHC data set, indicating the
theoretical uncertainty in the calculation of $\Mh$ of $\sim 1.5 \gev$
by red bands. We also show, as dashed lines without red bands, the
central value of the pre-LHC results (also discarding the LEP constraint).

One can see that in the CMSSM, VCMSSM and mSUGRA the heavier preferred spectra of the
post-2010-LHC fits result in somewhat higher best-fit predictions for 
$\Mh$. Only in the NUHM1, where the minimum was very shallow pre-LHC,
does the best-fit value come out slightly lower. 
Now all four models predict, excluding the LEP constraint, best-fit values for $\Mh$ {\em above} the
SM LEP limit of $114.4 \gev$~\cite{Barate:2003sz,Schael:2006cr}. 
One other significant effect of the 2010 LHC data on the one-parameter
$\chi^2$ function in the NUHM1 is seen in the region $\Mh < 110 \gev$.
We recall that in the NUHM1 the LEP constraint is weakened at low $\Mh$ 
because the $Z-Z-h$ coupling may be reduced: the 2010
LHC data help to close this loophole. Now most of the preferred
$\Mh$ region in the NUHM1 is indeed above $\sim 114 \gev$,
a tendency that was visible already in~\cite{mc5}.

\subsection*{\it Spin-independent dark matter scattering}

\begin{figure*}[htb!]
\resizebox{8cm}{!}{\includegraphics{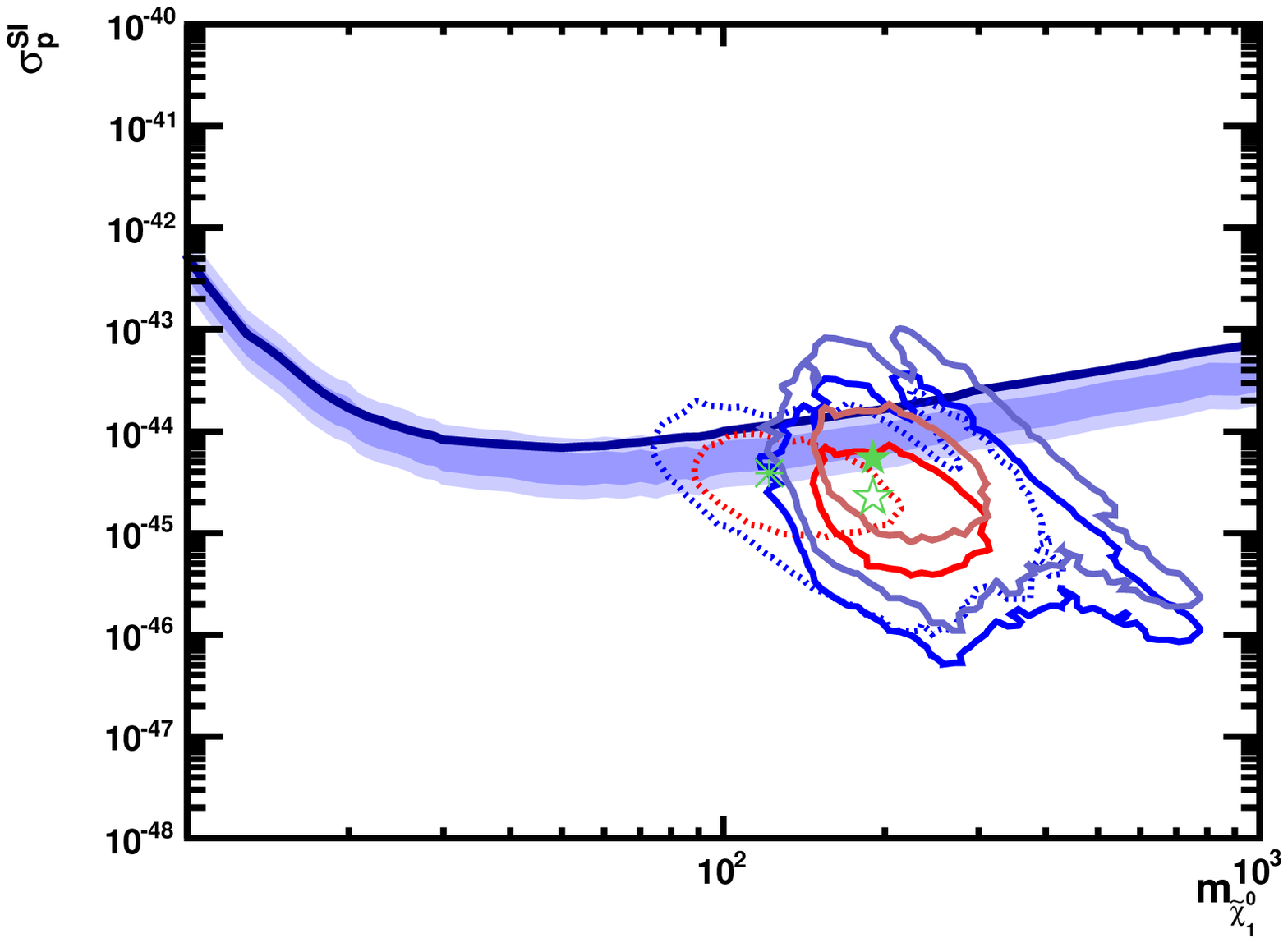}}
\resizebox{8cm}{!}{\includegraphics{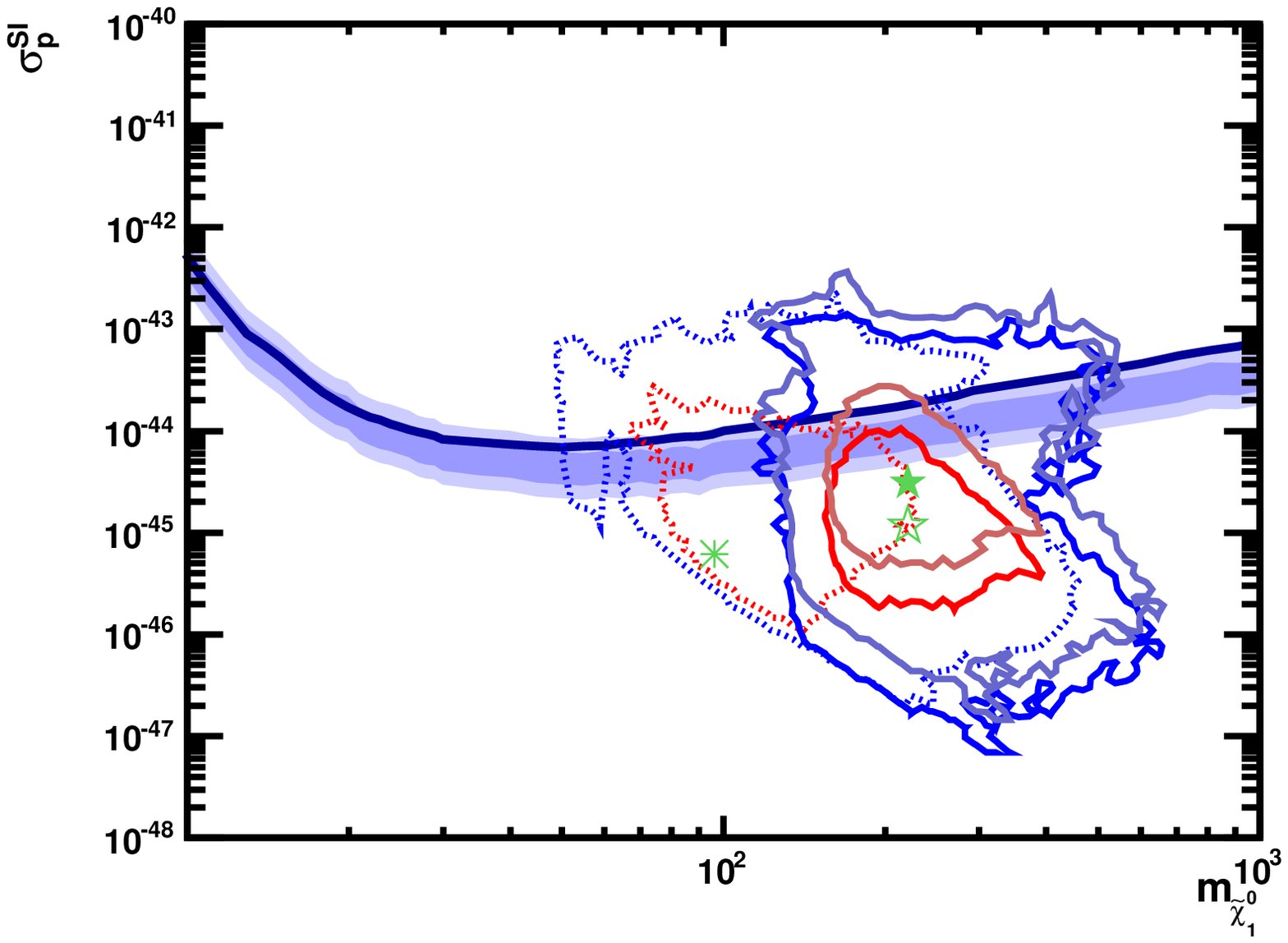}}
\resizebox{8cm}{!}{\includegraphics{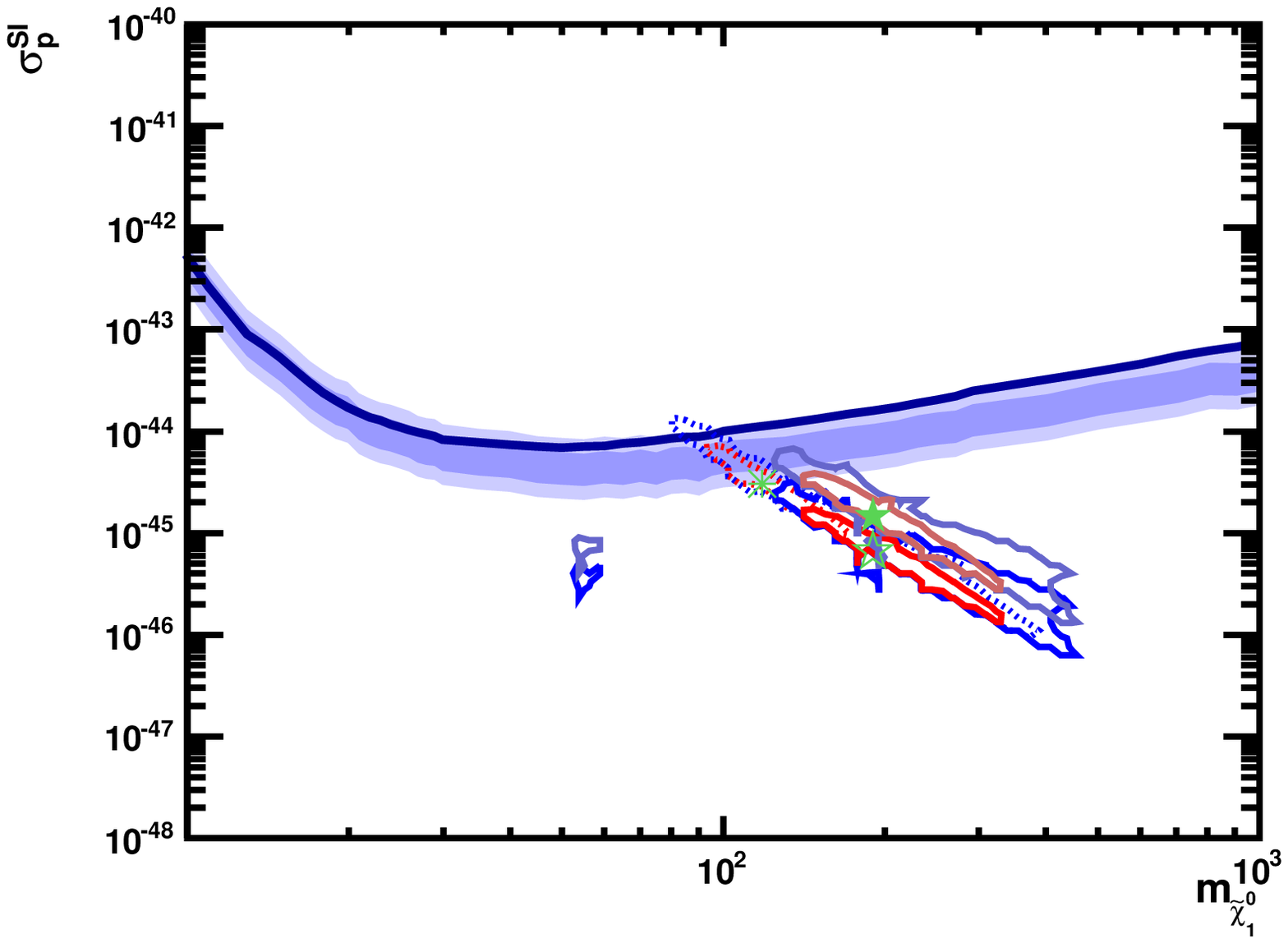}}
\resizebox{8cm}{!}{\includegraphics{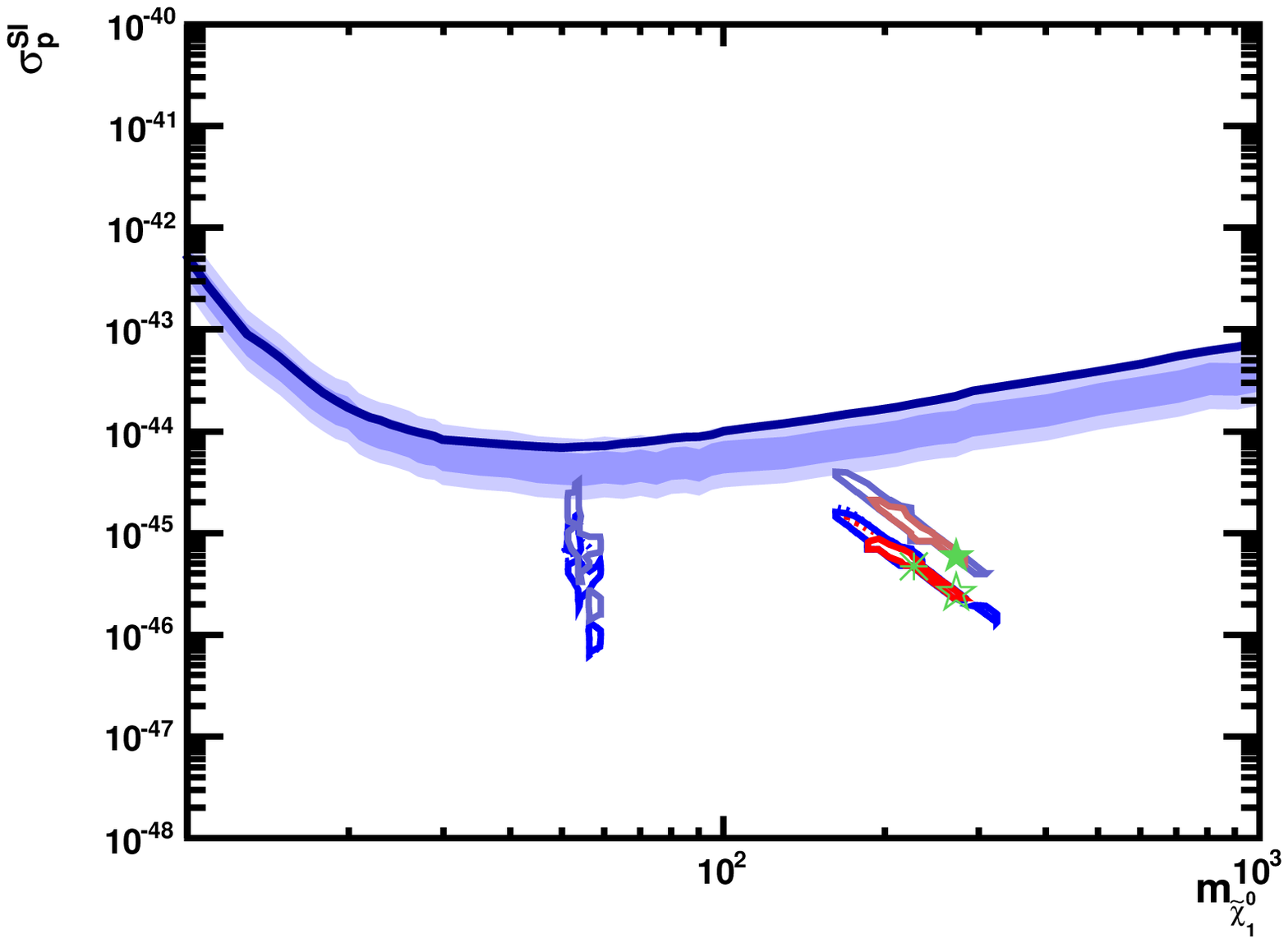}}
\vspace{-1cm}
\caption{\it The correlation between the spin-independent dark matter
scattering cross section $\ssi$ 
and $\mneu{1}$ prior to the inclusion of the current Xenon100 results
in the CMSSM (upper left panel), in the NUHM1 (upper right panel), 
in the VCMSSM (lower left panel) and in mSUGRA (lower right panel).
In each panel, we show  the 68 and 95\%~CL contours (red and blue,
respectively), the dotted curves correspond to our pre-2010-LHC results,
and the solid lines include the 2010 LHC results. 
Results assuming $\Sigma_{\pi N} = 50 \mev$ are
shown as brighter coloured curves and 
$\Sigma_{\pi N} = 64 \mev$ as duller coloured curves,
in each case disregarding uncertainties.
The green `snowflakes' (open stars) (filled
stars) are the best-fit points in the corresponding models. Also shown is the 90\% CL
Xenon100 upper limit~\protect\cite{Xenon100new} and its expected sensitivity band.
}
\label{fig:0err}
\end{figure*}

As a preface to discussing the importance of the uncertainties in the hadronic matrix elements
used in the calculation of $\ssi$, we first display results that ignore these uncertainties.
In Fig.~\ref{fig:0err}, we show our previous pre-LHC, pre-Xenon100 results in the 
$(\mneu{1}, \ssi)$ plane assuming $\Sigma_{\pi N} = 50 \mev$ as dotted curves, and  
post-LHC but still pre-Xenon100 results (again assuming $\Sigma_{\pi N} = 50 \mev$) as dashed curves
(red for 68\% CLs and blue for 95\% CLs), as calculated using {\tt SSARD}~\cite{SSARD}. 
We also show the corresponding predictions with the 
higher value $\Sigma_{\pi N} = 64 \mev$ as duller coloured curves. The current Xenon100 results
were not used in making these predictions, and we display separately the 95\% CL limit on the cross section as a function of 
$\mneu{1}$ as well as the sensitivity bands from \cite{Xenon100new}. We see three important effects in these
plots. One is that the 2010 LHC results push the predicted region in the $(\mneu{1}, \ssi)$ plane to higher masses,
but not to very much lower values of $\ssi$. The second effect is that the new Xenon100 constraint
intersects the regions favoured in our pre- and post-2010-LHC analyses of the CMSSM and NUHM1. The third effect is that of
the value of $\Sigma_{\pi N}$, which changes the predicted range of $\ssi$ by a factor $\sim 3$.
The combination of these two latter effects means that any combination of accelerator and
Xenon100 results must take careful account of the uncertainty in $\Sigma_{\pi N}$.

\begin{figure*}[htb!]
\resizebox{8cm}{!}{\includegraphics{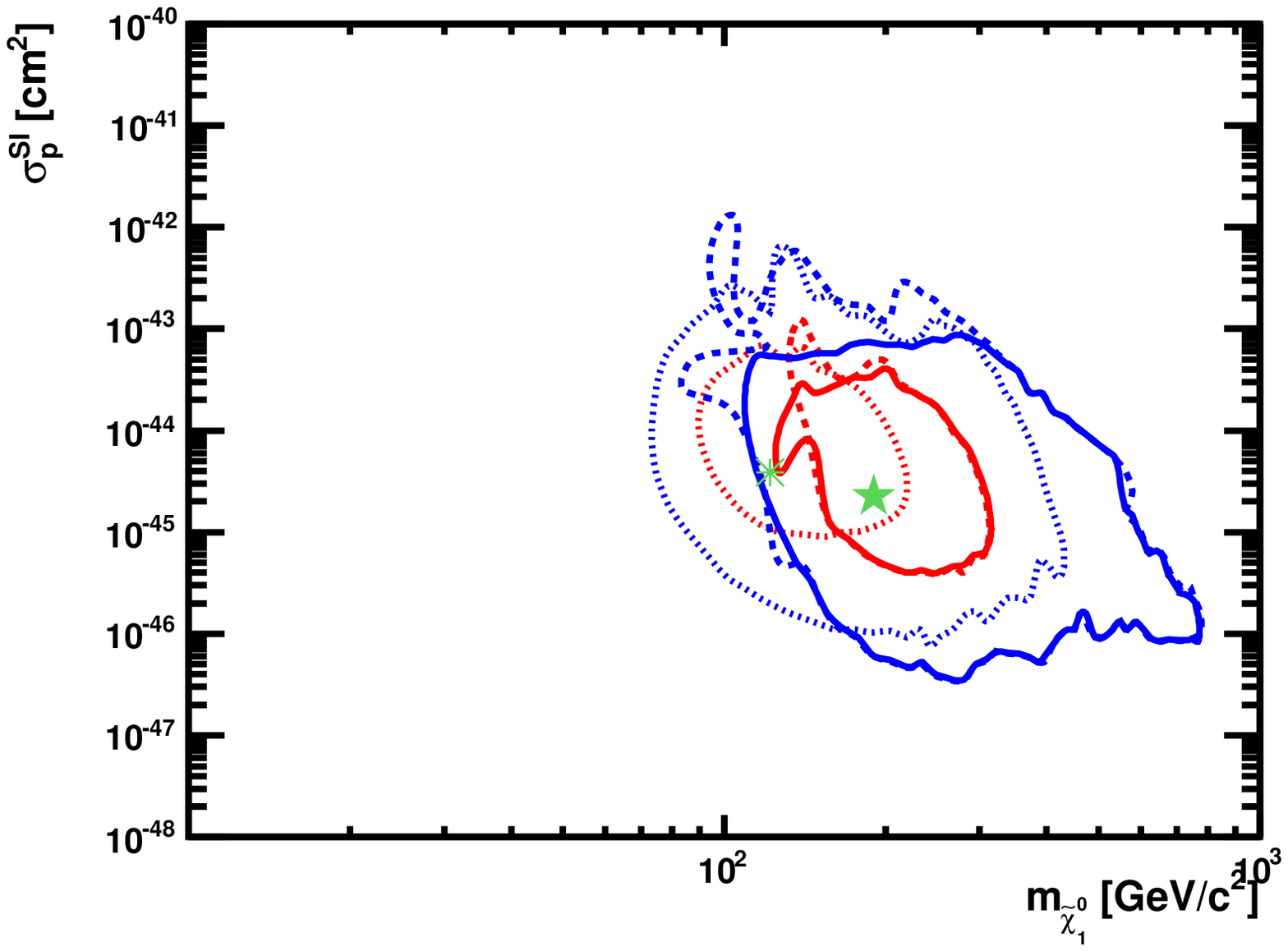}}
\resizebox{8cm}{!}{\includegraphics{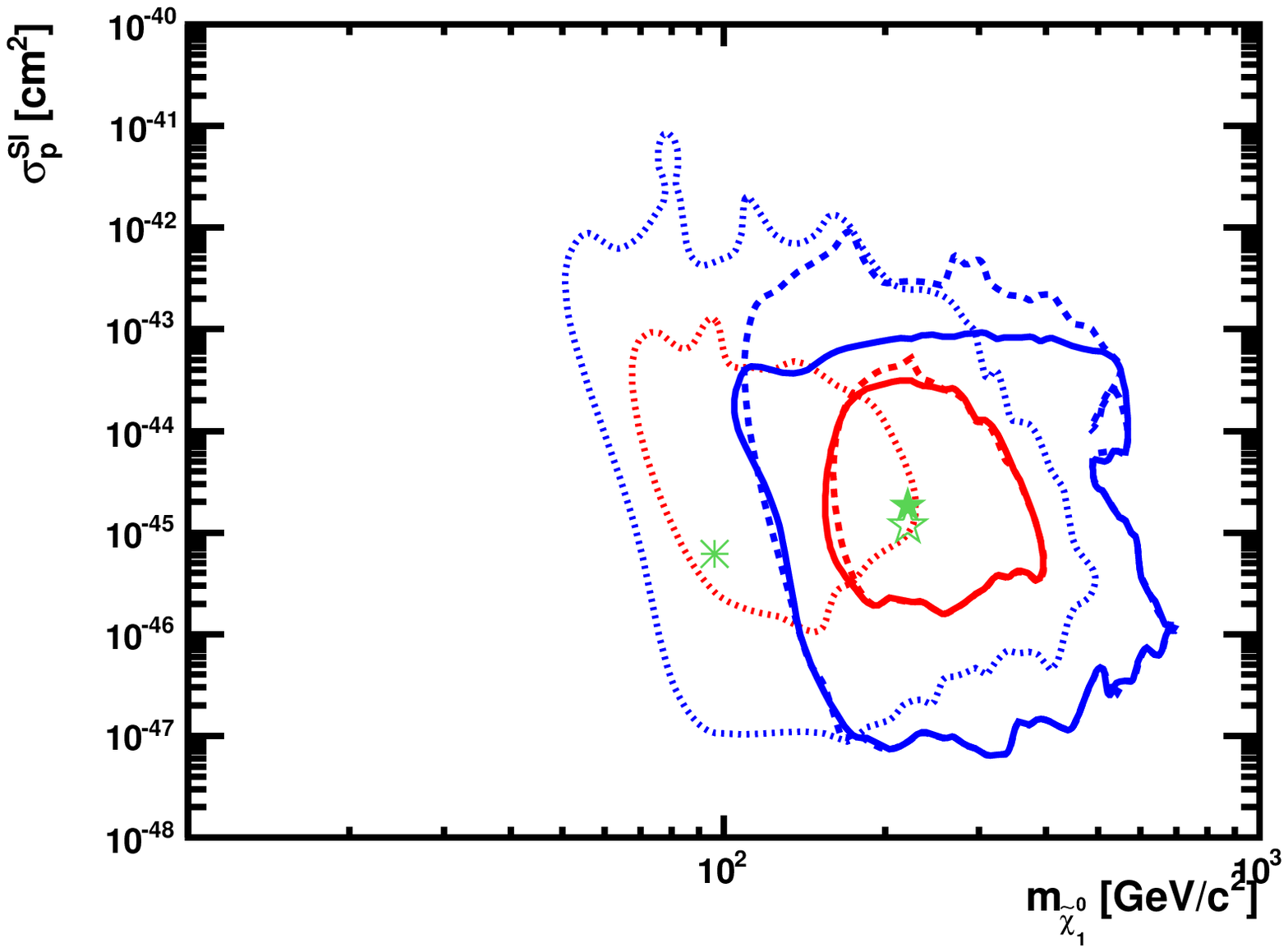}}
\resizebox{8cm}{!}{\includegraphics{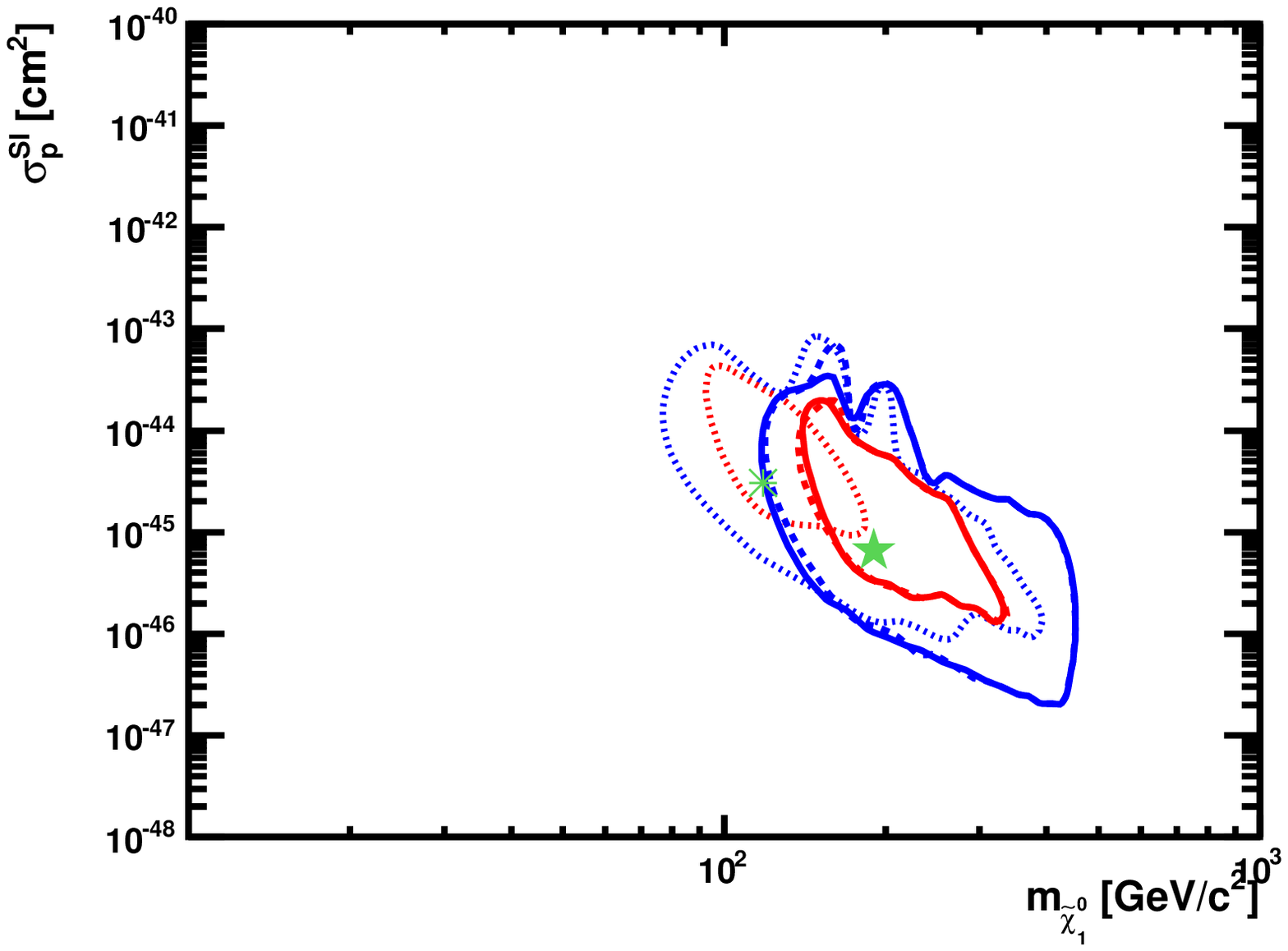}}
\resizebox{8cm}{!}{\includegraphics{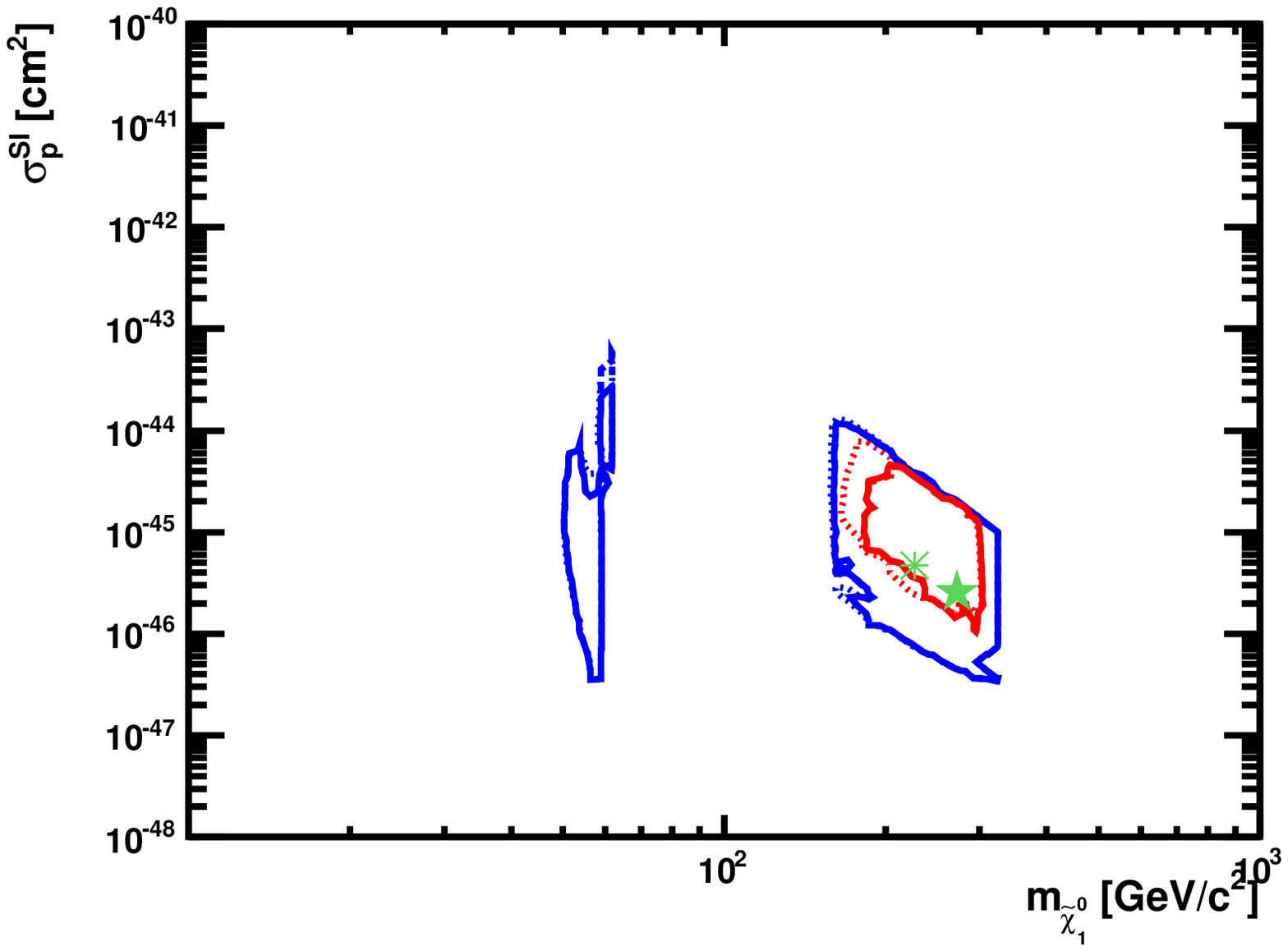}}
\vspace{-1cm}
\caption{\it The correlation between $\mneu{1}$ and the spin-independent dark matter
scattering cross section $\ssi$ 
calculated assuming a $\pi$-N scattering $\sigma$ term
$\Sigma_{\pi N} = 50 \pm 14 \mev$
in the CMSSM (upper left panel), in the NUHM1 (upper right panel), 
in the VCMSSM (lower left panel) and in mSUGRA (lower right panel).
In each panel, we show as solid (dashed) lines the 68 and 95\%~CL contours (red and blue,
respectively) after (before) applying the
Xenon100~\protect\cite{Xenon100new} constraint. The green filled (open)
stars are the best-fit points in each model including (excluding) the
Xenon100 data. Also shown are best fit and 68 and 95\%~CL contours 
obtained from the pre-2010-LHC data set excluding the Xenon100
result (green `snowflake' and dotted lines).
}
\label{fig:mneu1ssi}
\end{figure*}

We now discuss the combination of the LHC and Xenon100 constraints in the
$(\mneu{1}, \ssi)$ planes when the uncertainties in 
the hadronic matrix element $\Sigma_{\pi N}$ are included, as shown in Fig.~\ref{fig:mneu1ssi}.
As usual, the dotted lines are pre-LHC and Xenon100, the dashed lines are
post-2010-LHC but pre-Xenon100, and the solid lines incorporate also the
Xenon100 constraint, with our default assumption $\Sigma_{\pi N} = 50 \pm 14 \mev$~%
\footnote{These planes cannot be compared directly to those in~\cite{mc4,mc5}, because here
we use the {\tt SSARD} code~\cite{SSARD} to evaluate $\ssi$. This allows a more complete treatment of
different contributions to the scattering rates than does {\tt MicrOMEGAs}, 
including important uncertainties in the hadron scattering matrix elements~\cite{Ellis:2008hf}.
These lead, in particular, to larger ranges of $\ssi$ for fixed values of $\mneu{1}$.
We note in passing that {\tt MicrOMEGAs}~\cite{Belanger:2006is} uses
$\Sigma_{\pi N} = 55 \mev$ as a default.}.
In the absence of the Xenon100 constraint,
the LHC would have allowed values of $\ssi$ as 
large as $\sim 10^{-43}$~cm$^2$ at the 95\% CL in the
CMSSM and NUHM1, as seen in Fig.~\ref{fig:0err}, 
whereas only values below $\sim 10^{-44}$~cm$^2$
would have been expected at the 95\% CL in the VCMSSM and mSUGRA.
Since Xenon100 imposes $\ssi < 5 \times 10^{-44}$~cm$^2$ for
$\mneu{1} \sim 200 \gev$, this constraint
has significant impact in the CMSSM and NUHM1, as one could expect.

\begin{figure*}[htb!]
\resizebox{8cm}{!}{\includegraphics{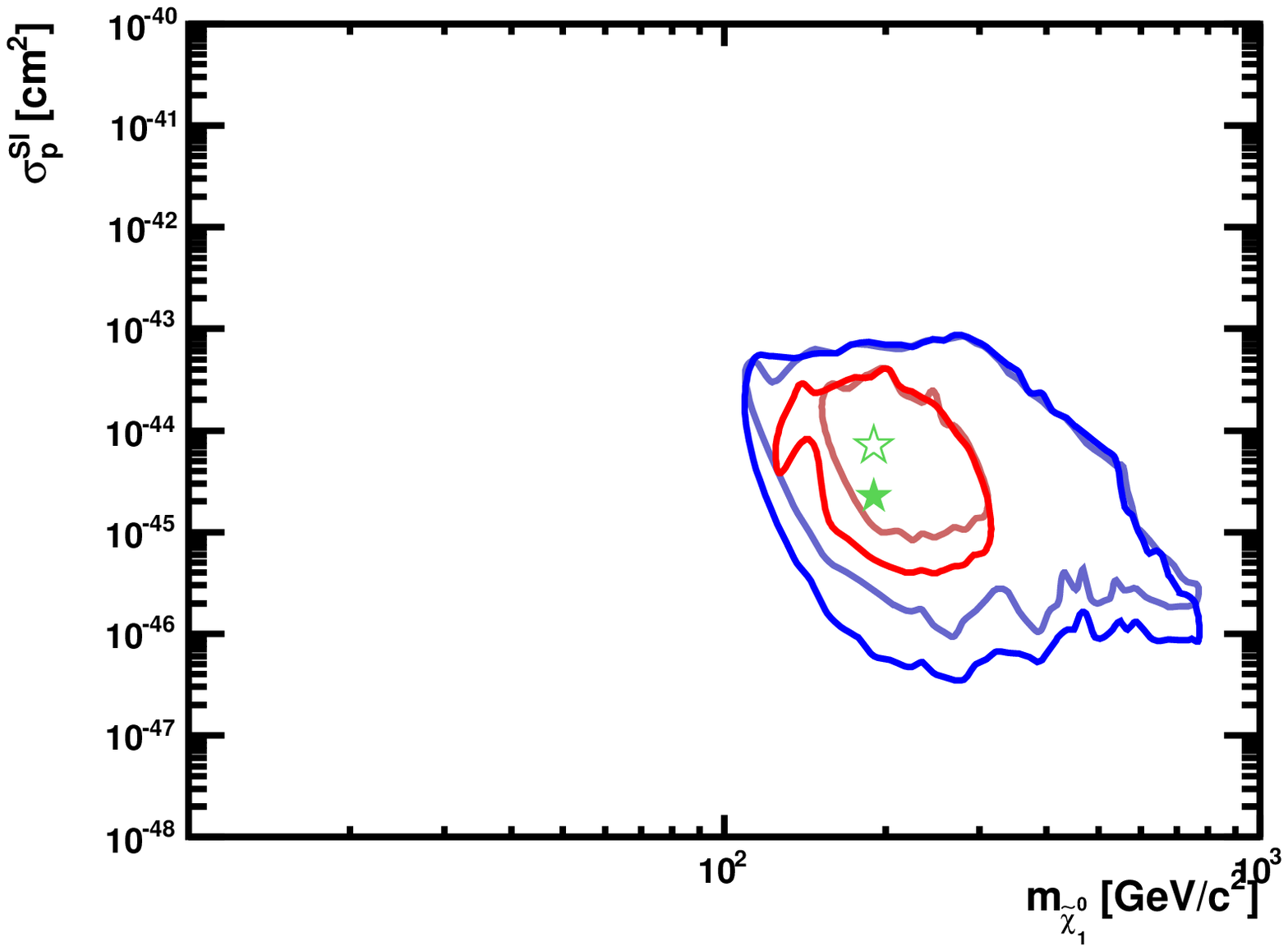}}
\resizebox{8cm}{!}{\includegraphics{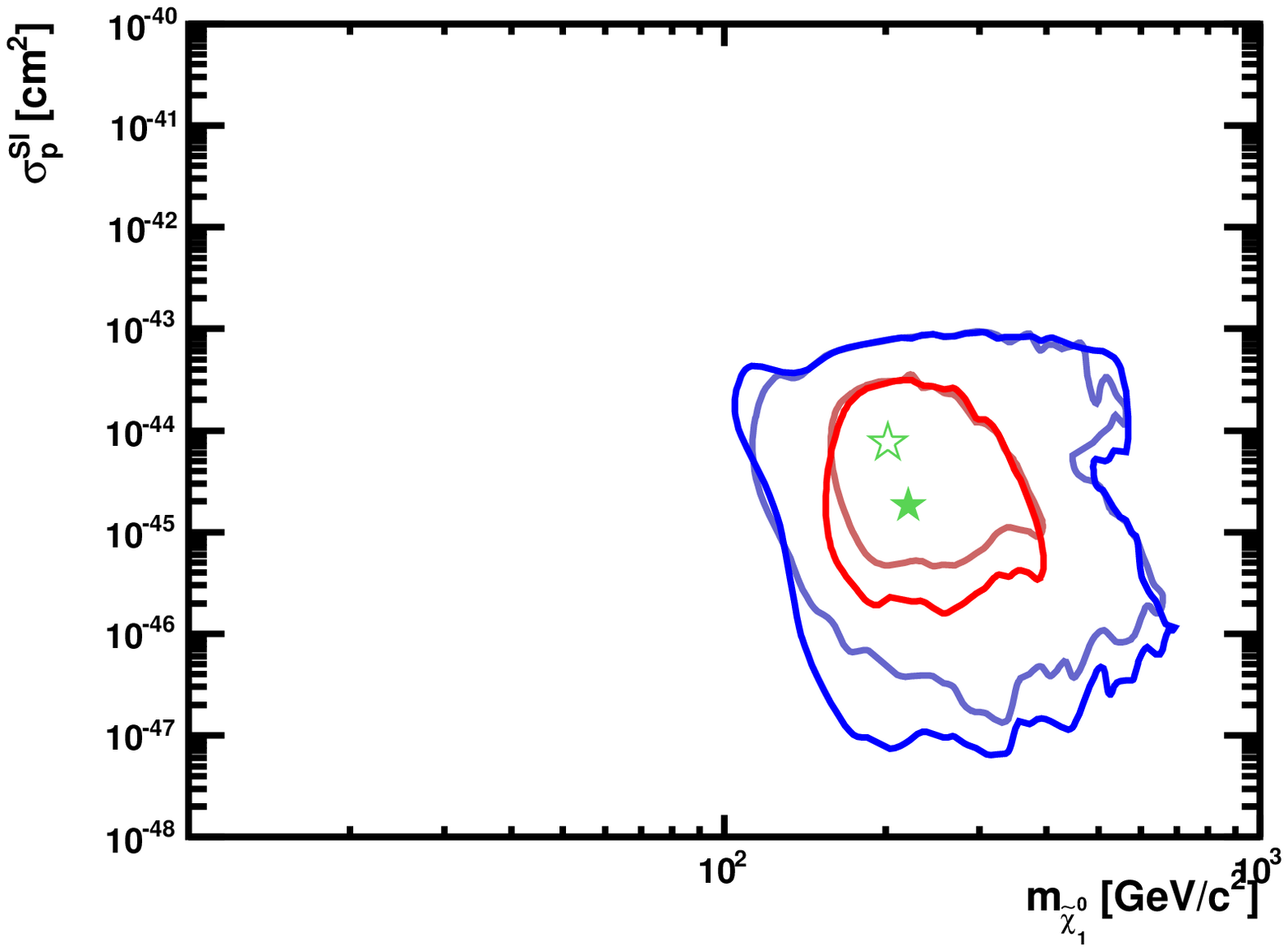}}
\resizebox{8cm}{!}{\includegraphics{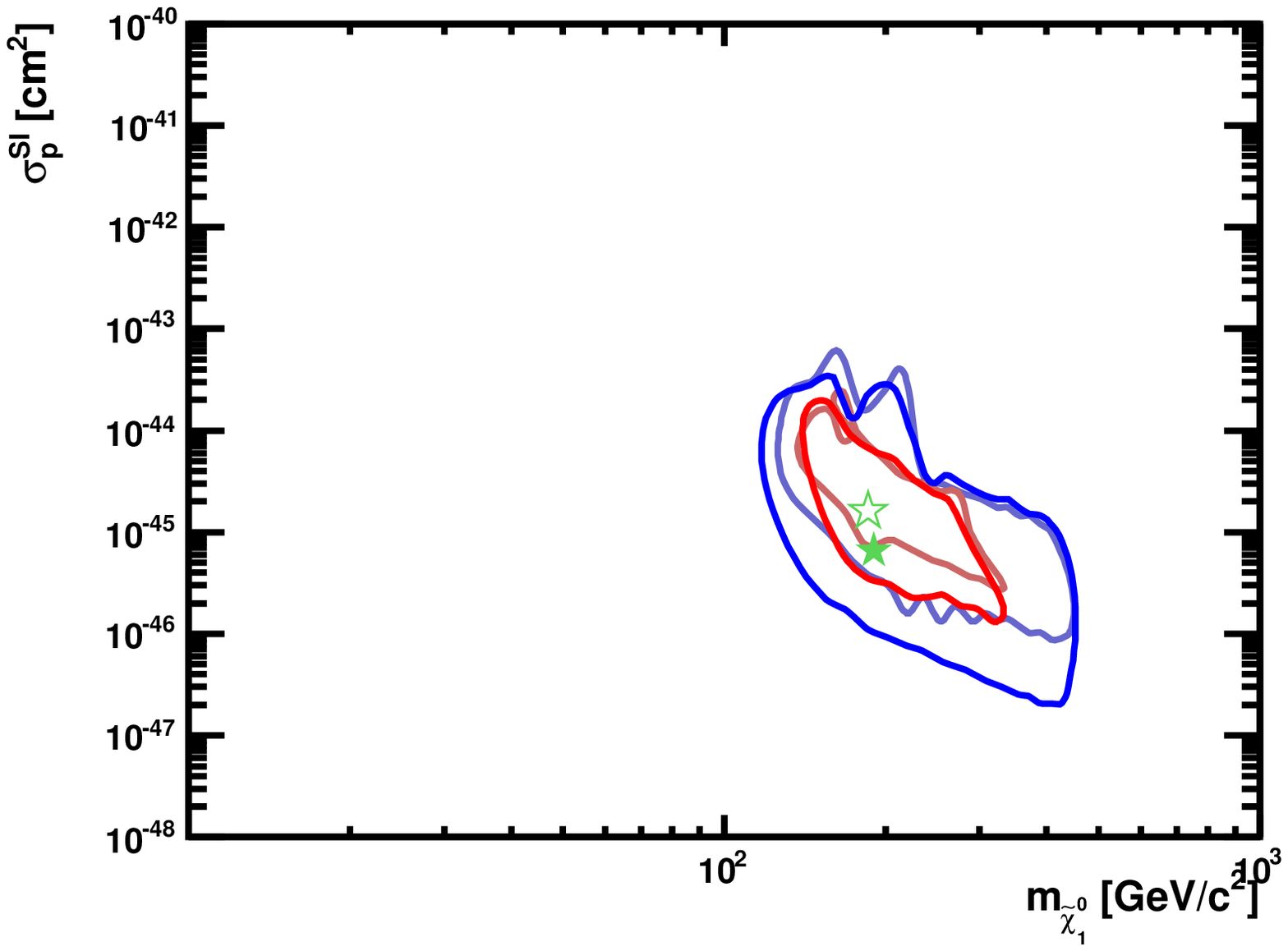}}
\resizebox{8cm}{!}{\includegraphics{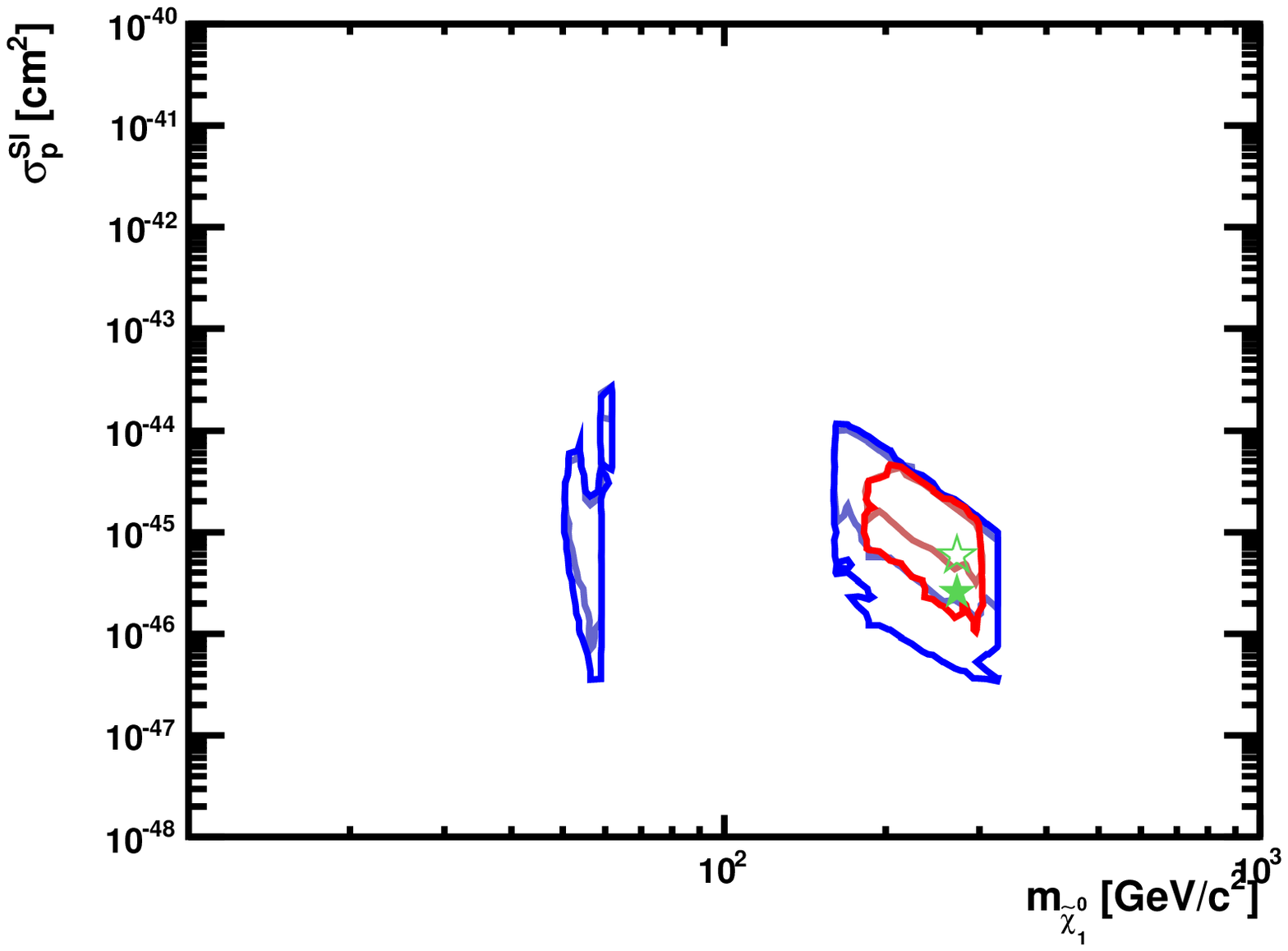}}
\vspace{-1cm}
\caption{\it The correlation between the spin-independent dark matter
scattering cross section $\ssi$ 
and $\mneu{1}$ after including the current Xenon100 results
in the CMSSM (upper left panel), in the NUHM1 (upper right panel), 
in the VCMSSM (lower left panel) and in mSUGRA (lower right panel).
In each panel, we show  the 68 and 95\%~CL contours as solid red and blue
lines, respectively. Results assuming $\Sigma_{\pi N} = 50 \pm 14 \mev$ 
are shown as brighter coloured curves, and those for 
$\Sigma_{\pi N} = 64 \pm 8 \mev$ are shown as duller coloured curves.
The green filled (open) stars are the best-fit points in each case.
}
\label{fig:5064}
\end{figure*}

In Fig.~\ref{fig:5064} we compare our predictions for $\ssi$ after incorporation of the 2010
LHC data set and the Xenon100 constraint, for two different choices of
$\Sigma_{\pi N} = 50 \pm 14 \mev$ (our default choice, shown in brighter colours) and 
$64 \pm 8 \mev$ (a less conservative choice, shown in duller colours). 
As usual, the upper
left panel shows predictions for the CMSSM, the upper right panel for is the NUHM1, the 
lower left panel shows the VCMSSM, and the lower right panel is for mSUGRA, and the
68\% (95\%) CL regions are indicated by solid red (blue) contours. 
In all models, we find that the upper limits on $\ssi$ are rather independent
of the value assumed for $\Sigma_{\pi N}$. However, the lower bounds on
$\ssi$ are quite different for our default assumption $\Sigma_{\pi N} = 50 \pm 14 \mev$
and the comparison choice $\Sigma_{\pi N} = 64 \pm 8 \mev$, differing by a factor $\sim 3$.
This means the interpretation of future direct dark matter search constraints
will be hamstrung by this uncertainty.

\subsection*{\it Spin-dependent dark matter scattering}

\begin{figure*}[htb!]
\resizebox{8cm}{!}{\includegraphics{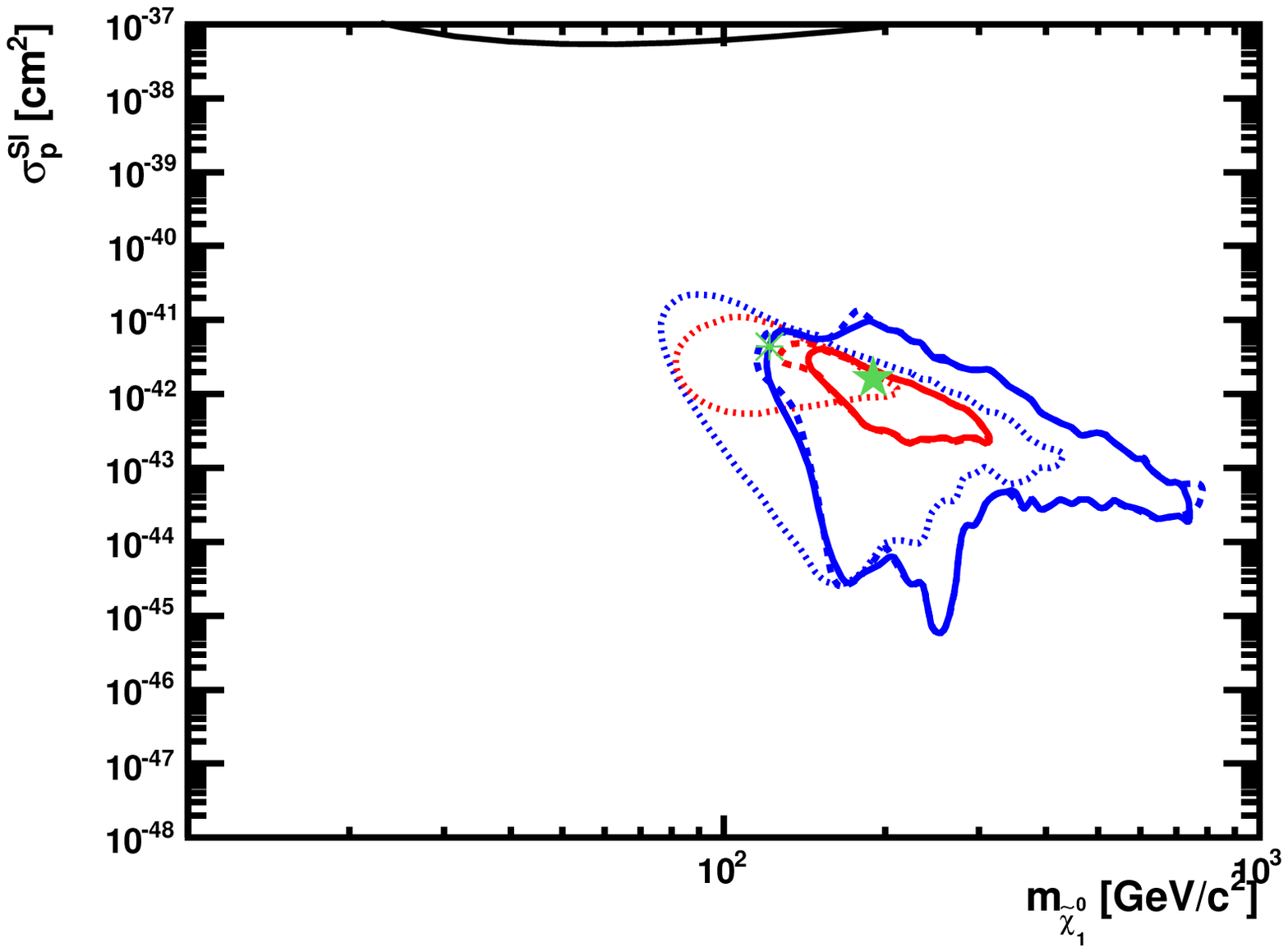}}
\resizebox{8cm}{!}{\includegraphics{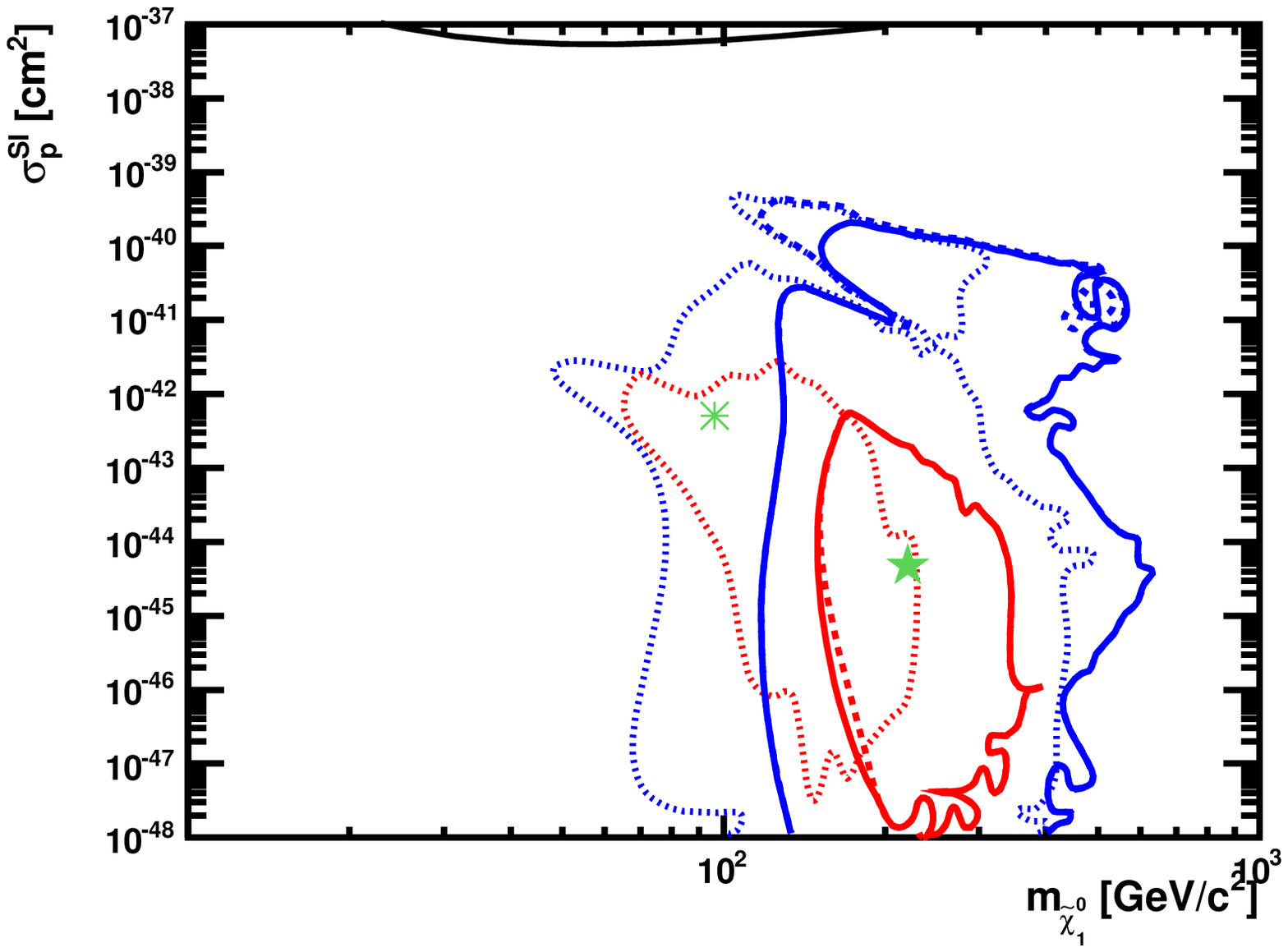}}
\resizebox{8cm}{!}{\includegraphics{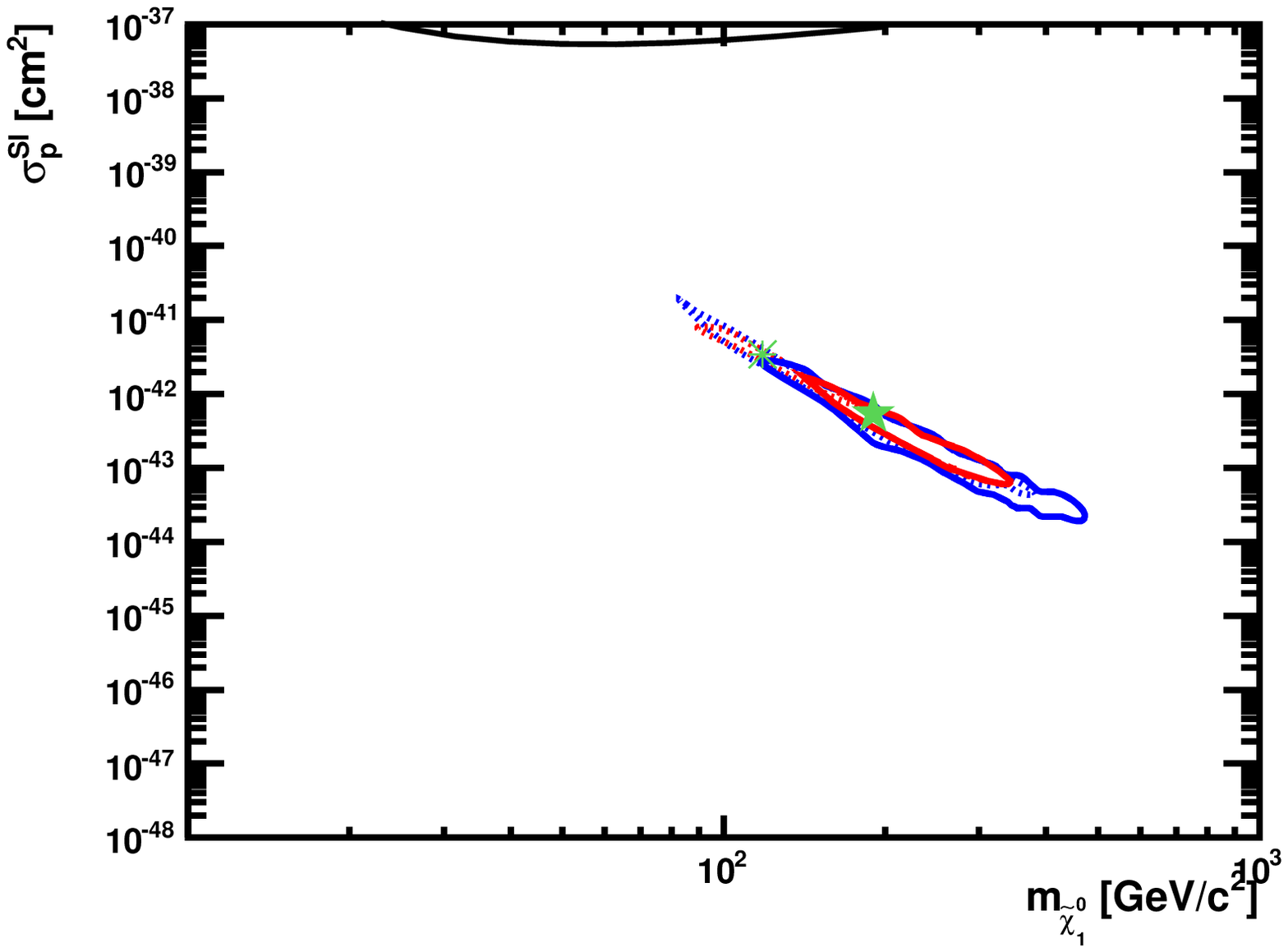}}
\resizebox{8cm}{!}{\includegraphics{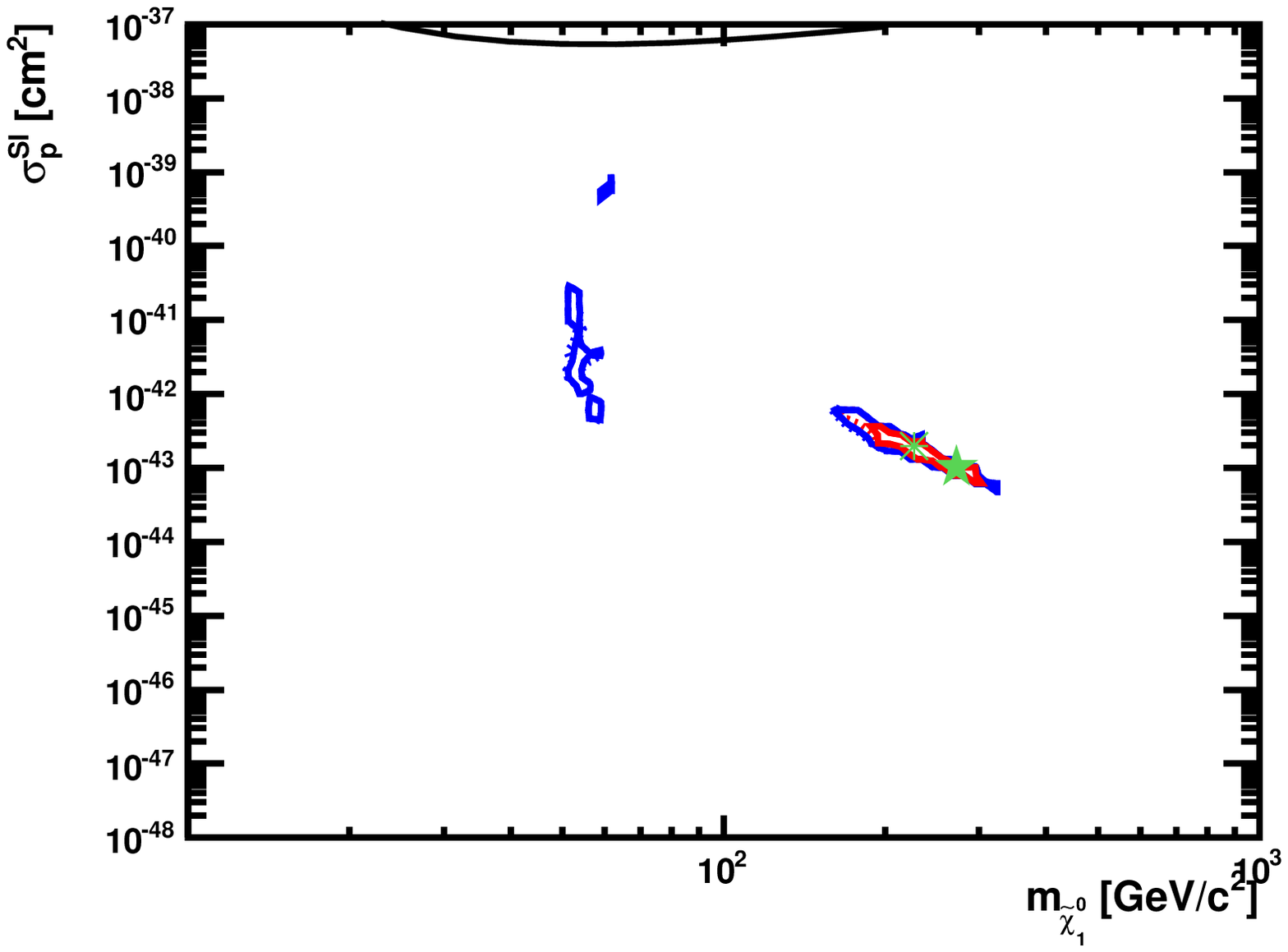}}
\vspace{-1cm}
\caption{\it The correlation between the spin-dependent dark matter
scattering cross section $\ssd$ 
(calculated assuming $\langle N | {\bar s}\gamma_\mu s | N(s) \rangle = - (0.09 \pm 0.03) \times s_\mu$,
where $s_\mu$ is the nucleon spin vector) and $\mneu{1}$
in the CMSSM (upper left panel), in the NUHM1 (upper right panel), 
in the VCMSSM (lower left panel) and in mSUGRA (lower right panel).
In each panel, we show  the 68 and 95\%~CL contours (red and blue,
respectively) including 2010 LHC data before and after applying the
Xenon100~\protect\cite{Xenon100new} constraint (solid and dashed lines, respectively). The green
filled (open) stars are the best-fit points obtained with these data sets. Also shown are best fit and the 68 and 95\%~CL contours 
obtained from fits to the pre-2010-LHC data set excluding the Xenon100
result (`snowflake', dotted lines). We also show in each panel (solid black line) the 90\% CL upper limit on $\ssd$
provided by the COUPP Collaboration~\protect\cite{COUPP}.
}
\label{fig:mneu1ssd}
\end{figure*}

The {\tt SSARD} code also provides as an output the spin-dependent
LSP-proton cross section,
$\ssd$, and we display in Fig.~\ref{fig:mneu1ssd} the predictions for $\ssd$ from our
likelihood analysis. We see that the range of $\ssd$ is much wider in the NUHM1 than
in the other models, with both larger and smaller values being possible.
Apart from the supersymmetric model parameters and the local
galactic dark matter density, which we fix here to be 0.3~GeV/cm$^3$, the principal
uncertainty in calculating $\ssd$ is the hadronic spin-dependent scattering matrix element,
which is dominated by the error in the strange axial-current matrix element, which we take
to be $\langle N | {\bar s}\gamma_\mu s | N(s) \rangle = - (0.09 \pm 0.03) \times s_\mu$,
where $s_\mu$ is the nucleon spin vector. Proportionally, the uncertainty
induced in $\ssd$ is far smaller than that induced in $\ssi$ by the error in $\Sigma_{\pi N}$.
As we see in Fig.~\ref{fig:mneu1ssd}, the most stringent direct experimental upper limit on $\ssd$
due to the COUPP Collaboration~\cite{COUPP} (solid black line) lies above $10^{-38}$~cm$^2$, 
significantly higher than our predictions in any of the CMSSM, NUHM1,
VCMSSM and mSUGRA. More stringent upper limits on $\ssd$ are sometimes quoted based on
experimental upper limits on energetic solar neutrinos that could be generated
by LSP annihilations inside the sun~\cite{IceCube}. These upper limits often assume that the LSPs are mainly
captured by spin-dependent scattering, which is not the case in general, and are
in equilibrium inside the sun, which is also not the case in general~\cite{Savagenu}. They also make simplifying assumptions
about the annihilation final states that are not in general valid in the specific models studied here.
Even with these assumptions, the upper limits lie above the ranges we
predict in the CMSSM, VCMSSM and mSUGRA and barely touch the NUHM1
range. Therefore, a direct confrontation of these models
with data on energetic solar neutrinos still lies in the future.

\section{Summary and Discussion}

We have explored in this paper the implications of the 2010 LHC data for some of the simplest realizations of the MSSM, namely the CMSSM, the NUHM1,
the VCMSSM and mSUGRA. In addition to the most sensitive available
ATLAS and CMS searches for jets + $\ETslash$, we have incorporated the
constraints imposed by searches for the heavy MSSM Higgs bosons
$H/A \to \tau^+ \tau^-$ and the constraints imposed by LHCb, CDF and D\O\
on \bmm, and we have also explored the impact of the direct Xenon100 search
for dark matter scattering.

We have found that the ATLAS 0L and CMS MHT analyses shift the
preferred regions in the $(m_0, m_{1/2})$ planes as compared to the 
situation based on the initial CMS $\alpha_T$ and ATLAS 1L searches
by amounts similar to those observed when comparing the results
incorporating the initial CMS $\alpha_T$ and ATLAS 1L searches with the 
pre-LHC situation. As a consequence, the preferred value of $\mgl$ 
has been shifted upwards to  1~TeV and beyond in the CMSSM, NUHM1 and 
VCMSSM. On
the other hand, the picture in mSUGRA is not changed significantly by the
newer ATLAS and CMS searches.

The CMS limits on heavy Higgs production and our compilation of LHCb,
CDF and D\O\ constraints on \bmm\ have impacts on the parameter spaces
of the NUHM1, but do not affect significantly the favoured regions of the
CMSSM, VCMSSM and mSUGRA.

The Xenon100 results have an impact on the model parameter spaces
that would be significant if $\Sigma_{\pi N}$ were large, $\sim 60 \mev$.
However, the current uncertainty in $\Sigma_{\pi N}$ does not permit a strong
conclusion to be drawn, and we emphasize again the importance of
experimental and theoretical attempts to reduce this uncertainty.

The adventure of the LHC search for SUSY has only just
begun in 2010. 
The negative results of the searches to date are not in serious
tension with the ranges of parameter spaces favoured pre-LHC in the
models we have studied. The favoured regions yet to be explored
offer good prospects for the SUSY searches during the 
LHC run in 2011/12. However, it is worthwhile to
consider whether the exclusion by the LHC
of very light squark and gluino masses may already have messages for future
experimental studies of supersymmetry (if it exists).

There is much discussion about the possible next large collider
project to follow the LHC, with high-energy lepton colliders among the
favourites. A key question is the centre-of-mass energy of such a
collider, and indications from the LHC are eagerly awaited. Any
definitive statement on the impact of LHC results must surely wait 
at least until the
end of the 2011/12 LHC run, and will require analyses that are less
model-specific than the results presented up to now. 
In this respect it has to be kept
in mind that the LHC searches are mainly sensitive to the 
production of coloured particles, whereas
lepton colliders will have a high sensitivity in particular for the 
production of colour-neutral states, such as
sleptons, charginos and neutralinos (and high-precision measurements
furthermore provide an indirect sensitivity to quantum effects of new
states). In this sense anything inferred
from the coloured sector on the uncoloured sector depends
on the underlying model assumptions, and in particular on 
assumptions about a possible universality of soft SUSY-breaking at the
GUT scale.

The upward shifts for the preferred
values of $m_{1/2}$ and, to a lesser extent, $m_0$, 
that we have found in the CMSSM, NUHM1 and VCMSSM
upon inclusion of the 2010 LHC and Xenon100 constraints 
translate within those models into corresponding shifts in the 
production thresholds of supersymmetric particles at $e^+e^-$ colliders.
It has to be noted, however, that together with this upward shift of the 
preferred mass values we observe a significant decrease in the fit
probabilities of those simple models, see Table~\ref{tab:compare}.
This indicates a slight tension in those models between the preference 
for rather light colour-neutral states arising in particular from 
\gmt\ and the search limits from the direct searches for 
coloured SUSY particles at the LHC. The mSUGRA scenario yields a 
significantly worse description of the data than the other considered 
models already for the pre-LHC data set, and inclusion of the 2010 LHC
and Xenon100 constraints has only a small impact on the preferred fit
values and the fit probabilities.
If the upcoming LHC results lead to
a further increase of the excluded mass regions for coloured
superpartners, the CMSSM, NUHM1 and VCMSSM scenarios 
could eventually get under pressure. Such a tension could be avoided in
realisations of SUSY with a larger splitting between the coloured and
the colour-neutral part of the spectrum (for instance in GMSB-type
scenarios), such that the masses 
of squarks and gluinos are in the TeV range, while sleptons, neutralinos
and charginos can still be light.

Alternatively, the spectrum could be compressed,
decreasing the splitting between the coloured and colour-neutral sparticles,
leading to softer jets from gluino and squark decays, and hence less
stringent constraints form searches for jets + missing transverse energy
at the LHC~\cite{stealth}.

\subsubsection*{Noted Added}

\begin{figure*}[htb!]
\resizebox{8cm}{!}{\includegraphics{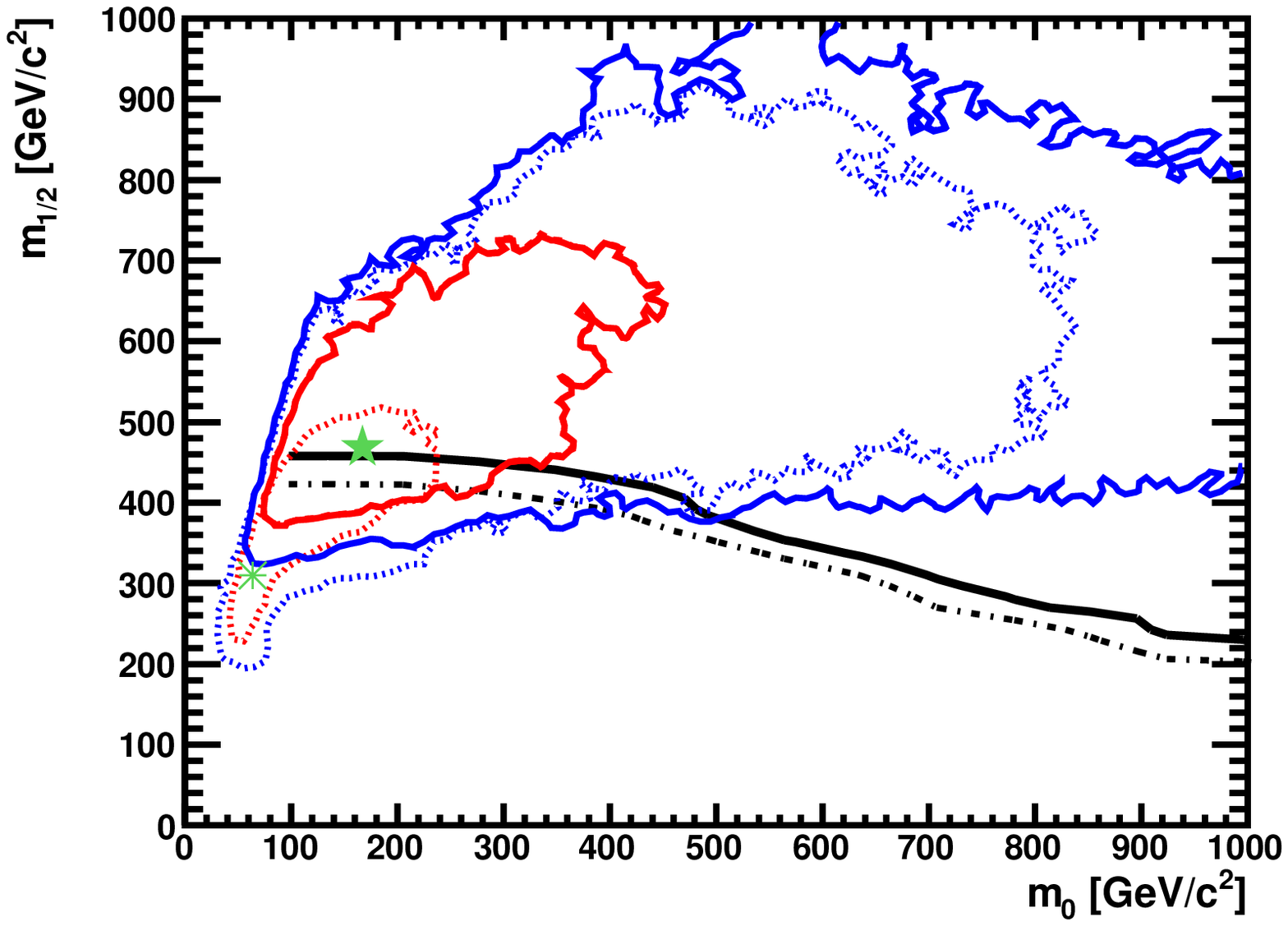}}
\resizebox{8cm}{!}{\includegraphics{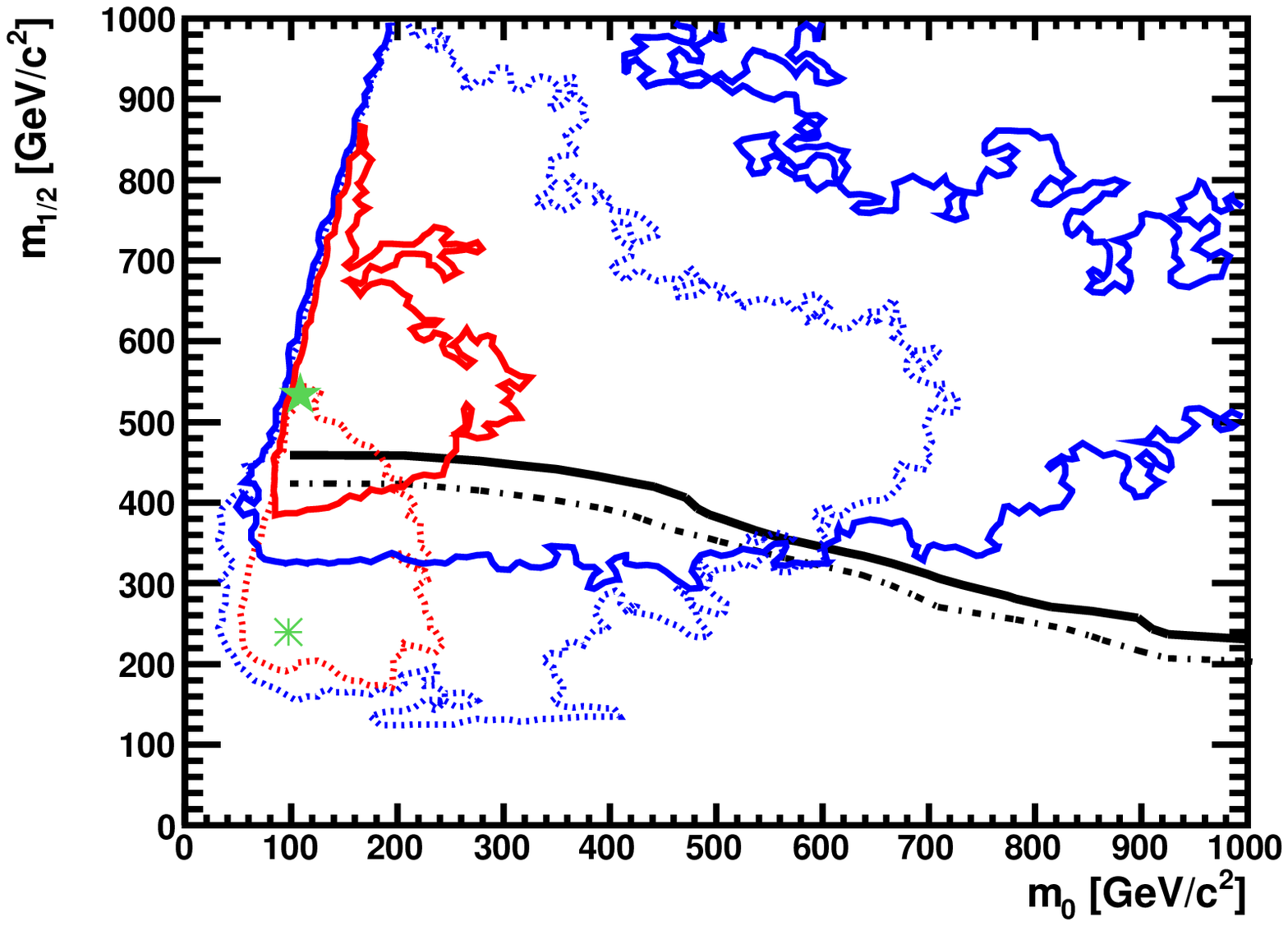}}
\resizebox{8cm}{!}{\includegraphics{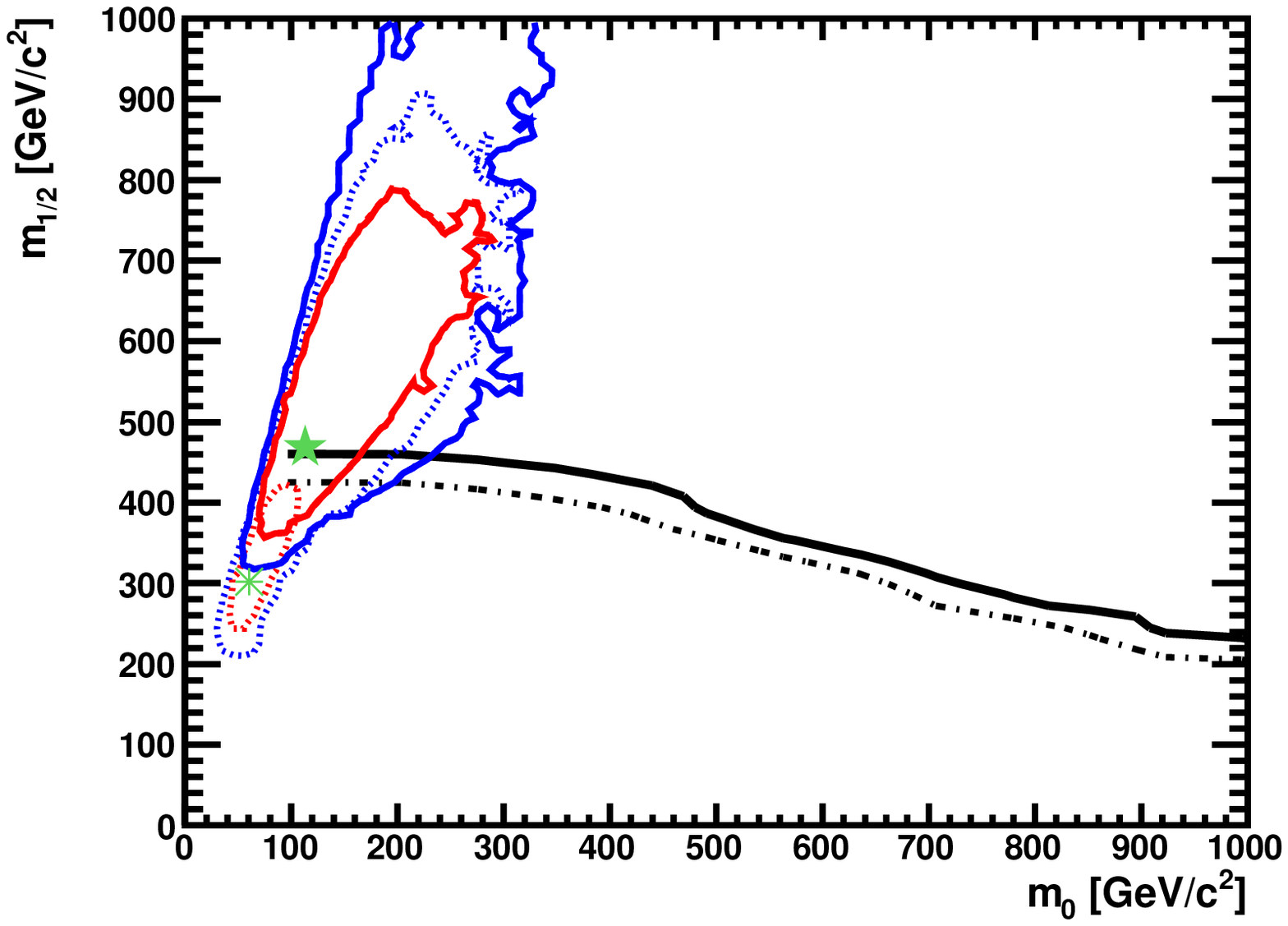}}
\resizebox{8cm}{!}{\includegraphics{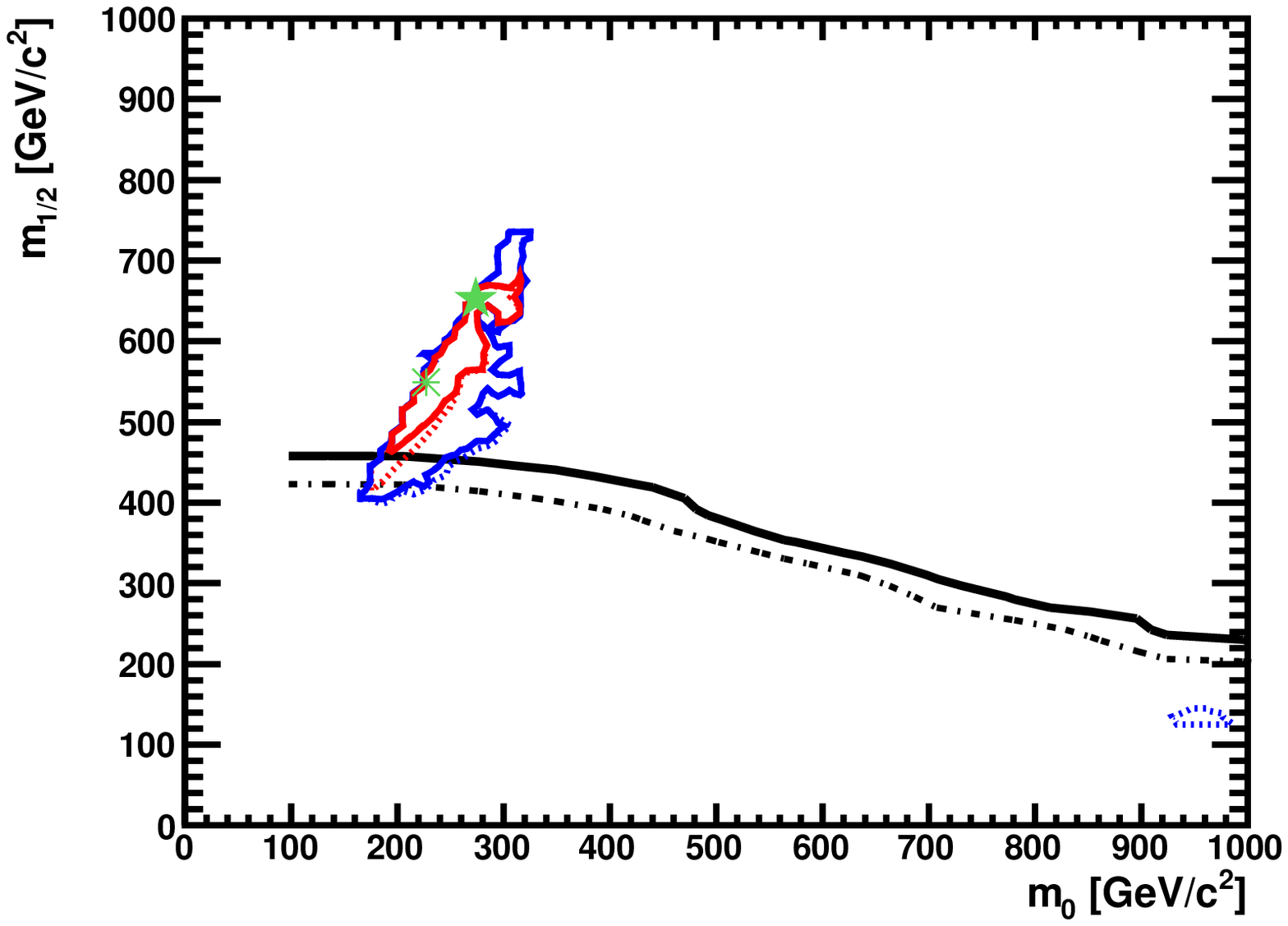}}
\vspace{-1cm}
\caption{\it The $(m_0, m_{1/2})$ planes in the CMSSM (upper left), the
NUHM1 (upper right), the VCMSSM (lower left) and mSUGRA (lower
right) as shown in Fig.\protect\ref{fig:m0m12}, now superposed by the 
preliminary 95\% CL limits obtained by the ATLAS Collaboration~\protect\cite{ATLAS165}
using a PCL approach (solid black lines) and a CL$_s$ approach (dash-dotted black lines).
}
\label{fig:165}
\end{figure*}

The analysis of this paper provides a baseline with which 2011 LHC data
can be confronted.
While completing this work, we became aware of preliminary results 
from an analysis of events with $\ge 2$ jets, missing transverse
energy and no detected leptons obtained with 165/pb of 2011 ATLAS data~\cite{ATLAS165}. 
In Fig.~\ref{fig:165} we superpose on the $(m_0, m_{1/2})$ planes shown previously in
Fig.~\ref{fig:m0m12} the preliminary 95\% CL limits obtained using a PCL
approach (solid black line) and a CL$_s$ approach (dash-dotted black line).
We see that in the CMSSM (upper left) and VCMSSM (lower left)
the new preliminary PCL 95\% contour runs very close to
the best-fit point we find with the combined 2010 LHC data. It runs somewhat
further away from the NUHM1 best-fit point, and outside our 68\% CL contour
for mSUGRA. The preliminary CL$_s$ contour runs further
below our best-fit points, still through our 68\% CL regions for the CMSSM,
NUHM1 and VCMSSM but below our 68\% CL region for mSUGRA.
We defer incorporating this result into our analysis until a final version is
published that enables its contribution to the global likelihood function to
be modelled. However, this preliminary result already highlights the potential
of the 2011 LHC run to probe deeper into supersymmetric parameter space.\\

\subsubsection*{Acknowledgements}

The work of O.B., M.J.D. and J.E. is supported partly by the London
Centre for Terauniverse Studies (LCTS), using funding from the European
Research Council 
via the Advanced Investigator Grant 267352. 
The work of S.H. was supported 
in part by CICYT (grant FPA 2010--22163-C02-01) and by the
Spanish MICINN's Consolider-Ingenio 2010 Program under grant MultiDark
CSD2009-00064. 
The work of K.A.O. was supported in part
by DOE grant DE--FG02--94ER--40823 at the University of Minnesota.
K.A.O. also thanks SLAC 
(supported by the DOE under contract number DE-AC02-76SF00515) and 
the Stanford Institute for Theoretical Physics
for their hospitality and support. M.J.D. thanks CERN for hospitality during the completion of this work.


\end{document}